\documentclass[structabstract]{aa}  

\usepackage{graphicx}
\usepackage[breaklinks, colorlinks, urlcolor=Cerulean, citecolor=Cerulean,linkcolor=Purple]{hyperref}
\usepackage[usenames,dvipsnames,svgnames,table]{xcolor}
\usepackage{array,booktabs,mathtools,caption}
\usepackage{epsf}
\usepackage{pstricks}
\usepackage{pst-plot}
\usepackage{pstricks-add}
\usepackage{epstopdf}
\usepackage{psfrag,xcolor}
\usepackage{amsmath} 
\usepackage{algorithm}
\usepackage{enumitem}
\makeatletter
\def\BState{\State\hskip-\ALG@thistlm}
\makeatother
\usepackage{natbib}

\def\Mpc{\,$h_{70}^{-1}$\,Mpc\;}
\def\Mpcp{\,$h_{70}^{-1}$\,Mpc}

\usepackage{longtable}
\usepackage[normalem]{ulem}

\usepackage{txfonts}
\begin{document}
\title{Identification of filamentary structures in the environment of superclusters 
	of galaxies in the Local Universe}
\subtitle{}

\author{Iris Santiago-Bautista \inst{1,2}
	\and C\'esar A. Caretta \inst{2}
	\and H\'ector Bravo-Alfaro \inst{2}
	\and Etienne Pointecouteau \inst{1}
        \and Heinz Andernach \inst{2}}

\institute{
	IRAP, Universit\`e de Toulouse, CNRS/CNES/UPS, Toulouse, France \\
	\email{isantiago@irap.omp.eu}
	\and
	Departamento de Astronom\'ia, DCNE-CGT, Universidad de Guanajuato, 
	CP 36023, Guanajuato, Mexico \\
	\email{isantiago@astro.ugto.mx}
}

\date{Draft, 2020}

\abstract
{The characterization of the internal structure of the superclusters
	of galaxies (walls, filaments and knots where the clusters 
        {are located})
	is paramount for understanding the formation of the Large Scale
	Structure and for outlining the environment where galaxies evolved
	in the last gigayears.}
{(i) To detect the compact regions {of high relative density}
        (clusters and rich groups of galaxies);
	(ii) to map the elongated structures {of low relative
        density} (filaments{, bridges and tendrils} of galaxies);
	(iii) to characterize {the} galaxy populations {on
        filaments} and study the environmental effects they {are
        subject to}.}
{We employed optical galaxies with spectroscopic redshifts from the SDSS-DR13
	inside {rectangular} boxes encompassing {the volumes of} 
        a sample of 46 superclusters of galaxies, up to $z$ = 0.15.
	A virial approximation was applied to correct the positions of the 
        galaxies in the redshift space for the ``finger of God'' projection 
        effect. 
        Our methodology implements different classical pattern recognition 
        and machine learning techniques (Voronoi tessellation, hierarchical 
        clustering, graph-network theory, minimum spanning trees, among 
        others), pipelined in {the Galaxy Systems-Finding 
        algorithm and the Galaxy Filaments-Finding algorithm}.}
{{We detected in total 2\,705 galaxy systems (clusters and 
        groups, of which 159 are new) and 144 galaxy filaments in the 46 
        superclusters} of galaxies. 
        The filaments we detected have a density contrast above 3, with a 
        mean value around 10, a radius of about 2.5 \Mpc and 
        {lengths} between 9 and 130 \Mpcp. 
	Correlations between the galaxy properties (mass, morphology and 
        activity) and the environment in which they reside (systems, 
        filaments and the dispersed component) suggest that {galaxies 
        closer to} the skeleton of the filaments {are more massive 
        by} up to 25\% {compared to those} in the dispersed component; 
        70\% of the galaxies in the filament region present early type 
        morphologies and the fractions of active galaxies (both AGN and SF) 
        seem to decrease as galaxies approach the filament.}
{Our results support the idea that galaxies in filaments are subject to
	environmental effects leading them to be more massive (probably due to
	larger rates of both merging and gas accretion), less active both in
	star formation and nuclear activity, and prone to the density-morphology
	relation. These results suggest that preprocessing in large scale 
        filaments could have significant effects on galaxy evolution.
}

\titlerunning{Identification of filamentary structures in the Local Universe}
\authorrunning{Santiago-Bautista et al.}
\maketitle

\section{Introduction}

The Large Scale Structure (LSS) of the Universe is composed 
of a network of groups and clusters of galaxies, elongated 
filaments, widely spread sheets, and voids 
\citep[e.g.][]{Peebles1980,Davis1982,Bond1996}.   
Both, the $\Lambda$CDM cosmological model 
\citep[e.g.][]{Bond1983, Doroshkevich1984} 
and recent numerical N-body simulations 
(e.g. \emph{Millennium}, \citeauthor{Springel2005}, \citeyear{Springel2005}; 
\emph{Bolshoi}, \citeauthor{Klypin2011}, \citeyear{Klypin2011}; 
\emph{Illustris}, \citeauthor{Illustris2014}, \citeyear{Illustris2014}),
reinforce that these structures are assembled under the effect of gravity,
generated by the total matter content. 
Since the baryonic matter follows, to first order, the distribution of 
the dark matter, the galaxies and gas populate these substructures 
accordingly \citep[e.g.][]{Eisenstein2005}. 
Moreover, there is increasing evidence that the galaxy properties 
(for instance mass, activity, morphology, luminosity, 
surface brightness, orientation, etc.) 
correlate with the LSS environment in which they are 
located \citep[e.g.][]{smargon2012,Scoville2013,Poudel2016,Kuutma2017,
Chen2017,Wang2018} or, more specifically, with the internal structure 
of the supercluster \citep[e.g.][]{Einasto2008,Gallazzi2009,Gavazzi2010,
Cybulski2014,Guglielmo2018}.
Furthermore, theoretical studies \citep[e.g.][]{cen1999} suggest that 
from one half to two thirds of the baryonic matter in the Universe is 
hidden in the filamentary structures of the LSS. 
Therefore, the characterization of the LSS (e.g. topology, density, 
temperature, dynamical state, matter distribution and its evolution along 
time) is an important step to place constraints on the current cosmological 
models.

Galaxy clusters are well studied through their gas component 
since they are the densest regions of the LSS.
However, the gas in filaments is most likely in a not so hot 
(T~$\sim~10^5$--$10^7$ K, or 0.01--1~keV) and relative 
low-density gas phase called Warm Hot Intergalactic Medium (WHIM).
There is already some evidence of such gas from X-ray emission observed within 
pairs of close clusters \citep[e.g.][]{Ursino2015, Alvarez2018}. 
In addition, the WHIM between pairs of clusters has been observed through 
the Sunyaev-Zel'dovich effect \citep[SZ, e.g.][]{Planck_filaments}. 
\cite{Tanimura2018} carried out statistical analyses using Planck 
SZ observations in the regions of superclusters. Their results show evidence
of inter-cluster gas of temperature T~$\sim 8 \times 10^6$ K.
Also, \citet{Eckert2017} presented deep X-ray observations of the galaxy 
cluster Abell 2744, the  analysis of which suggests a gas fraction of 
5\% to 15\% for the filaments that surround the cluster and a
plasma temperature of 1--2~$\times~10^7$ K. 
Therefore, the characterization of these structures through 
observables like X-ray emission or 
Sunyaev-Zel'dovich effect is still challenging due to the low density and 
temperature of the WHIM.

An alternative is to analyze the galaxy distribution at large scales.
Recently, with the availability of large sky area databases 
such as the Two Degree Field Redshift Survey \citep[2dFRS,][]{2dF2001}, 
the 2MASS Redshift Survey \citep[2MRS,][]{huchra2012} 
and the Sloan Digital Sky Survey \citep[SDSS,][]{DR13}, 
the development of accurate structure detection algorithms has become an 
even more important concern for astronomy.
Visually, the galaxy distribution shows filamentary ridge-like 
structures that connect massive clusters and groups.
However, the identification of these structures through a computational 
algorithm is not easy to achieve. 
A good algorithm should first produce an identification that resembles 
the human visual perception. It also should deliver 
quantitative results and be founded in a robust and well-defined numerical 
theory.
All of this must be done in an acceptable amount of time with reasonable 
computational resources.

Currently, there are several filaments-finding algorithms that have been 
tested on the basis of N-body simulations. 
For example, \citet{Aragon2007} present the ``multi-scale	
morphology filter method'' (\emph{MMF}) that divides cosmic structure
into nodes (clusters), filaments and walls by using a smoothing over a
range of scales (from a Delaunay tessellation reconstruction, \emph{DFTE}) 
and a morphological response filter.
Another approach presented by \cite{Aragon2010} makes use of a
watershed segmentation techniques to trace the spines of the filaments. 
Also, \cite{Cautun2013} propose an algorithm that takes into account 
the density, tidal field, velocity divergence and velocity shear of the 
galaxies, called \emph{NEXUS}. 
Other examples are the algorithms by 
\citet{Gonzales-Padilla2010},
which uses the binding energy for selecting the filament members; and
the \emph{DisPerSE} algorithm, by \citet{Sousbie2011},
based on the Morse theory -- both utilize Delaunay-Voronoi tessellation 
based on density estimations.

On the other hand, several attempts were made to trace the 
distribution of the real cosmic web using the SDSS database.
For example, the algorithm by \citet{Bond2010}, called ``smoothed Hessian 
major axis filament finder'' (\emph{SHMAFF}), was applied to the SDSS-DR6 
after removing the ``finger-of-God'' (FoG) effect.
\cite{Platen2011} compared three different reconstruction techniques,
namely the \emph{DFTE}, the ``natural neighbor field estimator'' 
(\emph{NNFE}) and a \emph{Kriging} interpolation, and searched for voids 
also in DR6.
They found that DFTE works quantitatively better than the others while the 
\emph{Kriging} and \emph{NNFE} have a better performance in 
producing visually appealing reconstructions than \emph{DFTE}.
\cite{Smith2012} applied their ``multi scale probability mapping'' 
(\emph{MSPM}), which combines probability and scale density information 
with a ``Friends-of-Friends'' (\emph{FoF}) algorithm, over the SDSS-DR7 
galaxies. 
Their method allowed them to recover structures from clusters to 
filaments of up to $\sim$10\,$h^{-1}$\,Mpc.
\cite{Tempel2014} applied a \emph{Bisous} model on the 
SDSS-DR8 spectroscopic galaxies to trace the filament spines. 
Their method adjusts cylinders to the galaxy positions applying a 
stochastic metric.
The ``subspace constrained mean shift'' (\emph{SCMS}) approach, which uses a
``kernel density estimator'' (\emph{KDE}), was applied by 
\cite{Chen2016_method} to DR7 and by \citet{Chen2015orig} to DR12.
This method allows the identification of high density regions by smoothing 
the galaxy distribution. They apply this technique over slices of 0.05 in
redshift for the SDSS sky area.
Moreover, \cite{Alpaslan2014} found, for the GAMA survey, that there are fine 
filaments embedded inside the SDSS voids. These structures, 'tendrils', have 
a lower density than the SDSS filaments and appear to be morphologically 
distinct, they are more isolated and span shorter distances. 
{A comprehensive review and comparative analysis of these
algorithms can be found in \citet{Libeskind2018}.}

Another approach to analyze the LSS structure is to study the superclusters 
of galaxies. 
{These are traditionally defined as concentrations
of galaxy clusters} \citep[e.g.][]{Abell1961, Einasto2001, chow2014},
{building up the cosmic web from a network of connected high 
density nodes}; {or directly from the distribution
of galaxies \citep[e.g.][]{Luparello2011, CostaDuarte11, Liivamagi2012}}. 
{They can also be defined kinematically by mapping galaxy peculiar 
velocity flows, a technique still restricted to the very nearby Universe
\citep{Tully2014, Dupuy2019}. This last method is the closest to a
purely gravitational potential based approach, and allows the 
identification of the ``basins of attraction'' that partition the Universe
in cells or cocoons \citep[e.g.][]{Dupuy2019, Einasto2019}.}
{For this work we adopted the supercluster} 
{second-order clustering definition for determining 
the superclusters of the sample.}
These systems are not virialized and the {contents of the} 
inter-cluster medium (dark matter halos, gas and galaxies) dynamically 
interact and organize by falling through the gravitational potential of 
the more massive structures, forming walls, filaments, groups and clusters.
As shown in \cite{Tanaka2007}, the possibility to find elongated chain-like 
structures increases in superclusters. 
{Also, following the classification of superclusters by
\citet{Einasto2014} in ``filament-type'' and ``spider-type'' ones, both
have filaments, in a linear or radial configuration respectively.}
Following this approach, \citet{Cybulski2014}, {for instance,} 
applied a combination of Voronoi tessellation and minimum spanning tree 
(\emph{MST}) techniques over the Coma supercluster region in order to search 
for bridges between clusters of galaxies.

{Motivated by the above} context, we developed a 
methodology\footnote{https:$//$gitlab.com$/$iris.santiagob89$/$LSS\_structures}
for the identification of structures {in the environment of superclusters} using the 
galaxies embedded in them. 
We restrict our study to the SDSS-DR13 area, and use 
only galaxies with spectroscopic redshifts for our analysis.
The approach we follow seeks to detect structures by using only the 
geometrical information of the galaxy distribution.
By using different pattern recognition methods we identify high to 
moderate-density galaxy systems, and low-density filaments connecting them.
This allows the identification of structures over a wide 
range of scales (1~$-$~100 Mpc), from groups to long filaments. 
Moreover, the identified structures are validated through comparisons 
with previously reported catalogs. 
We also carried out a qualitative validation 
through a kernel method which is one of the most used 
methodologies for the detection of overdensity regions.
Our aim was to investigate whether previous filament candidates 
in the sample, identified from chains of Abell/ACO clusters, are bona-fide 
structures and to characterize their galaxy populations.
Finally, we studied the relation between the galaxy properties and the 
supercluster environment in which they reside (e.g. systems, 
filaments and the dispersed component of the superclusters).

This paper is organized as follows: 
In Section \ref{data} we present the data for the sample of superclusters 
under analysis and the sample of galaxies from the SDSS survey. 
In Section \ref{methods} we describe in detail the implementation of 
mathematical tools and pattern recognition methods applied for the detection 
of high density regions (clusters and groups) and for the 
skeletonization of the low density filamentary structures.
In Section \ref{highden} we describe the algorithm for detecting
clusters and groups of galaxies inside superclusters' boxes, while in 
Section \ref{lowden} we present the algorithm {for finding the 
filaments and their skeletons}.
In Section \ref{aplic} we describe the application of the algorithms to
one of the superclusters, MSCC\,310, as an example of their use.
Section \ref{valid} is devoted to the validation and 
evaluation of the methodology and discussion of its results.
In Sections \ref{disc} and \ref{gal_prop} we present the results concerning 
the analyses of the galaxy properties as function of the 
supercluster environment. We also discuss these results and compare with previously reported results.
Finally, in Section \ref{concl} we present the conclusions of this work.
Through this paper we assume the Hubble constant 
$H_0=70~h_{70}$~km~s$^{-1}$~Mpc$^{-1}$, the matter density $\Omega_m=0.3$, 
and the dark energy density $\Omega_\Lambda=0.7$.

\section{The data}
\label{data}

\subsection{The superclusters and filament candidates}

We are interested in unveiling and studying LSS filaments, which can be defined
as chains of clusters connected by bridges of galaxies and probably by gas
and dark matter.
As mentioned previously, these elongated structures should most likely be 
found in superclusters since they probably just passed the quasi non-linear 
regime described by Zel'dovich's approximation (1970, see also the 
``sticking model'' by \citealt{Shandarin1989}).
In the current evolutionary stage of LSS, superclusters are basically a 
network of sheets, filaments and knots (clusters and groups) of galaxies, 
gas and dark matter, just starting a global gravitational collapse process.

Thus, we selected a sample of superclusters of galaxies from the Main 
SuperCluster Catalogue \citep[MSCC,][]{chow2014}, which are 
inside the SDSS region (in order to have a sample of galaxy data 
as homogeneous as possible).
The original MSCC is an all-sky catalog that contains 601 superclusters,
identified in a complete sample of rich Abell/ACO clusters, with updated
redshifts from 0.02 to 0.15, by using a tunable \emph{FoF} algorithm. 
From these superclusters, 166 are inside the SDSS Data Release 13 (DR13) 
region.
For this work we selected those superclusters with 5 or more clusters 
having their box volume (see below) inside the SDSS-DR13 
survey area.
In addition, we used as reference the list of filament candidates 
for MSCC superclusters by Chow-Mart\'inez et al. (in preparation), 
in order to select {the superclusters with the most promising}
filaments.
Roughly speaking, these filament candidates were identified 
as chains of at least three clusters, members of the superclusters, 
separated by less than 20 $h_{70}^{-1}$ Mpc from each other. 
The present work also intends to validate these filament candidates 
by searching for the bridges of galaxies that we expect to 
connect them. 
It is worth mentioning that some of the filament candidates may reveal 
to be only chance configurations, with no bridges of galaxies connecting 
the clusters of a chain. Also, some bridges may exist, but not 
necessarily along the straight lines connecting the clusters.

Our final sample consists of 46 superclusters of galaxies, which are 
listed in Table \ref{table1}.
The ID of the supercluster in MSCC is listed in column 1, with its proper
name in column 2, when it exists. 
Column 3 presents sky coordinates, RA ($\alpha$) and Dec ($\delta$),
of the supercluster mean position, while column 4 shows its mean redshift. 
Columns 5 and 6 list the richness (number of member clusters) and the
number of filament candidates found previously in each supercluster.
The IDs of the Abell/ACO member clusters are listed in column 7.

\begin{table*}
	\caption{\label{table1} Sample of MSCC superclusters used in the 
present work.}

\centering
\scalebox{0.9}{

\small \addtolength{\tabcolsep}{-1pt}
  \begin{tabular}{r c r r r c l}
\hline\hline                 
SCl ID & Name & RA, Dec    & $\bar{z}$ & N$_{Cl}$ & N$_{fil}$ & Abell/ACO \\
(MSCC) &      & [deg, deg] &           &          &           & clusters \\

(1)    & (2)  & \multicolumn{1}{c}{(3)}& (4) & \multicolumn{1}{c}{(5)} & \multicolumn{1}{c}{(6)} & (7) \\
\hline
55  &      &  17.75,    15.44 &     0.0614 &    5 & 1 & A0150  A0152A A0154B A0158B A0160B \\
72  &      &  25.17,     0.64 &     0.0802 &    5 & 1 & A0181A A0208A A0237A A0267B A0279A \\
75  &      &  28.09,    -5.15 &     0.0937 &    7 & 1 & A0256A A0256B A0266  A0269  A0274A A0274B A0277  \\
76  &      &  28.35,    -2.61 &     0.1299 &   16 & 3 & A0211  A0233e A0255  A0256C A0261B A0265  A0267C A0268B A0271  A0274C A0279B \\
       &       &                              &                    &        &		& A0281  A0285  A0295D A0303C A0308e \\
175 &      & 125.29,    17.07 &     0.0942 &    6 & 1 & A0635A A0650B A0651A A0657A A0658A A0659  \\
184 &      & 130.10,    30.24 &     0.1056 &    6 & 1 & A0671B A0690C A0694  A0695B A0699B A0705A \\
211 &      & 147.87,    64.88 &     0.1191 &    8 & 1 & A0764  A0802  A0804B A0845  A0871e A0906e A0975  A1014A \\
219 &      & 153.99,    19.14 &     0.1155 &    5 & 1 & A0938B A0942A A0952A A0991B A0994A \\
222 &      & 155.14,    49.21 &     0.1382 &   10 & 2 & A0915B A0927A A0950A A0965A A0990  A1002A A1003C A1003D A1004  A1040C \\
223 &      & 155.24,    62.94 &     0.1399 &    5 & 1 & A0917  A0947A A0962A A1025A A1025B \\
229 &      & 156.14,    33.03 &     0.1423 &    7 & 2 & A0924  A0951  A0982  A1007B A1036  A1045  A1053B \\
236 &      & 156.76,    10.38 &     0.0328 &    6 & 1 & A0938A A0957A A0999A A1016A A1020A A1142A \\
238 &      & 156.98,    39.55 &     0.1068 &   21 & 4 & A0967A A0971A A0971B A0972A A0995A A0997A A0997B A0997C A1010B A1021B \\
		&		&								&					&			&	&A1021C A1021D A1026B A1028A A1031A A1031B A1033  A1040A A1050A A1054A A1055  \\
248 &      & 159.49,    44.26 &     0.1246 &    5 & 1 & A1040B A1050B A1054B A1056  A1074A \\
264 &      & 165.29,    12.20 &     0.1161 &    8 & 1 & A1105C A1116A A1129A A1141A A1147A A1157  A1201B A1209A \\
266 &      & 165.91,    11.85 &     0.1273 &    8 & 1 & A1131  A1137B A1141B A1147B A1152  A1159  A1183A A1209B \\
272 &      & 167.83,    41.33 &     0.0760 &    6 & 1 & A1173  A1174A A1187  A1190  A1193A A1203  \\
277 &      & 169.41,    49.67 &     0.1103 &    7 & 1 & A1154  A1202B A1218B A1222  A1225  A1227A A1231A \\
278 & Leo  & 169.37,    28.46 &     0.0333 &    6 & 1 & A1177B A1179B A1185A A1228A A1257A A1267A \\
283 &      & 170.79,    20.34 &     0.1379 &   12 & 3 & A1177C A1188  A1230B A1232B A1242A A1243B A1247e A1251  A1268  A1272  A1274  A1278  \\
295 & Coma  & 173.63,    23.11 &     0.0223 &    5 & 1 & A1100A A1177A A1179A A1367  A1656  \\
310 & UMa  & 175.91,    55.23 &     0.0639 &   21 & 3 & A1212  A1270  A1291A A1291B A1291C A1318A A1318B A1324A A1324B A1349A A1349B\\ 		
		&             &									&					&		&	& A1377  A1383  A1396A A1396B A1400A A1400B A1400C A1436  A1452  A1457A \\
311 &      & 176.12,     9.93 &     0.0833 &    8 & 1 & A1337A A1342A A1358A A1362B A1372A A1379  A1385A A1390  \\
314 &      & 177.07,    -2.01 &     0.0788 &    6 & 1 & A1364A A1376A A1386A A1389A A1399A A1404A \\
317 &      & 177.42,    -1.59 &     0.1278 &   13 & 2 & A1373A A1373B A1376C A1386D A1386E A1386F A1389C A1389D A1392  A1399C A1407 \\
		&		&								&					&		 &		& A1411  A1419B \\
323 &      & 179.66,    27.26 &     0.1396 &   12 & 1 & A1384A A1403A A1403B A1413B A1420C A1425B A1431B A1433C A1444C A1449B \\
		&		&								&						&		&	& A1455C A1495  \\
333 &      & 181.43,    29.34 &     0.0813 &    9 & 1 & A1423A A1427  A1431A A1433A A1444B A1449A A1455B A1515A A1549A \\
335 &      & 182.42,    29.50 &     0.0732 &    6 & 1 & A1444A A1455A A1478A A1480B A1486A A1519A \\
343 &      & 183.88,    14.31 &     0.0809 &    5 & 1 & A1474  A1481A A1499A A1526C A1527A \\
360 & Dra  & 190.94,    64.41 &     0.1055 &   11 & 1 & A1518A A1539A A1544A A1559  A1566  A1579A A1621  A1640A A1646  A1674A A1718A \\
386 &      & 199.50,    38.33 &     0.0715 &    5 & 1 & A1680A A1691  A1715A A1723B A1749B \\
407 &      & 208.55,    26.70 &     0.1364 &    6 & 1 & A1797B A1817C A1817e A1818C A1819  A1824  \\
414 & Boo  & 211.31,    27.32 &     0.0709 &   24 & 3 & A1775A A1775B A1781B A1795  A1797A A1800  A1817A A1818A A1831A A1831B \\
			&		&								&					&		&		&A1832A A1863A A1869A A1869B A1873B A1873C A1874A A1886A A1898A \\
			&		&								&					&		&		&A1903A A1908A A1909A A1912B A1921A \\
419 &      & 212.33,     7.17 &     0.1122 &    5 & 1 & A1850  A1862  A1866A A1870  A1881  \\
422 &      & 213.21,    28.95 &     0.1430 &    9 & 2 & A1832B A1840B A1854  A1867A A1874B A1891B A1903C A1908B A1912E \\
430 &      & 216.72,    25.64 &     0.0982 &    6 & 1 & A1909B A1910A A1912A A1912C A1926A A1927  \\
440 & BooA & 223.17,    22.28 &     0.1170 &    9 & 1 & A1939B A1972  A1976  A1980  A1986  A1988B A2001A A2006  A2021C \\
441 &      & 223.22,    28.40 &     0.1249 &    5 & 1 & A1973A A1982D A1984  A1990A A2005B \\
454 &      & 228.28,     7.33 &     0.0456 &    6 & 1 & A2020A A2028A A2033B A2040B A2055A A2063B \\
457 &      & 228.59,     6.98 &     0.0789 &    6 & 1 & A2028B A2029  A2033C A2040C A2055B A2063C \\
460 &      & 229.70,    31.17 &     0.1142 &    9 & 1 & A2025D A2034A A2049A A2056C A2059B A2062  A2067B A2069  A2083B \\
463 & CrB  & 232.18,    30.42 &     0.0736 &   14 & 2 & A2056A A2056B A2059A A2061A A2065  A2067A A2073A A2079A A2079B \\
		&				&							&						&		&	&A2089  A2092A A2106A A2122A A2124  \\
474 & Her  & 241.56,    16.22 &     0.0363 &    5 & 2 & A2147  A2151  A2152A A2153A A2159A \\
484 &      & 245.57,    42.39 &     0.1364 &    7 & 1 & A2158B A2172  A2179  A2183  A2196  A2198D A2211A \\
579 &      & 351.82,    14.79 &     0.0427 &    5 & 1 & A2572  A2589  A2593A A2593B A2657  \\
586 &      & 354.20,    23.67 &     0.1274 &    5 & 1 & A2611e A2619B A2627  A2647e A2650e \\
\hline
   \end{tabular}  }
\tablefoot{The superclusters have five or more Abell cluster members, with 
$z \leq 0.15$, and inside the SDSS-DR13 region. Superclusters with proper 
names are indicated in column 2.}

\end{table*}

For the Abell/ACO clusters and for the galaxies in the superclusters
box volumes ({see Section \ref{boxes}}), we first transformed 
their radial-angular coordinates to rectangular coordinates as follows:

\vspace{-20pt}
\begin{align}
X&=D_C~\cos\left(\delta \right) \cos\left( \alpha \right), \label{transformation1}\\
Y&=D_C~\cos\left(\delta \right) \sin\left( \alpha \right), \label{transformation2}\\
Z&=D_C~\sin\left(\delta \right),\label{transformation3} 
\end{align}
where $D_C$ is the co-moving distance as obtained using the spectroscopic redshift
and the cosmological parameters indicated above.

\subsection{The SDSS galaxies}
\label{galaxies}

The main galaxy sample of SDSS-DR13 \citep{DR13} is a suitable database to 
search for filamentary structures on the LSS since:
(i) it covers a large sky area (14,555 square degrees), containing various 
MSCC superclusters;
(ii) it contains homogeneous photometric and spectroscopic data for galaxies 
with an astrometric precision of 0.1 arcsec rms and uncertainty in radial 
velocities of about 30 km~s$^{-1} $ \citep{Bolton2012};
(iii) it is roughly complete to the magnitude limit of the main 
galaxy sample (r$_{Pet} =17.77$), which corresponds to an average $z \sim 0.1$, 
going (inhomogeneously) deeper for data releases after DR7 \citep{DR7};
(iv) at the limit of our sample, $z = 0.15$, the SDSS spectra are complete for 
galaxies brighter than M$_r \sim -21$.

SDSS-DR7 joins the SDSS-I/II spectra for one million galaxies and quasars.
It has $\sim$6\% incompleteness due to fiber collisions 
\citep{Strauss2002} and another $\sim$7\% incompleteness 
attributed to pipeline misclassification \citep{Rines2007}.
These spectra are included in the final data release of the SDSS-III 
\citep{DR12}. 
The Baryon Oscillation Spectroscopic Survey (BOSS) is part of the SDSS-III 
observations and has obtained spectra for another 1.4 million galaxies. 
The BOSS observations are divided in two main samples, LOWZ ($z<0.4$) and 
CMASS ($0.4<z<0.7$).
The SDSS-DR13 \citep{DR13} includes spectra for more than 
2.6 million galaxies and quasars.

Although photometric redshifts are available for SDSS galaxies, for 
this work we have selected those objects listed on the SpecObj sample 
with spectroscopic redshifts available (downloaded from the SkyServer 
web service) and denoting an extragalactic object (that is, galaxies 
and low-$z$ quasars).
The SpecObj table contains the best and unique spectra for 
the same location within 2 arcsec called ``sciencePrimary" objects.
We considered galaxies within a redshift range from 0.01 to 0.15 
and selected spectra with quality flag ``good'' or ``marginal''.  
Since the kind of study presented here relies on the galaxy 
distance 
measurement, we restricted our analysis to galaxies with spectroscopic 
redshift due to its higher accuracy. 
However, galaxies with photometric redshift can be included to the sample 
in further analyses to test if their addition increase the filament signal 
of detection. 

For the present work we also made use of value-added sub product catalogs 
such as the MPA-JHU catalog \citep{Brichmann, Kauffmann03, Tremonti}.
They calculated different galaxy properties (stellar mass, metallicity, 
activity type classification, star forming rate,
among others) using the spectra from the SDSS-DR8 galaxies \citep{DR8}.
As explained by \cite{Tremonti}, the galaxy properties in the MPA-JHU 
catalog are calculated by processing the galaxy spectrum in a way that 
even the weaker emission lines are detectable.   
In order to analyze the morphological distribution of the galaxies in the 
different supercluster environments we employed the morphological classification provided 
by \cite{Huertas-Company}.
They calculate a probabilistic morphological classification, for the 
SDSS-DR7 spectroscopic galaxies, 
by applying deep learning techniques that make use of their photometry. 
They also compare their automated classification with a sample of the 
Galaxy Zoo \citep{Lintott2008,Lintott2011} visual classification. 
They show that their classification in early and late types are in good 
agreement with the visual classification.

\subsection{The superclusters' boxes}
\label{boxes}

For each supercluster in Table \ref{table1} we selected all the SDSS galaxies
(according to the above criteria) located inside the corresponding box volume.
These boxes were defined in rectangular coordinates, in a way that their walls 
were set at a distance of 20 $h_{70}^{-1}$ Mpc {beyond}  the 
center of the farthest clusters {in each direction, for} each 
supercluster. 
{This extension was applied in order to guarantee that
any connection of the supercluster with external structures could be
detected}.
{The box volumes of the} {superclusters vary from
($45~h_{70}^{-1}$~Mpc)$^{3}$ to ($157~h_{70}^{-1}$~Mpc)$^{3}$.
Compared to the typical sizes of the observed and} {simulated 
basins of attraction in \cite{Dupuy2019}} 
{[($50-100~h_{70}^{-1}$~Mpc)$^{3}$], the boxes we use here 
are a little bit larger, as expected, implying that we are sampling the 
LSS in a general way, not restricting the analysis to the densest 
parts of the superclusters. Our sampling of the superclusters may be
compared to the one by \citet{Krause2013}, that is, more embracing than 
the sampling done by, for example, \citet{Kopylova2006} and 
\citet{Liivamagi2012}.}

In Table \ref{table2} we list the box volume (column 2), 
the number of galaxies inside this volume (column 3), 
its mean volume and surface (sky projected) number densities 
(columns 4 and 5), 
the baseline density (column 6), 
the number of galaxies with surface density above the baseline density 
(column 7),
the segmentation parameter $f$ and the number of \emph{HC} groups 
(see Section \ref{HC} for details) (columns 8 and 9), 
the remaining FoG corrected groups of richness 
between 5 and 9 galaxies (column 10) 
and of richness higher or equal to 10 galaxies (column 11),
the ranges of radius and velocity dispersion for this final 
list of groups (columns 12 and 13), for each supercluster.

In particular, the superclusters MSCC\,236, MSCC\,314 and MSCC\,317 lie 
close to the limits of the SDSS region: although all their member clusters 
are inside, their boxes were reduced to a margin of 10 $h_{70}^{-1}$ Mpc 
in only one direction.

\begin{figure}[!t]
	\includegraphics[width=9cm]{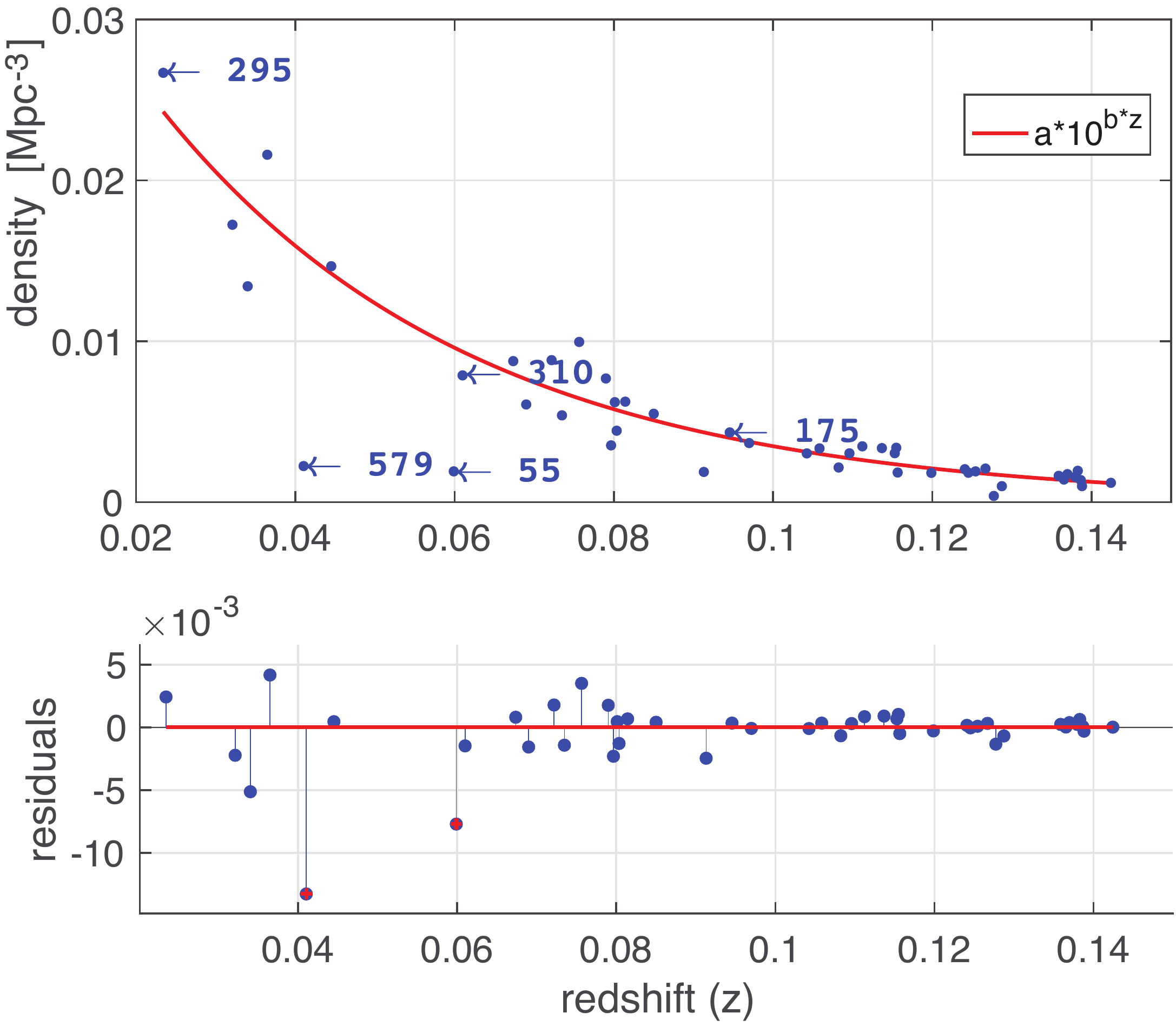}
	\caption{Distribution of mean volume densities ({see
fourth column of Tab. \ref{table2}}), for the 46 superclusters in our 
sample, as function of redshift (blue points). 
The red line corresponds to the best fit of a power law function. 
Residuals of the fitting are shown in the bottom panel.
MSCC\,579 and MSCC\,55 were excluded from the fitting. }
	\label{zvsdens}
\end{figure}

Figure \ref{zvsdens} shows the diminution of mean volume density of the
boxes with redshift, due to Malmquist bias. 
The fitted function will be used as the selection function for the SDSS 
galaxies {considered} in this work.
{It may be noted that the mean densities of
the superclusters MSCC\,55 and MSCC\,579 lie far below the fit.
This is also due to the positions of these superclusters close to 
the border of SDSS coverage: their samplings seem 
sparse and irregular. In fact, for MSCC\,579 one can clearly see
the shape of the cones of observation through the galaxy distribution.}
For this reason, {the analysis of these superclusters
and of the three cited above} must be taken with caution. 

It is worth noting that, since we have only the radial velocity component 
available (redshift), the transformation from radial-angular coordinates to 
rectangular coordinates is more complicated for the galaxies. 
Their peculiar velocity may bias their redshift-space coordinate, especially 
when they are members of clusters and groups of galaxies, being subject to the FoG effect. 
Thus, for the galaxy data used to detect the filaments, we first applied a 
correction, to be described below, which redefined their individual $D_C$ 
in equations 1-3.

\section{Mathematical tools}
\label{methods}

In what follows, we considered the $N$ galaxies in each supercluster volume 
as a set of points ${x_1,x_2,...,x_N} \in X$, all part of a sample $X$. 

\subsection{Voronoi Tessellation (VT)}
\label{VT}

The Voronoi tessellation \citep{Voronoi1908}
of a sample $X$, Vor($X$), can be defined as the subdivision of a 
2D plane or a 3D space into cells with the property that the seed point 
$x_i \in X$ is located in the cell $v_i$ if and only if the 
Euclidean distance 
$D_E$($x_i$, $v_i$) $<$ $D_E$($x_i$, $v_j$) for each $v_j \in X$ with 
$j \neq i$.
In other words, the \emph{VT} divides the space into polygonal cells, centered 
on the seed points (in our case, galaxies), in a way that 
the cell walls are equidistant to all nearest seeds
\citep[e.g.][]{Platen2011}. 
Therefore, the density at each galaxy position $x_i$ is determined as 
$d_i=1/v_i$, with $v_i$ being the volume (or area) of the cell enclosing 
the object $x_i$.
\citet{Scoville2013} and \citet{Darvish2015}, for instance, use \emph{VT} to find the high 
density regions in sky slices while \cite{Cybulski2014} apply \emph{VT} to 
identify the filamentary structures in the Coma cluster region.

\subsection{Hierarchical Cluster Analysis (HC)}
\label{HC}
Hierarchical clustering is a machine learning method whose objective is to 
group objects with similar properties. It has been used in different areas of 
science such as artificial intelligence, biology, medicine and business. 
In general, it can be used to carry out pattern recognition analysis, allowing 
to regroup, segment and classify any kind of data.
This method is equivalent to a reduction of the dimensionality of the data and 
reduces considerable the computing time.
\textcolor{black}{
In astronomy, the most popular application of \emph{HC} has been for the detection
of substructures inside galaxy clusters, following the algorithm developed
by \citet{SG96}. This algorithm considers the positions, redshifts and 
potential binding energy between pairs of galaxies to detect substructures
\citep[see also][]{Guennou2014}.}

Since we are interested in finding galaxy structures on
scales larger than the ones for substructures and for structures that 
may be less strongly
gravitationally bounded, we
chose to use an agglomerative hierarchical clustering analysis method that
\textcolor{black}{considers only the positions and redshifts or 3D estimated
positions of galaxies.}
A detailed description of the \emph{HC} algorithm can be found in 
\cite{Theodoridis2009,Theodoridis2010} and \cite{Murtagh2011}. 
For our analysis we chose Ward's minimum variance clusterization criteria, 
described in detail by \cite{Murtagh2014}. 
In general, Ward's method works by merging the groups following the 
criterion:
\begin{equation}
\Delta D(c_1,c_2)=\frac{|c_1||c_2|}{|c_1|+|c_2|} ||c_1-c_2||^2.
\end{equation}
\noindent where $\Delta D$ is a term that measures the distance between 
two groups $c_1$ and $c_2$, respectively.

In our case, initially each point is considered as a group, sub-cluster or 
singleton, then each group can be agglomerated with a neighbor that has the 
minimum $\Delta D$ distance. 
The agglomeration continues until all points are grouped together. 

The results of the \emph{HC} clusterization can be represented by a dendrogram 
or hierarchical tree.  
A dendrogram represents, in a graphical form, the connections between 
elements and groups at different levels of agglomeration.
The height of each connection line in the tree corresponds the distance between 
two elements or centroids connected.
This representation also allows visualizing the principal branch structures 
where the singletons are the final leaves. 
The number of desired groups $N_{cut}$ is, therefore, obtained by cutting the 
hierarchical tree at a certain level. The exact value of this level depends
on the characteristics of the sample or, more properly, on the underlying
physics that is used to define the groups.

Each created group can be represented by a 2D/3D Gaussian model, $P_j(x)$. 
This allows to classify the groups by their Gaussian properties, e.g. centroid 
(mean position, $C_j$), richness (number of members, $N_j$) and compactness 
(covariance, $\sigma_j$).

\subsection{Graphs}
\label{graphs}

Graph theory-based algorithms have shown to be a suitable tool to analyze 
complex networks. Some of the most common subjects where these algorithms are 
applied successfully are social networks, computer vision, statistics, business 
and transportation networks.

A graph is a representation of the connections in a network.  
It is composed of ``nodes'' and ``edges'', where each node represents an 
object, and the edges represents the connections between each two nodes. 
Also, the edges can have weights that represent the strength of the 
connection.
An undirected graph has edges that do not have direction. 
We can define an undirected graph 
as $G=(U,E,W)$, with $n$ nodes (or vertices) 
$u_i \in U$, $m$ edges $e_{kl} \in E$ and a weight set $W$ with a $w_{kl}$ for 
each edge $e_{kl}$.
The information of a graph can be represented by a square adjacency matrix. 
The values of the matrix entries indicate the weight of the connection between 
nodes.
Hence, the adjacency matrix $A$ of the graph $G$ is defined as:\\
\begin{center}
	$
	[A]_{kl}= 
	\begin{cases}
	1\qquad {\rm if}\ (u_{k}, u_{l})\in E \\ 
	0\qquad {\rm otherwise} 
	\end{cases}
	$
\end{center}
where $u_k$ and $u_l$ are nodes in G.
{One can refer to \citet{ueda1997} for a discussion about the use of 
graph theory approach for quantifying the LSS of the Universe.}

\subsection{Minimum Spanning Tree (MST)}
\label{MST}
A spanning tree connects all nodes in a graph in a way that does not produce 
cycles. A graph can contain several unconnected spanning trees. 
Since the edges in a graph can have weights, the minimum spanning tree
algorithm \citep{Graham1985} searches for a spanning tree that 
minimizes the total weight.
This algorithm traces a tree-like continuous path for a group of edges and 
nodes in an optimal way. 
In particular, the Kruskal minimum spanning tree algorithm analyzes the edges 
in sequence, sorting them by weight. 
At the beginning, the shortest edge is analyzed and this would be the first 
tree branch. 
Then, the nodes are added to the tree under three conditions:
(i) only one node is added to the tree; 
(ii) a node is added based in the number of connected edges; 
(iii) their edges cannot be connected to another existing node in the tree.  
The process continues with the following edges in the graph 
until all connected edges are analyzed. 
Finally, the tree is extracted from the graph and the process begins again 
with the remaining nodes until all are tested. 
As its name remarks, the result is a forest of optimized 
independent trees.

\subsection{Dijkstra's shortest path}
\label{Dij}
Dijkstra's algorithm \citep{Dijkstra1959} is a classical method for searching 
the shortest path between two nodes in a graph. 
We define a path 
of length $e_{kl}$ between two nodes $u_k$ and $u_l$ as a sequence of 
connected nodes $u_1, u_2,..., u_n$ if 
$k \ne l ~\forall~ k,l \in {1,...,n}$.
In general Dijkstra's algorithm works as follows.

First, an origin is selected by taking the node at the beginning of the path,
$u_0$. 
Then, a distance value is assigned to all nodes: set as zero for the origin, 
$s(u_0)$, and as infinity for all the other nodes, $s(u_i)=\inf$.
Next, all nodes are marked as unvisited and $u_0$ is marked as current $a$.
The algorithm then calculates the distance from the current node $a$ to all 
the unvisited nodes connected by the edges $e_{i}$ as  $s_{new} = s(e_{ai})+w_{ai}$; 
here $s(e_{ai})$ is the distance from $a$ to the node $u_i$ and $w_{ai}$ is 
the weight of the edge $e_{i}$.
If $s(e_{ai})+w_{ai}<s(e_{i})$, then the distance is updated and the connected 
node label is updated as the current $a$.
After visiting all neighbors of the current node, they are marked as 
visited. A visited node will not be checked again; then the recorded distance 
$s(e_{ai})$ is final and minimal.
Finally, if all nodes have been visited, the algorithm stops. 
Otherwise, the algorithm sets the unvisited nodes with the smallest distance 
(from the initial node $u_0$, considering all nodes in the graph) as the next 
``current node'' and continues from the second step.
A detailed description of the algorithm can be consulted in \cite{Ray2014}. 

\subsection{Kernel Density Estimator (KDE)}
\label{KDE}
As mentioned before, \emph{VT} is used to measure the local 
density at each point position. 
However, in some cases, it fails on the identification of large 
overdensity regions, as mentioned by \textcolor{black}{\cite{Cybulski2014}.}
An alternative for the \emph{VT} method is to apply kernel density estimators.
In general, \emph{KDE} methods work by adjusting a kernel function over each observation 
in the sample.
However, the choice of the optimal kernel model and its intrinsic parameters is 
still under investigation in the pattern recognition community. 
Also, there are several attempts to apply adaptive Gaussian model kernels, 
in other words, to change the size of Gaussian model as a function of different 
parameters, for instance the distance to the nearest neighbor \citep{Chen2016_method} 
or a weighting function \citep{Darvish2015}.

For this work we used the results from the \emph{VT} method (see Section \ref{VT}) 
as the input parameters for the \emph{KDE}.
We start by fitting an ellipsoid inside each \emph{VT} cell.
Thus, instead of choosing a fixed bandwidth for the kernel, we employ the 
eigenvalues and eigenvectors of the ellipsoids to calculate a Gaussian kernel 
$\phi_{\boldsymbol{\Sigma}}$ centered at $\mu$ with covariance matrix 
$\boldsymbol{\Sigma}$ for each observation.
Therefore, each n-dimensional kernel is represented as:

\begin{equation}
\phi_{\boldsymbol{\Sigma}}(\boldsymbol{\mathrm{x}}-\mu)=(2\pi)^{d/2}|\boldsymbol{\Sigma}|^{-1/2}e^{-1/2(\boldsymbol{\mathrm{x}}-\mu)^T\boldsymbol{\Sigma}^{-1}(\boldsymbol{\mathrm{x}}-\mu)}.
\end{equation}
Then the \emph{KDE} can be estimated as:
\begin{equation}
\hat{p}_{\mathrm{KDE}}(\boldsymbol{\mathrm{x}})=\sum_{i=1}^{N}\alpha_i \phi_{\boldsymbol{\Sigma}_{i}} \left(\boldsymbol{\mathrm{x}}-\boldsymbol{\mathrm{x}}_i\right),
\end{equation}
where $\alpha_i$ is a weight factor calculated from the \emph{VT} cell volume ($v_i$) as $1/v_i$.

The identification of the overdensity regions is done through the projection 
of KDE kernels in 2D planes superposing a regular rectangular grid to the data. 
Thus, the density estimation is obtained at each grid intersection by calculating 
the average density of all kernels that overlap at that point. 
Then, observations closer to an evaluating point will contribute more to the 
density estimation than points farther away from it. 
Consequently, the density will be higher in areas with many observations than 
in areas with few observations.

\subsection{Transversal profiles}
\label{profile_extraction}

The distribution of galaxy properties in filaments is analyzed by 
constructing transversal profiles.
These profiles are calculated by setting up a series of concentric 
cylinders with axes orientated along the filament skeletons.
Then, a bin is considered to be the volume within two concentric cylinders 
of radius $R_{cy}$ and $R_{cy} + \Delta R_{cy}$.
The occurrence of a galaxy proxy in each bin is determined with 
respect to the galaxy distance {from the filament skeleton} $D_{ske}$.
The total count of galaxies per bin is weighted by the bin volume, in a 
similar way as making a normalized histogram. 
In order to compare samples of different sizes, a normalization is applied 
by dividing the number of events in a bin by the total number of galaxies in 
the sample.

\section{Galaxy Systems-Finding algorithm (GSyF)}
\label{highden}

\subsection{Detection of high density regions}
\label{systems}

We first searched for the high {relative} density regions, 
clusters and groups of 
galaxies (which we will refer generically as galaxy systems), inside the
studied superclusters, since these systems are the natural nodes 
for filaments. 
This was also necessary for correcting the FoG effect and having the data 
prepared for the application of the filaments-finding algorithm 
(see next Section). 
Furthermore, the detection of galaxy systems allows the identification
of new ones possibly not known before (especially poorer galaxy groups),
and the improvement of the membership estimation of the superclusters 
themselves. 

A description of the algorithm, including the strategy
we used for optimizing its parameters using simulated mock 
volumes, is presented in \citet{Santiago-Bautista2019a}.
Here we review the main steps of this algorithm.
First we calculate the local surface density at the position of each 
galaxy in the projected area of the supercluster by applying 
the \emph{VT} method (section \ref{VT}). 
The \emph{VT} individual area of the galaxy can be directly converted to a 
surface density estimation ($d_i = 1/a_i$), in this case in units
of deg$^{-2}$.
It is worth to note that the boxes we have considered for the 
superclusters in our sample comprehend slices in redshift-space in a 
range between 0.02 $\leq \Delta z \leq$ 0.07.

In order to identify the galaxy systems, we start by applying the \emph{HC} 
method (section \ref{HC}), but only to the $N_{gal}$ galaxies with 
densities above a baseline density, $d_{bas}$, which should be analogue 
to a background density. In a certain sense, this separates supercluster 
galaxies from void galaxies (that is, under-dense regions).
This baseline is calculated from the mean density by 
randomizing the galaxies in each sky projected area.
Since the distribution of points in the space is not isotropic, 
it is not possible to set directly a background density from 
the projected positions of the galaxies. 
Therefore, it is necessary to simulate an isotropic distribution of the points, 
in order to set the baseline value \citep[see, e.g.][]{Cybulski2014}.
A set of 1\,000 randomization of the points positions is generated, each 
with the same sample number over the same area.   
Then, the mean surface density is calculated by:

\begin{equation}
d_{bas}=\frac{1}{m} \sum_{j=1}^{m}\frac{1}{n} \sum_{i=1}^{n}d'_{i,j},
\end{equation}

\noindent where $d'_{i,j}=1/a'_{i,j}$ corresponds to the inverse of the area 
of the point $x'_i$ for the randomization $j$.

Since the distribution of galaxies is not homogeneous among the different
boxes, we calculate independent baseline values for each supercluster, see 
Table \ref{table2}.
A density contrast ($\delta_i$) is then calculated as:

\begin{equation}
\delta_i=\frac{d_i-d_{bas}}{d_{bas}}.
\end{equation}

That is, the $N_{gal}$ galaxies to which we apply the \emph{HC} 
are the ones with a density contrast, $\delta_i > 0$ 
\citep[see][for an evaluation of the negligible effect of 
slightly changing the density threshold]{Santiago-Bautista2019a}.

Then, we apply \emph{HC} to the set of parameters (RA, Dec, $1000z$) for these 
galaxies (the factor of 1000 is the weight for $z$ values 
to be comparable to the sky coordinates values).
The number of groups taken from the analysis is defined as a cut of the \emph{HC} 
tree, fixed to $N_{cut} = N_{gal}/f$, with a segmentation parameter $f$, 
which is the expected mean number of elements per group.
Currently, the selection of the optimal number of groups in clusterization 
methods is still a topic under investigation in the pattern recognition 
community, which includes the \emph{HC} algorithm.
A specific value of $f$ was calculated for each supercluster ($3\leq f \leq36$) 
according to the optimization process described in 
\citet{Santiago-Bautista2019a}. 
This strategy was adopted because a physically motivated value for $f$ 
would depend on many parameters, like the density of galaxies in each box, 
the sampling of these galaxies with respect to the real distribution, the 
redshift, among others which are difficult to estimate for our data.

Finally, we select only those systems with a number of galaxies, $N_{j}$, 
larger than two.
These pre-identified systems are, then, subject to the next step of 
refinement: the iterative estimation of the dynamical parameters virial 
mass and radius.

\subsection{FoG correction}
\label{FoG}

After identifying the galaxy systems 
we proceed to refine the galaxy membership and correct the galaxy 
positions for the FoG effect by using a virial approximation.
We apply a simplified version of the algorithm presented by \cite{Biviano2006} 
for the estimation of the virial mass and radius. We do not apply the 
surface pressure term correction based on the concentration parameter. 
Avoiding such a correction can lead to an overestimation of the virial radius, 
however, for this geometric analysis, only a virial approximation is enough.

The virial parameters calculation algorithm works as follows:
First we take the projected center and mean velocity of the system from 
the results of \emph{HC} ($C_j$). The projected center is then set at the position 
of the brightest $r$-band magnitude member galaxy (BMG) within 1~$\sigma_j$ 
from the \emph{HC} center, while the \emph{HC} mean velocity is used directly.
Those galaxies expected to belong to the system are selected among all 
galaxies in the sample (those with spectroscopic redshift in SDSS-DR13) 
that are projected inside a cylinder of radius $R_{a}$, hereafter, aperture.
\cite{Biviano2006} show that the dynamical analyses are similar for 
different aperture sizes. 
We chose an aperture of $R_{a}$ = $1~h_{70}^{-1}$~Mpc.
Then, for the line-of-sight direction, we select galaxies with a difference 
in velocity up to $S_{a} = \pm$3000 km~s$^{-1}$ with respect to the mean cluster velocity.
This would correspond to three times the velocity dispersion of a rich cluster.
A robust estimation of mean velocity, $v_{LOS}$, and 
velocity dispersion, $\sigma_v$, for the galaxies inside the cylinder 
is obtained by using Tukey's biweight method \citep{Beers1990}. 
An approximation of the mass $M_a$ in the aperture is computed as:
   \begin{equation}
   M_a=\frac{3\pi}{2G}~\sigma_{v}^{2}~R_{h}, \label{mass} 
   \end{equation}
\noindent where $G$ is the gravitational constant, the 3$\pi$/2 is the 
deprojection factor and $R_{h}$ is the projected harmonic radius.

We calculate the virial radius, $R_{vir}^3 = (3/4\pi)(M_{vir}/\rho_{vir})$, 
by assuming a spherical model for nonlinear collapse, that is, 
by taking the virialization density as 
$\rho_{vir}=18 \pi^2 [3 H^2(z)]/[8 \pi G]$,
and $M_a$ as an estimation for $M_{vir}$,
we have:
   \begin{equation}
   R_{vir}^3 = \frac{\sigma_v^2~R_h}{6 \pi~H^2(z)}
   \end{equation}
Then, the aperture $R_a$ is updated to the calculated $R_{vir}$ value,
the mean velocity to $v_{LOS}$, and $S_a$ to $\sigma_v$, defining a new
cylinder.
This process is repeated iteratively until the radius $R_{vir}$ converges.
$M_{vir}$ is finally calculated at the end of the iteration process.

The correction for the FoG effect is carried out by adjusting 
the position of the $N_{mem}$ galaxies inside the final 
cylinder.
This is done by scaling their comoving distances 
along the cylinder to the calculated virial radius.

\begin{table*}
	\centering
		\small\addtolength{\tabcolsep}{-2pt}
		\caption{\label{table2}{Properties of 
the supercluster boxes and of the galaxy systems detected inside them
by GSyF {algorithm}.}}
	\scalebox{0.9}{
		 \begin{tabular}[]{r r r r r r r r r r r c r}
				\hline\hline                 
SCl ID & {$V$}     & $N$              & $d = N/V$    & $d_{surf} = N/A$ & $d_{bas}$    & $N_{gal}$     & $f$ & N$_{HC}$     & \multicolumn{2}{c}{N$_{FoG}$}   & $R_{vir}$ &  $\sigma_v$ \\
(MSCC) & [10$^3 h_{70}^{-3}$ Mpc$^{3}$] & ({DR13}) & [$h_{70}^{3}$ Mpc$^{-3}$] & [deg$^{-2}$] & [deg$^{-2}$] & $d_i>d_{bas}$ & & $N_{j}\ge 3$ & $N_{mem} <10$ & $N_{mem}\ge 10$ & [\Mpc] & [km~s$^{-1}$] \\
(1)    & (2)               & (3)              & (4)          & (5)              & (6)          & (7)           & (8) & (9)          &  (10) & (11)                    & (12)      & (13) \\
				\hline
 55  &   424.3  &   812  &  0.0019  &   77.6  &   8.3  &   468  & 27  &   57  &  11  &   5  &  1.1  -  2.4  &  245  -  806  \\   
 72  &   549.7  &  1941  &  0.0035  &  232.0  &  22.1  &  1341  & 18  &  228  &  30  &  22  &  0.9  -  2.4  &  184  -  689  \\   
 75  &   854.6  &  1607  &  0.0019  &   95.8  &  22.6  &   877  & 15  &   69  &   9  &   7  &  1.5  -  3.4  &  335  -  1144  \\   
 76  &  2628.3  &  2617  &  0.0010  &  110.1  &  26.6  &  1536  & 27  &  204  &  14  &  11  &  1.3  -  3.6  &  214  -  1051  \\   
175  &   577.4  &  2504  &  0.0043  &  116.3  &  25.8  &  1315  &  6  &  172  &  22  &  10  &  1.2  -  2.7  &  220  -  735  \\   
184  &   692.9  &  2101  &  0.0030  &   70.5  &  20.7  &  1003  &  3  &  137  &  10  &   9  &  1.4  -  2.7  &  278  -  739  \\   
211  &   814.7  &  1484  &  0.0018  &   28.6  &  11.5  &   654  &  6  &   53  &   2  &   3  &  1.9  -  2.1  &  406  -  496  \\   
219  &   628.4  &  1913  &  0.0030  &  118.2  &  19.5  &  1273  &  6  &  175  &  13  &  10  &  1.4  -  3.7  &  292  -  1151  \\   
222  &   955.9  &  1865  &  0.0020  &   97.3  &  20.3  &   885  &  3  &  123  &  10  &   8  &  1.3  -  4.1  &  245  -  1207  \\   
223  &   777.0  &   776  &  0.0010  &  302.7  &  13.8  &   247  &  3  &   45  &   1  &   3  &  2.1  -  2.1  &  475  -  475  \\   
229  &  1352.4  &  1855  &  0.0014  &   45.2  &  22.3  &   745  &  3  &  106  &   6  &   2  &  2.2  -  2.2  &  498  -  498  \\   
236  &   643.6  &  8636  &  0.0134  &   52.9  &  10.3  &  4733  &  3  &  309  &  93  &  73  &  0.5  -  1.8  &  105  -  703  \\   
238  &  3861.8  &  8328  &  0.0022  &   74.4  &  20.3  &  4860  &  6  &  832  &  38  &  74  &  0.7  -  4.0  &  112  -  1293  \\   
248  &   690.1  &  1263  &  0.0018  &   72.4  &  17.4  &   564  &  3  &   75  &   5  &   2  &  1.7  -  2.8  &  366  -  730  \\   
264  &   923.7  &  1704  &  0.0018  &   59.6  &  24.5  &   626  &  3  &  105  &  11  &  12  &  1.3  -  3.5  &  245  -  1026  \\   
266  &   458.5  &   958  &  0.0021  &   49.5  &  28.8  &   318  &  3  &   55  &   6  &   3  &  1.6  -  3.1  &  320  -  823  \\   
272  &   138.4  &  1379  &  0.0100  &  135.3  &  28.0  &   654  &  3  &   87  &  10  &   5  &  1.1  -  2.4  &  219  -  699  \\   
277  &   905.3  &  2748  &  0.0030  &   76.0  &  20.2  &  1329  &  6  &  179  &  17  &   9  &  1.4  -  2.5  &  278  -  675  \\   
278  &   459.3  &  7920  &  0.0172  &   52.3  &  10.3  &  4116  &  6  &  222  &  80  &  35  &  0.5  -  1.9  &  112  -  711  \\   
283  &  1478.8  &  2320  &  0.0016  &   70.3  &  20.2  &  1379  & 12  &  239  &  10  &  17  &  1.5  -  3.4  &  295  -  907  \\   
295  &   535.5  & 14308  &  0.0267  &   48.5  &   7.2  &  7422  &  6  &  272  & 114  &  46  &  0.4  -  2.0  &   74  -  909  \\   
310  &  1558.8  & 12286  &  0.0079  &   76.9  &  15.7  &  7529  &  6  & 1015  & 116  & 139  &  0.8  -  3.0  &  140  -  1182  \\   
311  &   958.8  &  5270  &  0.0055  &   91.8  &  22.4  &  3050  &  6  &  416  &  48  &  40  &  0.8  -  2.4  &  131  -  704  \\   
314  &    91.9  &   558  &  0.0061  &  135.2  &  27.0  &   289  &  3  &   49  &  10  &   4  &  1.2  -  2.3  &  254  -  659  \\   
317  &   438.6  &   840  &  0.0019  &  104.2  &  38.2  &   433  &  6  &   76  &  10  &   5  &  1.8  -  3.3  &  366  -  929  \\   
 323  &  1909.6  &  3330  &  0.0017  &   77.3  &  21.7  &  1764  &  6  &  304  &  17  &  21  &  1.5  -  3.7  &  295  -  1069  \\   
333  &   445.5  &  1968  &  0.0044  &   65.1  &  22.6  &   793  &  3  &  135  &  14  &  27  &  1.1  -  2.9  &  221  -  949  \\   
335  &   574.5  &  3099  &  0.0054  &   62.2  &  21.4  &  1285  &  3  &  211  &  29  &  38  &  0.8  -  3.0  &  144  -  973  \\   
343  &   427.9  &  2679  &  0.0063  &  105.8  &  19.2  &  1526  &  6  &  196  &  23  &  25  &  0.8  -  2.4  &  131  -  675  \\   
360  &   657.7  &  2199  &  0.0033  &   80.1  &  15.3  &   934  & 12  &  160  &  15  &  16  &  1.3  -  2.5  &  253  -  653  \\   
386  &   535.9  &  3256  &  0.0061  &   54.9  &  17.2  &  1600  &  9  &  257  &  33  &  40  &  1.0  -  2.7  &  211  -  852  \\   
407  &   800.0  &  1126  &  0.0014  &   48.9  &  22.8  &   481  & 12  &   79  &   5  &   5  &  1.5  -  4.0  &  280  -  1184  \\   
414  &  1245.9  & 10902  &  0.0088  &   93.0  &  23.1  &  6366  &  6  & 1066  & 144  & 161  &  0.8  -  3.2  &  140  -  1191  \\   
419  &   497.6  &  1723  &  0.0035  &   91.7  &  19.7  &  1103  &  6  &  196  &  25  &  20  &  1.2  -  3.3  &  211  -  976  \\   
422  &   884.6  &  1065  &  0.0012  &   41.9  &  24.2  &   382  &  3  &   62  &   2  &   6  &  2.1  -  2.3  &  474  -  526  \\   
430  &   437.0  &  1603  &  0.0037  &   88.6  &  22.9  &   871  &  3  &  121  &  20  &   9  &  1.3  -  2.4  &  281  -  647  \\   
440  &  1017.1  &  3442  &  0.0034  &   99.4  &  72.9  &   917  &  6  &  143  &  24  &  14  &  1.5  -  3.3  &  309  -  935  \\   
441  &   516.1  &  1058  &  0.0021  &   60.8  &  20.9  &   425  &  6  &   59  &   2  &   3  &  3.0  -  3.0  &  796  -  796  \\   
454  &   389.0  &  5704  &  0.0147  &   99.7  &  18.9  &  3231  &  6  &  524  &  84  & 106  &  0.7  -  1.9  &  142  -  610  \\   
457  &   529.6  &  4072  &  0.0077  &  129.0  &  22.7  &  2605  &  6  &  443  &  58  &  44  &  1.0  -  3.1  &  187  -  1038  \\   
460  &  1041.1  &  3499  &  0.0034  &  108.4  &  27.3  &  1925  &  3  &  335  &  35  &  23  &  1.3  -  3.6  &  238  -  1073  \\   
463  &   959.2  &  8466  &  0.0088  &  121.6  &  22.4  &  5278  &  3  &  898  & 113  & 113  &  0.6  -  3.1  &  103  -  1077  \\   
474  &   343.8  &  7424  &  0.0216  &  109.2  &  15.3  &  4506  &  9  &  166  &  64  &  26  &  0.6  -  2.6  &  122  -   1115  \\   
484  &   805.0  &  1319  &  0.0016  &   43.1  &  19.2  &   571  &  6  &   86  &   5  &   4  &  2.3  -  3.0  &  536  -  793  \\   
579  &   658.9  &  1477  &  0.0022  &  142.9  &   1.2  &  1234  &  3  &  149  &  19  &  23  &  0.6  -  1.9  &  128  -  659  \\   
586  &   962.1  &   373  &  0.0004  &   18.3  &  13.6  &    72  &  3  &    8  &   0  &   0  &  -99.0  -  -99.0  &  -99  -  -99  \\   
				\hline
			\end{tabular}
		}
\tablefoot{The value -99 is set when GSyF does not detect systems.}

\end{table*}

\begin{figure*}
   \centering
      \includegraphics[width=16cm]{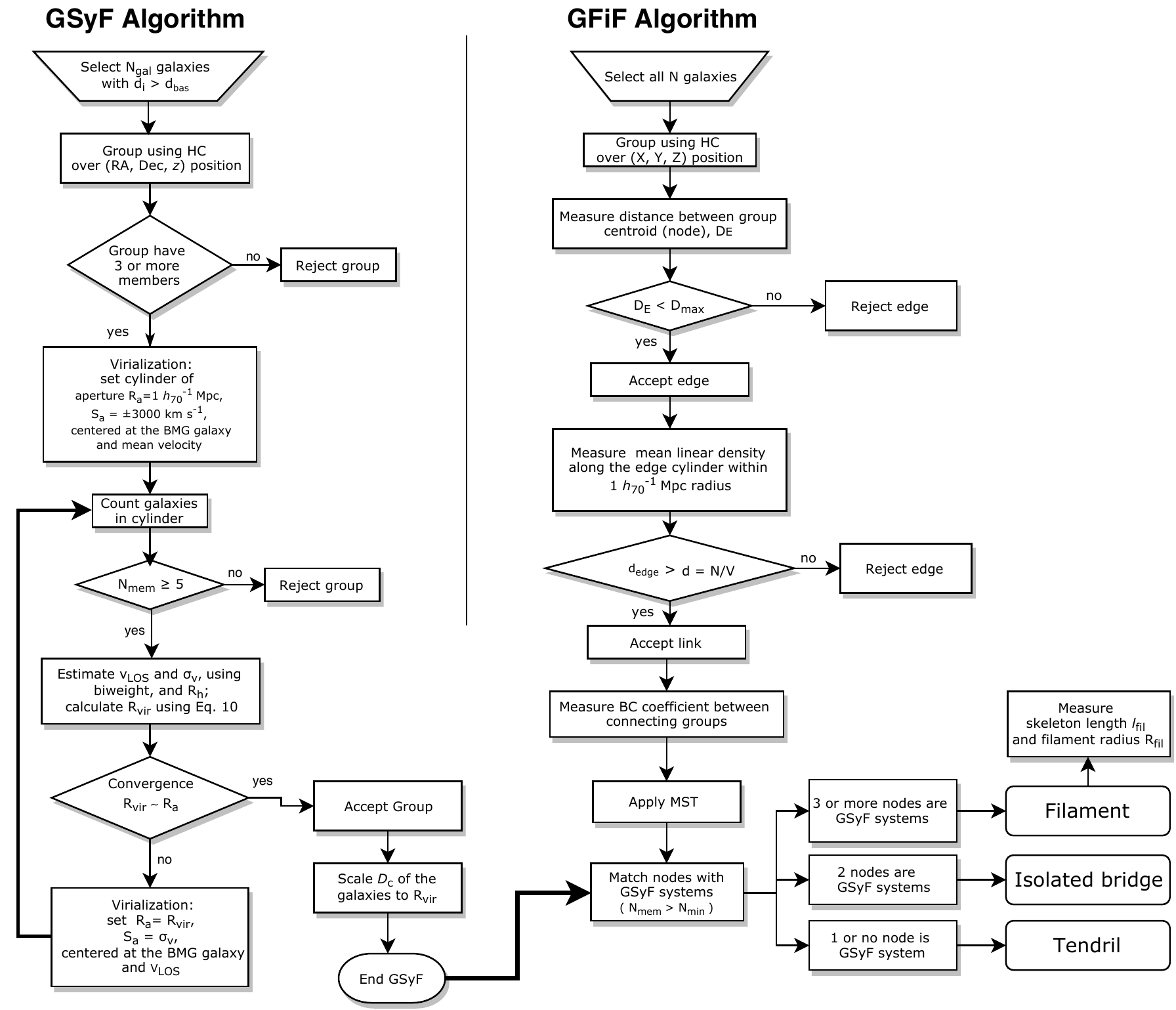}
      \caption{Flow chart of the GSyF (left side) and GFiF 
(right side) algorithms.}
      \label{flujograma}
\end{figure*}

{A schematic representation of the GSyF algorithm, 
including the FoG correction, can be found on the left side of 
Figure \ref{flujograma}.} 

\section{Galaxy Filaments skeleton-Finding algorithm (GFiF)}
\label{lowden}

\subsection{Detection of low density regions}

Since we are interested in the detection and analysis of elongated and
low {relative} density contrast structures, we apply again a combined \emph{VT}+\emph{HC} method to the
data, but now in the rectangular 3D space, with the positions of the galaxies 
corrected for the FoG effect. Thus, the \emph{VT} densities are now volume densities,
in units of Mpc$^{-3}$.
{At the beginning} 
of this analysis the \emph{HC} method is applied to all galaxies in the volume 
without density restrictions, that is, no baseline is applied. 
{Density restrictions are considered later as criteria for the construction of filaments.}
Another difference between this application of \emph{VT}+\emph{HC} and the one used for the 
GSyF methodology is a relaxed cut in the hierarchical tree.
Since we are interested in detecting more elongated representative 
structures, we tested values for the segmentation parameter $f$ 
between 10 and 40.
The direct effect of relaxing the cut is to allow the detection of 
groups at lower densities.

\begin{figure}[!t]
	\includegraphics[width=9cm]{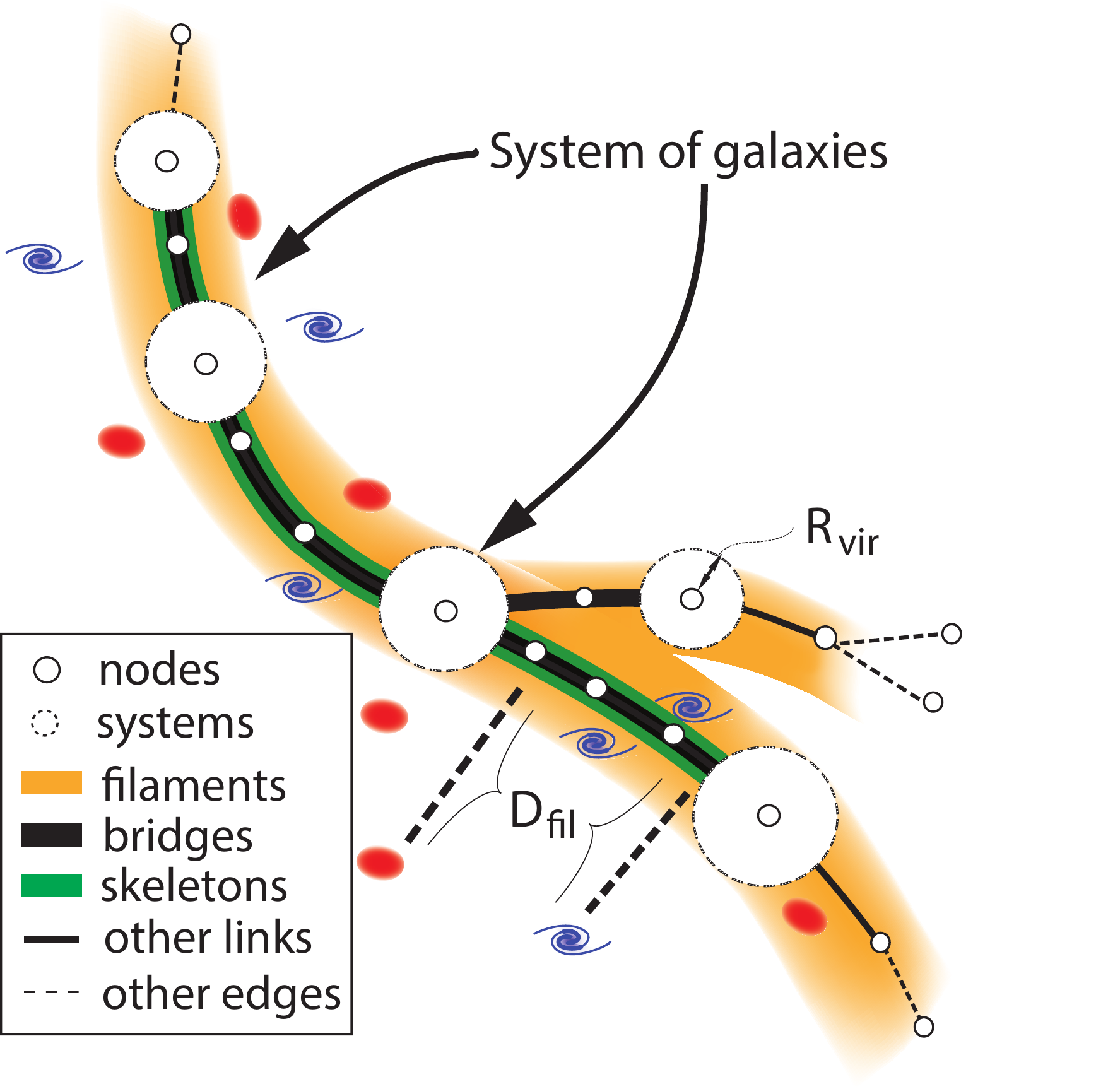}
	\caption{Representation of a filament. Graph nodes are represented 
by white circles and edges by dark lines. The five systems connected are 
represented by a dotted circle of radius $R_{vir}$. A bridge connecting two 
systems is represented as a bold black line. The distance from galaxies to 
the filament (bold dashed line) is measured perpendicularly respect to the 
edges. }
	\label{squema}
\end{figure}

Here we need to make some practical definitions in order 
to describe our strategy. 
\begin{itemize}
	\item {The nodes to which we will apply the method correspond, 
in the context we are working, to the \emph{HC} group centroids.}
	\item {An edge is defined as any connection between two nodes.}
	\item  The  real ``links'' between the systems are defined as 
the most promising edges, filtered according to their proximity and 
density contrast.
	\item {Spanning trees are extracted as described in section 
\ref{MST}, by cutting the graph in no-cycled optimal trees. }
	Some nodes inside a spanning tree may have been detected as 
galaxy systems by the GSyF algorithm.
	\item A ``bridge'' is defined as a sequence of links and nodes 
between two systems.
	\item A ``filament'' is identified if a spanning tree bridges 
three or more systems connected by bridges.
    \item If the spanning tree contains none or only one system, it is 
called ``tendril''.
	\item The ``skeleton'' is the medial line of a filament. The
	method for finding it, which intends to reduce the dimensionality 
	of the objects (in our case, galaxy filaments), is known as 
	skeletonization.
\end{itemize}
Figure \ref{squema} shows schematically these definitions.

\subsection{Chaining the filaments}
\label{chain}

Once we have applied \emph{HC} we measure the Euclidean distance $D_E$ of each 
group' centroid (node) against all its group neighbors.
These connections (edges) can be represented by an undirected graph as 
described in section \ref{graphs}.
The weights $W$ of the edges are set by the Bhattacharyya coefficient, $BC$,  
defined as:

\begin{equation}
   BC(P_1,P_2)=\sum_{x\in X}\sqrt{P_1(x)P_2(x)}.
\end{equation}

The Bhattacharyya coefficient quantifies the amount of overlapping between 
two distributions $P_1(x)$ and $P_2(x)$. 
Thus, the orientation of the two groups weights the connection between 
them.

In the next step, we filter the edges by two criteria:
First we select the edges corresponding to a $D_E$ smaller than a 
threshold, $D_{max}$ (hereafter, linking length).
Secondly, we consider an edge as a real link of galaxies based on 
the following:
 i) we define a cylinder along the edge with radius of 1~$h_{70}^{-1}$~Mpc;
 ii) we measure the linear density of galaxies along the 
cylinder;
 iii) if the mean linear density of the cylinder is above 
$d = N/V$ (Table \ref{table2}) we take the cylinder as a link of galaxies 
connecting the two nodes.
	
Each ensemble of connected links is a tree in the 
forest graph.
We then apply Kruskal's \emph{MST} technique (section \ref{MST}) on the 
forest graph to identify independent trees and their dominant branches. 

To proceed we need to match the list of detected 
spanning trees with the list of detected GSyF systems.
However, due to the effect of losing sampled galaxies 
with increasing redshift (see Figure \ref{zvsdens}), the richness of 
the detected systems depends on the redshift. In other words, 
to have a comparable richness for two similar system, for instance one 
at $z =$ 0.03 and the other at $z =$ 0.13, we have to apply a correcting 
factor to the richness of the second one. 
To overcome this limitation, we apply the following lower
limit for the richness of the systems at the supercluster redshift:
$\log_{10}N_{min}= a\log_{10}z+b$, with $a=-1.0$ and $b=-0.2$.
This leads to a lower richness limit of $N_{min} =$ 30 to 5 
galaxies per system, from the nearest and farthest supercluster in our 
sample respectively.

Having now the systems and the bridges between them 
(instead of the nodes and edges in the previous step) we can identify the filaments.
As stated above, and following the definition by 
Chow-Mart\'inez et al. (in preparation),
we search for the filaments which have at least 3 galaxy
systems connected by bridges.

\begin{figure}[!t]
	\includegraphics[width=9cm]{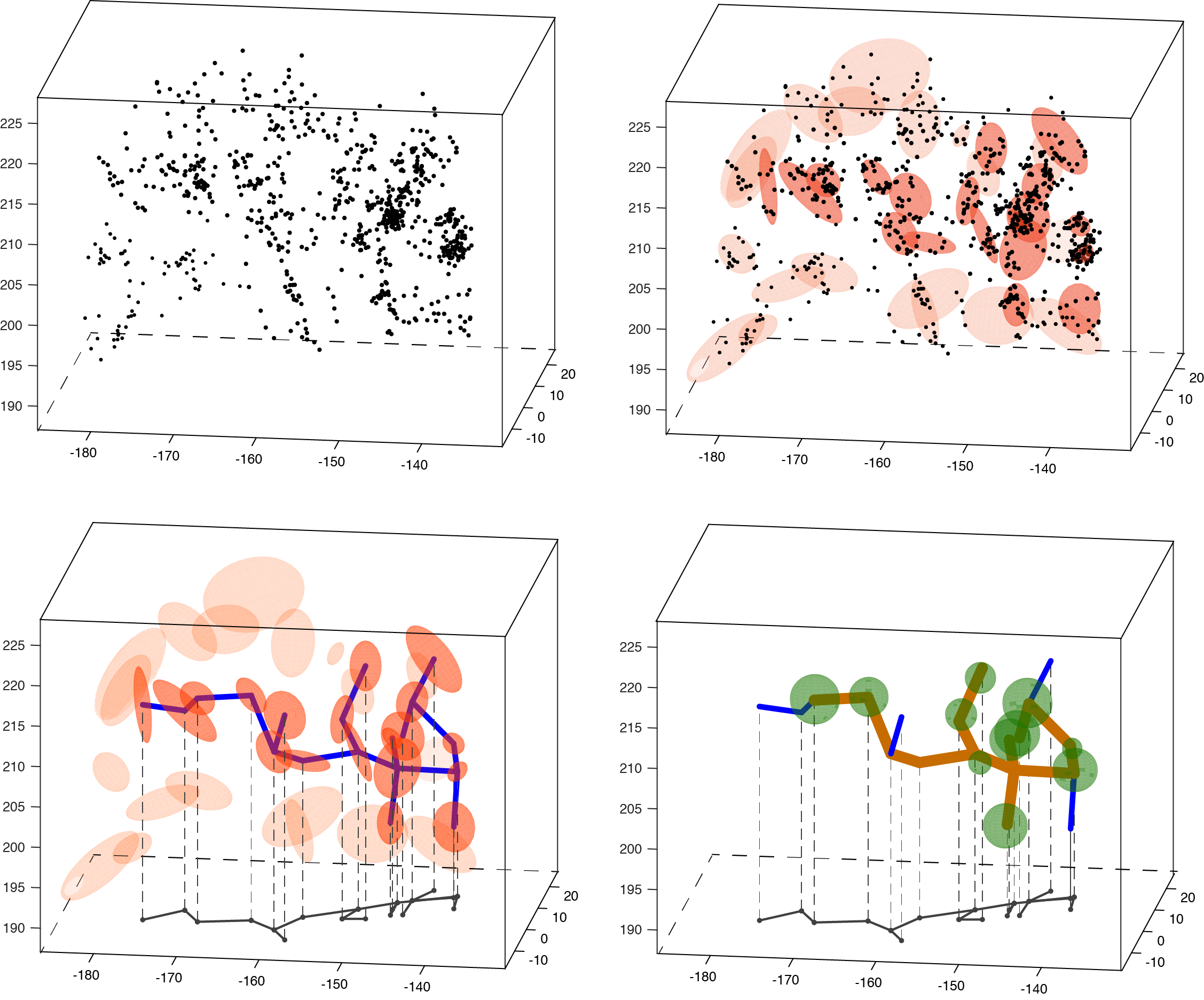}
	\caption{Illustration of the steps of the GFiF algorithm. In the
first box (top-left) one can see the distribution of galaxies. In the 
second one (top-right) the \emph{HC} groups are marked, with denser
red colors representing the richer \emph{HC} groups. The filtered edges (links) 
among the groups of the spanning tree are displayed in the third box 
(bottom-left).
The last box (bottom-right) presents the systems (green circles), bridges 
(brown lines) and other links (blue lines) found among the groups of the 
preceding step.}
	\label{illustr}
\end{figure}

Although isolated bridges (that is, connecting only one 
pair of systems) and  tendrils (connections between nodes with no system embedded) are important and are also a sub-product of the 
algorithm, we will focus our discussion hereafter only on the filaments.
The connecting edges of these filaments are then refined using  
Dijkstra's algorithm (section \ref{Dij}).
This refinement allows the identification of the filament skeleton, 
i.e. the principal branch connection.
According to the pattern recognition literature, a skeleton 
represents the principal features of an object such as topology, 
geometry, orientation and scale. 
Figure \ref{illustr} shows schematically the steps of the GFiF
algorithm.

The results of the filaments-finding algorithm depend on several parameters,
in particular the number of \emph{HC} groups, $N_{cut}$ (or, equivalently, $f$),
and the linking length $D_{max}$. 
Therefore, it is necessary to carry out a search for the optimal combination 
of these parameters.
In addition, in order to find the longest filaments possible inside the
supercluster volume, we search for the linking length that maximizes 
the number of filaments in the box, that is just before they begin to
percolate.
The optimization for these parameters 
is described in detail in \citet{Santiago-Bautista2019a}.
{We
found that the optimized parameter $f$ decreases with $z$ from about
20 to 10 galaxies in the range covered by our sample, that is, 
depending on the sampling, a smaller density is found with
increasing $z$. 
The $D_{max}$, in turn, increases with $z$, in a way
to compensate the decrease in $f$.}

{A schematic representation of the GFiF algorithm
can be found on the right side of Figure \ref{flujograma}.}

\section{Detection algorithms in action} 
\label{aplic}
In order to illustrate the detection algorithms presented above, we now 
describe their application to one of the superclusters in our sample, 
MSCC\,310, the \emph{Ursa-Majoris} Supercluster 
(see Tables \ref{table1} and \ref{table2}).
This supercluster contains 21 Abell clusters, with redshifts in the range 
from 0.05 to 0.08, and it is one of the largest in volume
in our sample: it occupies an area in the sky of about 1\,700 deg$^2$, 
equivalent to a volume of {($116~h_{70}^{-1}$~Mpc)$^{3}$} 
(including the 20~$h_{70}^{-1}$~Mpc added to the box limits from the farthest 
clusters).

The volume contains $N = 12\,286$ SDSS galaxies with spectroscopic 
redshift. 
This corresponds, to a mean surface density of 76.9 gal.deg$^{-2}$ or  
0.008 gal.$h^{3}_{70}~$Mpc$^{-3}$, see Table \ref{table2}.
The list of parameter values used is shown in Table \ref{parameter_m}.

\subsection{Application of GSyF to MSCC\,310}

\begin{figure}[!t]
	\includegraphics[width=10cm]{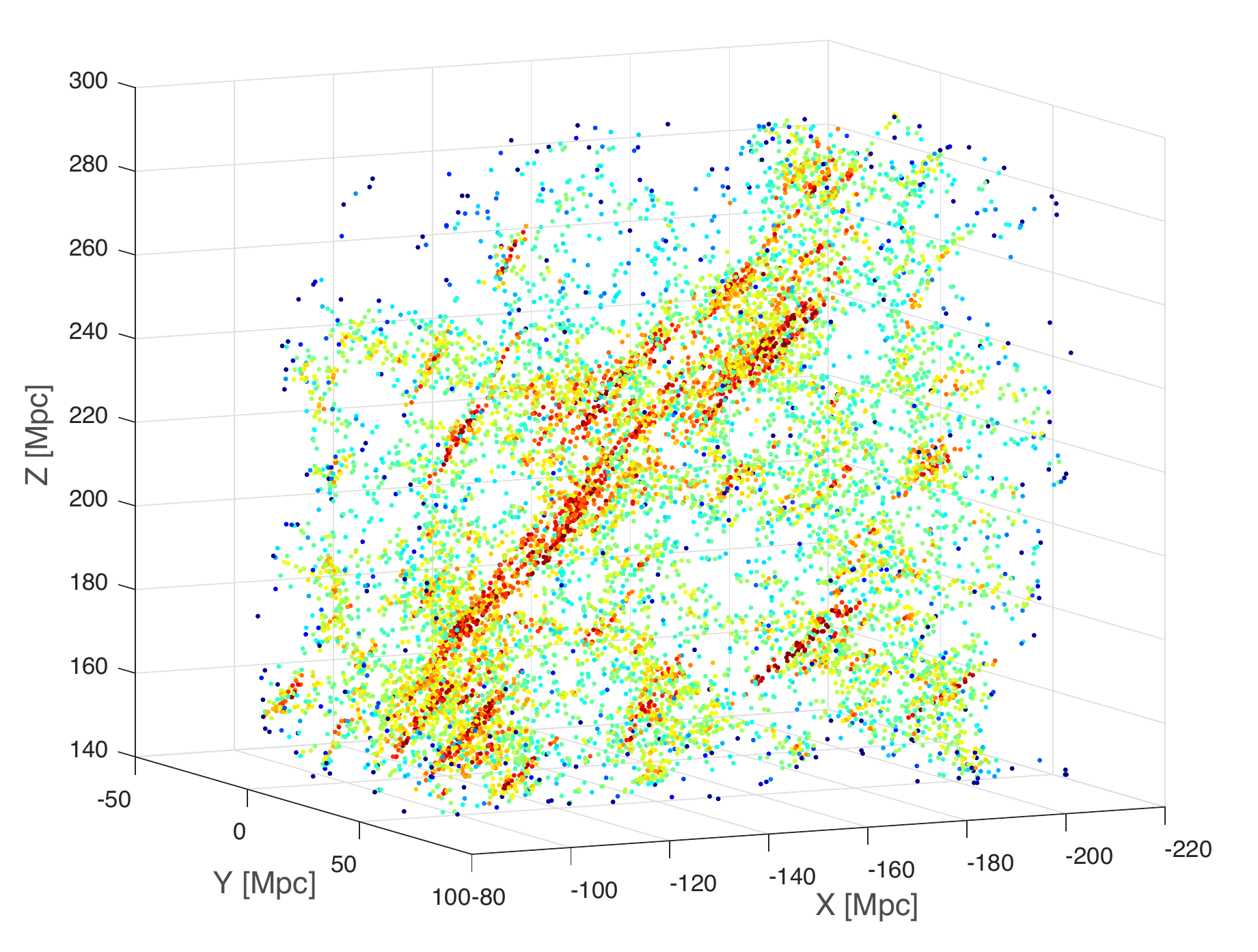}
	\includegraphics[width=10cm]{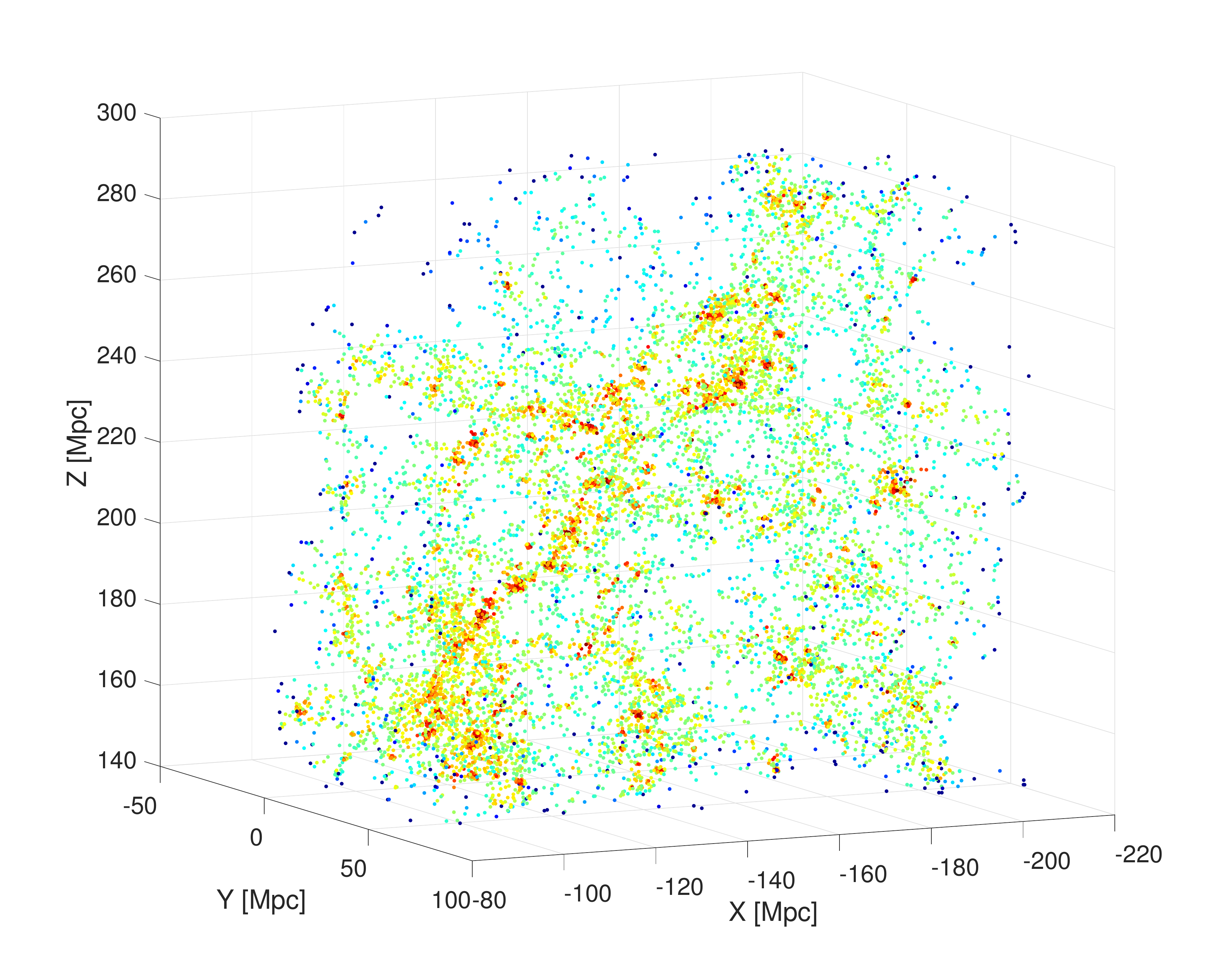}
	\caption{3D galaxies for the MSCC\,310 supercluster volume. 
          Top, galaxy positions before the application of the FoG correction. 
          Bottom, Galaxy positions after correction for FoG effects. 
          The color represent density as calculated from 3D \emph{VT}. 
          The highest density is represented in red while green to blue 
          represents lower densities. }
	\label{FoG-execution}
\end{figure}

First we applied the \emph{VT} algorithm to the projected distribution of 
MSCC\,310 galaxies and calculated their $d_i$. 
Then we made 1\,000 simulations to estimate $d_{bas}$ (15.7~deg$^{-2}$,
for this case).
We applied the \emph{HC} algorithm to the $N_{gal} =$ 7\,529 galaxies with 
$\delta_i > 0$.
After calculating the best $f$ parameter from the 30
mock simulations ($f=6$, in this case), we have taken the 
$N_{cut} =$ 1\,140 groups generated from the \emph{HC} application.
As expected, these groups 
have, on average, $\sim 6$ members. 
Of these, we retained  1\,015 with $N_j \geq 3$.

The iterative virial refinement was initialized by assigning the center 
of each \emph{HC} group at the brightest r-band galaxy member close to its 
geometrical centroid (see Table \ref{systemsMSCC310}).
For MSCC\,310 groups, the mean difference between the geometrical 
center and brightest groups' galaxy projected position 
was found to be about 350~$h_{70}^{-1}$~kpc.
On average, the virial refinement needed six iterations to produce 
convergence to the virial radius.
This refinement resulted in 122 systems with $N_{mem} \geq 10$ for the 
MSCC\,310 volume. 
The refinement also detected 113 smaller systems with $5 \leq N_{mem} < 10$.

In Table \ref{systemsMSCC310} we list the properties of the first 
25 richest systems for the MSCC\,310 supercluster. 
Column 1 assigns a sequential number to the systems, while 
column 2 presents their richness. 
The coordinates of the brightest member of the system are indicated in 
columns 3, 4 and 5, while columns 6, 7 and 8 show the coordinates of the 
final position of the centroid.
The other calculated properties of the systems -- velocity dispersion, 
harmonic and virial radius -- are presented in columns 9, 10 and 11 
respectively. 
Column 12 denotes the cross-reference with Abell clusters.
The range of virial radii of the GSyF systems with $N_{mem} \geq 10$
in MSCC\,310 was $0.7-2.5~h_{70}^{-1}$~Mpc.
For groups with $5 \leq N_{mem} < 10$ the range of virial radius lies within 
$0.4-0.9$~$h_{70}^{-1}$~Mpc.
After the refinement, the projected central position of the systems changed, 
on average by 170~$h_{70}^{-1}$~kpc, while the redshift was refined for some 
cases up to $\Delta z\sim0.001$ or $\Delta\sigma_{v}\sim 300$km~s$^{-1}$. 

As an example, the richest system in MSCC\,310 is the cluster A1291\,A.
Its \emph{HC} initial centroid position (set as the position of the brightest 
galaxy in the \emph{HC} group: $\alpha=172.73$, $\delta=56.49$ and $z=0.0611$) 
changed by 13\Mpc  after 17 iterations of the virial refinement (the 
final centroid position corresponds to $\alpha=173.01$, $\delta=56.09$, 
$z=0.0535$).
This position is at 240~$h_{70}^{-1}$~kpc from the system's brightest 
galaxy detected for A1291\,A which has coordinates \citep[$\alpha=173.05$, 
$\delta=56.05$, $z=0.0585$, ][]{Lauer2014}. 

Finally, as described in section \ref{FoG}, we correct the $D_C$ of the 
member galaxies in each system by re-scaling their dispersion range to 
the $R_{vir}$ of the system.
An example of the MSCC\,310 volume, before and after the correction, 
is shown in Figure \ref{FoG-execution}.

\begin{table*}
   \caption{\label{systemsMSCC310} Main properties of the 25 richest systems identified in the volume of the supercluster MSCC\,310.}
   \begin{center}
		{\small \begin{tabular}[]{r r r r r r r r r c c l }
		\hline\hline                 
	System & $N_{mem}$ &  \multicolumn{3}{c}{C$_{BMG}$}&\multicolumn{3}{c}{C$_{FoG}$}  & $\sigma_v$    &             $R_{h}$  & $R_{vir}$           & cross-ref \\
	Nr.    &         & RA$_{J2000}$ & Dec$_{J2000}$ & $z$             &  RA$_{J2000}$  & Dec$_{J2000}$ & $z_{LOS}$                  & [km~s$^{-1}$] &  [$h_{70}^{-1}~$Mpc]  & [$h_{70}^{-1}~$Mpc] & ACO Nr.\\
	(1) & (2) & (3) & (4) & (5) & (6) & (7) & (8) & (9) & (10) & (11) & (12)\\ 
	\hline		
 1 & 123 & 173.09667 & 55.96744 & 0.0515 & 173.01489 & 56.09476 & 0.0535 & 1182 & 1.09 & 2.54 & A1291A\\
 2 & 103 & 174.01464 & 55.07526 & 0.0571 & 174.17793 & 55.19984 & 0.0587 & 1103 & 1.28 & 2.56 & A1318A\\
 3 &  95 & 180.26970 & 56.37019 & 0.0648 & 180.07048 & 56.20431 & 0.0649 &  762 & 1.04 & 1.87 & A1436\\
 4 &  94 & 167.09625 & 44.15030 & 0.0587 & 167.10957 & 44.07194 & 0.0590 &  644 & 0.81 & 1.53 & A1169\\
 5 &  91 & 176.83909 & 55.73018 & 0.0515 & 176.80652 & 55.69322 & 0.0518 &  712 & 0.80 & 1.64 & A1377\\
 6 &  82 & 177.19063 & 54.51936 & 0.0601 & 177.05852 & 54.64399 & 0.0604 &  845 & 0.95 & 1.94 & A1383\\
 7 &  67 & 168.84947 & 54.44412 & 0.0695 & 168.90590 & 54.50901 & 0.0700 &  659 & 0.91 & 1.63 & \\
 8 &  61 & 175.27722 & 55.18836 & 0.0593 & 175.25812 & 55.29493 & 0.0609 & 1103 & 1.30 & 2.58 & A1349A\\
 9 &  61 & 163.40237 & 54.86794 & 0.0716 & 163.54104 & 54.84312 & 0.0722 &  640 & 0.89 & 1.58 & \\
10 &  59 & 172.33060 & 54.12608 & 0.0689 & 172.44575 & 54.08347 & 0.0690 &  582 & 0.75 & 1.40 & A1270\\
11 &  54 & 158.24537 & 56.74813 & 0.0448 & 158.32527 & 56.82434 & 0.0454 &  459 & 0.69 & 1.16 & \\
12 &  52 & 180.22849 & 51.42263 & 0.0666 & 180.46535 & 51.65240 & 0.0649 & 1069 & 1.08 & 2.37 & A1452\\
13 &  50 & 152.32009 & 54.21099 & 0.0465 & 152.41409 & 54.42037 & 0.0460 &  415 & 0.68 & 1.08 & \\
14 &  46 & 183.70265 & 59.90619 & 0.0600 & 183.59629 & 59.90299 & 0.0599 &  443 & 0.75 & 1.17 & A1507B\\
15 &  43 & 168.06955 & 57.07599 & 0.0471 & 168.12753 & 57.04832 & 0.0467 &  491 & 0.77 & 1.26 & \\
16 &  42 & 178.37743 & 52.68944 & 0.0716 & 178.59933 & 52.77017 & 0.0695 &  761 & 0.84 & 1.74 & \\
17 &  39 & 151.21598 & 54.56786 & 0.0470 & 150.99614 & 54.65456 & 0.0472 &  460 & 0.64 & 1.13 & \\
18 &  39 & 163.28303 & 56.33167 & 0.0772 & 163.35966 & 56.33847 & 0.0745 & 1003 & 0.89 & 2.13 & \\
19 &  36 & 172.42861 & 55.38047 & 0.0685 & 172.44796 & 55.42240 & 0.0684 &  534 & 0.54 & 1.19 & \\
20 &  36 & 162.94746 & 55.38567 & 0.0739 & 162.89992 & 55.34739 & 0.0737 &  367 & 0.68 & 1.00 & A1112A\\
21 &  34 & 182.19381 & 53.33371 & 0.0813 & 182.19494 & 53.31805 & 0.0821 &  573 & 0.66 & 1.33 & \\
22 &  33 & 181.31486 & 43.16902 & 0.0528 & 181.40583 & 43.20470 & 0.0526 &  504 & 0.67 & 1.23 & \\
23 &  31 & 177.05076 & 52.85209 & 0.0503 & 177.04947 & 52.59945 & 0.0505 &  556 & 0.67 & 1.31 & \\
24 &  31 & 178.57105 & 55.47082 & 0.0508 & 178.68334 & 55.20458 & 0.0513 &  584 & 0.82 & 1.45 & \\
25 &  31 & 151.31217 & 53.14899 & 0.0463 & 151.31677 & 52.99140 & 0.0451 &  431 & 0.68 & 1.11 & \\ 
	\hline
\end{tabular}
\tablefoot{The complete version of this table and the tables of systems 
of the other superclusters can be found {in the electronic version
of this paper.}}
}
\end{center}
\end{table*}

\subsection{Application of GFiF to MSCC\,310}

With the co-moving distances for the MSCC\,310 galaxies corrected for
the FoG effect, we proceeded to transform their sky coordinates to
rectangular ones
following equations \ref{transformation1}, \ref{transformation2} and 
\ref{transformation3}.

The $N_{gal}$ was now taken to be the total number of galaxies in the box 
of MSCC\,310, $N =$ 12\,286, for which we applied the GFiF method.
The \emph{VT} algorithm was then applied to calculate volumetric numerical 
densities. 
We used a segmentation parameter of $f=16$ for the \emph{HC} algorithm.
From that, we identified 768 low density groups and, for each pair, we 
calculated the $D_E$ distance between centers and the $BC$ weight.
As expected, the implementation of the \emph{HC} algorithm over all galaxies 
detected larger groups ($\sim$ 15 galaxies on average now) and more 
elongated, with a mean $\sigma_j$ of 1.8\Mpc compared with the mean 
$\sigma_j$ of 0.5\Mpc found with the application of GSyF.

\begin{figure}[!t]
	\includegraphics[width=9cm]{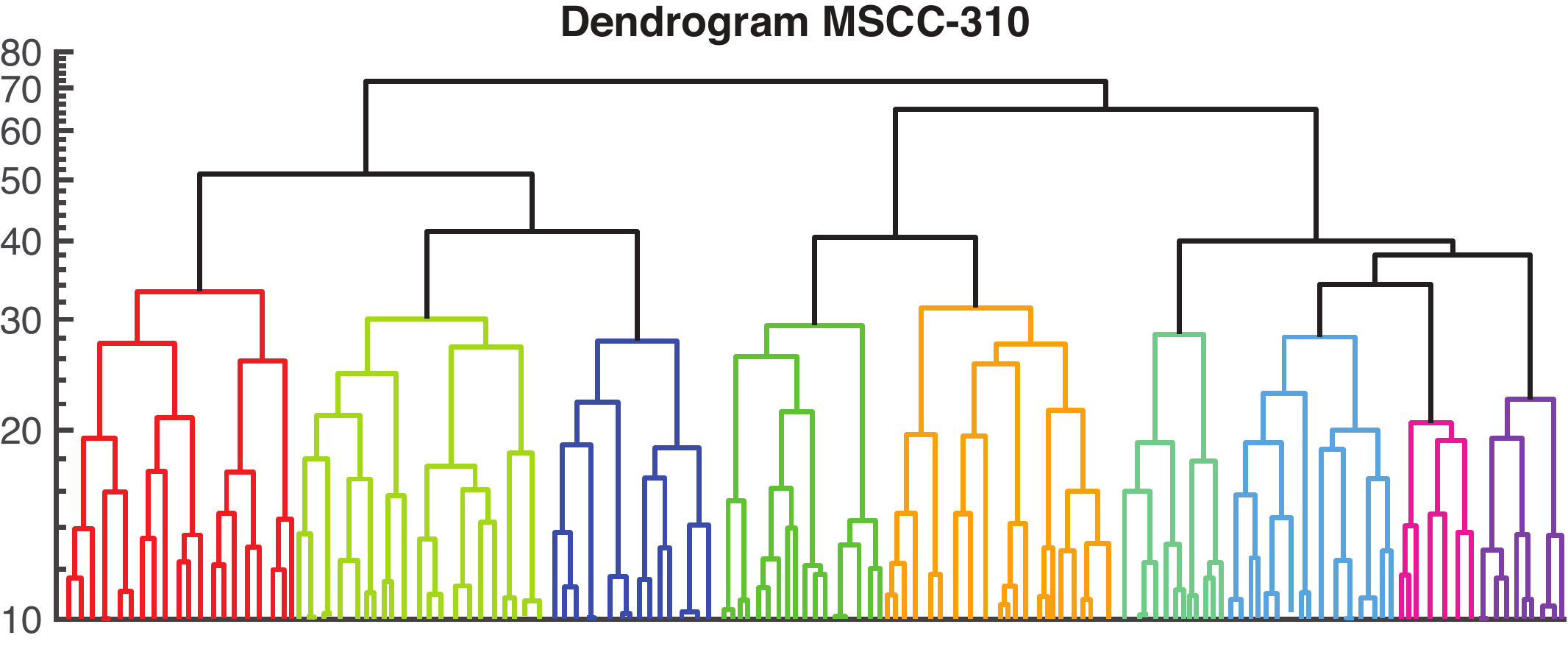}\\
	\includegraphics[width=9cm]{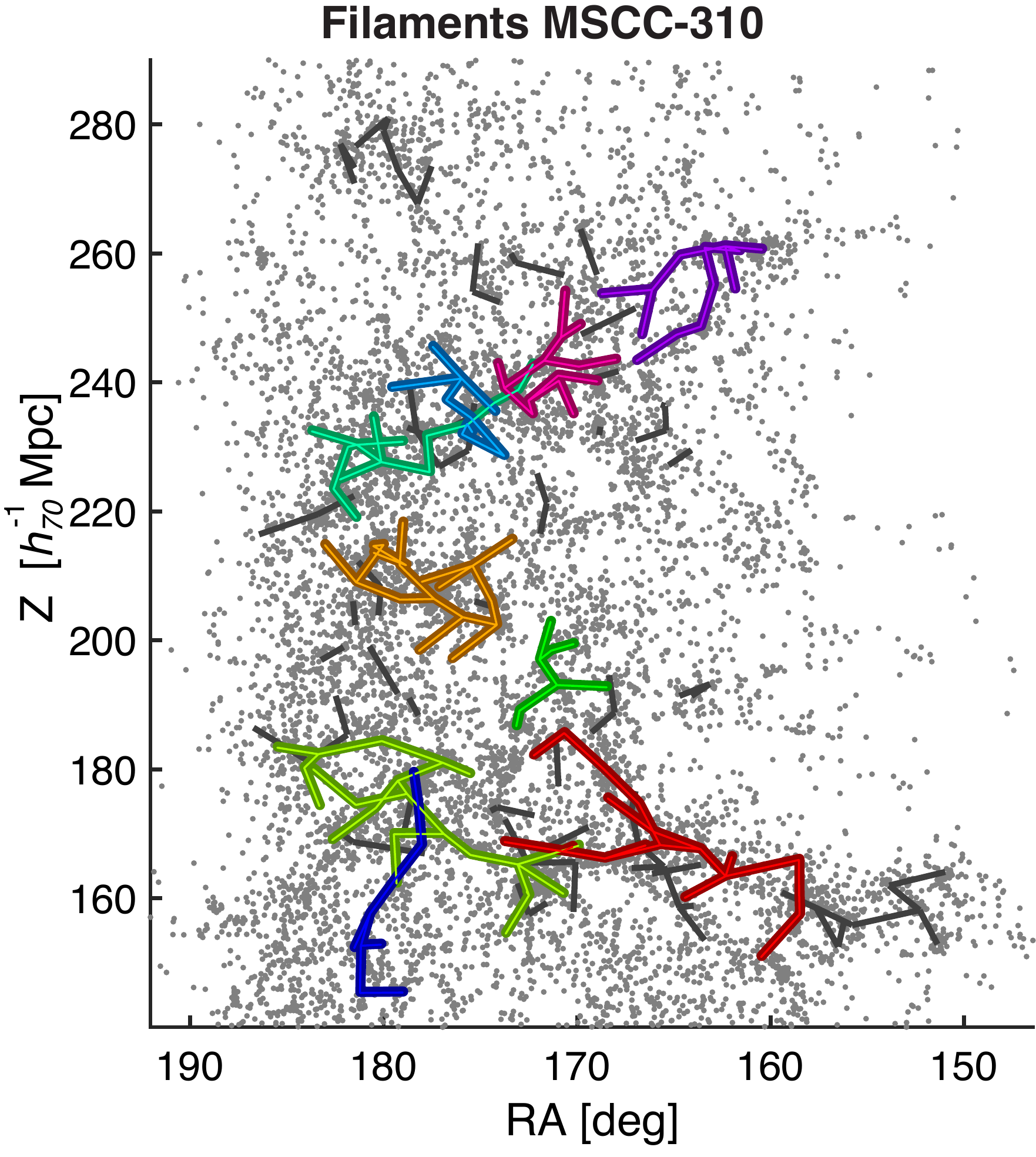}
	\caption{Results of GFiF algorithm for the MSCC\,310 supercluster 
volume. 
Upper panel: dendrogram with the 9 detected filaments represented by different 
colors.
The y axis of the dendrogram plot indicates the distance at each level of the 
tree.
Lower panel: RA$\times$Z distribution, where SDSS galaxies are represented by 
gray points and filaments are by lines according to the colors in the upper 
panel.
Tendrils are represented by gray color lines.}
	\label{filamentsMSCC310}
\end{figure}

In order to filter the connections, a linking length of $D_{max} = 8$\Mpc 
was used resulting on {334} edges. 
As described above, $D_{max}$ and $f$ were obtained by the optimization 
process described in \cite{Santiago-Bautista2019a}.
The second filter, the minimum mean linear density along 
the edge cylinders (in this case 0.008 gal.Mpc$^{-3}$), left 273 links
from the 316 connections smaller than $D_{max}$.     
This resulted in 34 trees, to which we applied \emph{MST}. 
Of these, only 9 are linking 3 or more systems of galaxies with a 
richness $N_{mem}$ above 11 galaxies, the rest are isolated bridges and 
tendrils. 
This result is shown in the dendrogram depicted in Figure 
\ref{filamentsMSCC310} (top panel) which shows 9 dominant filaments for 
the MSCC\,310 supercluster.

Concerning the systems embedded in the structures, from the
359 \emph{HC} groups (nodes) in the spanning trees, 116 matched with the systems
with $N_{mem} \ge 10$ identified with GSyF.
From these, 61 were found to be in filaments (53\%),
26 (22\%) in bridges between pairs of systems, and 29 (25\%) not connected
by bridges, that is, relatively isolated.

The filaments detected by the GFiF algorithm in the MSCC\,310 supercluster 
and their main properties are listed in Table \ref{filaments_table}. 
Column 1 assigns a sequential number to the filament;
column 2 lists the number of systems detected by GSyF linked by the filament;
column 3 shows the number of galaxies attributed to the filament;
columns 4 to 6 are the mean, min. and max. redshift of the filament; 
column 7 corresponds to the mean number density inside the filament;
column 8 is the mean transversal radius of the filament measured 
at $10 \times d$;
columns 9 and 10 show the number of nodes that constitute the filament 
and the number of central skeleton nodes, respectively; 
column 11 is the length of the filament skeleton.

We can also observe the filaments inside the MSCC\,310 volume in Figure 
\ref{filamentsMSCC310} {(bottom panel)}. 
In this panel {the 9 filaments are plotted over 
the distribution of galaxies} in a RA\,[deg]$~\times~$Z\,[Mpc] plane. 
This projection allows the {recognition} of structures both 
in {one of the coordinates of the sky plane} and depth.
Filaments are depicted in colors, the same colors in both panels of this 
figure. {Isolated bridges (that connect two systems alone, 
without forming a filament)} {are represented only in the bottom 
panel and by black lines. Tendrils are not represented to avoid crowding.}
The longest paths for the filament {skeletons}, that is,
those that connect the farthest systems of each filament, 
range from 18 to 62\Mpc and connect 
up to 11 systems inside the MSCC\,310 volume. 
Moreover, we measured the paths between pairs of systems chained together
by bridges; such distances range from 5 to 24\Mpcp.

\begin{table*}
  \caption{\label{filaments_table} Main properties of the filaments 
    extracted through GFiF for the supercluster MSCC\,310.}
    \begin{center}
      {\small \begin{tabular}[]{c r r c c c c c r r c}
\hline\hline                 
Fil. & N$_{sfil}$ & N$_{gfil}$ & \multicolumn{3}{c}{redshift} & $d_{fil}$                 & $R_{fil}$           & \multicolumn{2}{c}{N$_{nod}$} & $\ell_{fil}$ \\
ID   & systems    & gals.      &     [mean, &min, &max]       & [$h_{70}^{3}~$Mpc$^{-3}$] & [$h^{-1}_{70}~$Mpc] & filament  & skeleton          & [$h^{-1}_{70}~$Mpc] \\
(1)  & (2)        & (3)        & (4) & (5) & (6)              & (7)                       & (8)                 & (9)       & (10)              & (11)\\ 
\hline		
MSCC\,310-F1 & 10 & 499 & 0.0609 & 0.0518 & 0.0689 & 0.4180 & 2.81 & 18 & 11 & 61.6\\
MSCC\,310-F2 &  8 & 523 & 0.0502 & 0.0443 & 0.0588 & 0.5210 & 2.84 & 22 & 11 & 51.8\\
MSCC\,310-F3 &  7 & 407 & 0.0481 & 0.0427 & 0.0528 & 0.3486 & 2.50 & 19 & 10 & 49.0\\
MSCC\,310-F4 &  8 & 313 & 0.0656 & 0.0585 & 0.0710 & 0.4263 & 2.46 & 14 & 10 & 59.0\\
MSCC\,310-F5 &  7 & 325 & 0.0700 & 0.0642 & 0.0774 & 0.5303 & 2.60 & 13 &  7 & 47.6\\
MSCC\,310-F6 &  6 & 243 & 0.0725 & 0.0651 & 0.0791 & 0.2986 & 1.97 & 12 &  9 & 39.3\\
MSCC\,310-F7 &  4 & 219 & 0.0551 & 0.0479 & 0.0619 & 0.4062 & 2.76 &  7 &  5 & 20.7\\
MSCC\,310-F8 &  4 & 164 & 0.0546 & 0.0485 & 0.0617 & 0.2118 & 2.22 &  9 &  7 & 33.7\\
MSCC\,310-F9 &  4 & 124 & 0.0464 & 0.0437 & 0.0528 & 0.1464 & 1.13 &  9 &  6 & 17.9\\ 
\hline
		\end{tabular}
	}
\end{center}

\end{table*}
\begin{table*}
   \caption{\label{parameter_m} Glossary of parameters used by GSyF and 
                                GFiF algorithms
   }
   \centering
      \scalebox{0.9}{
         {\small \begin{tabular}[]{l l l }

\hline \hline                 
Param.& Description & MSCC\,310 \\

\hline \hline
\multicolumn{2}{c}{Properties of the rectangular box} \\
\hline
$N$                    & Total nr. of galaxies                           & 12\,286 \\
$V$	               & Total volume                                    & (116 $h_{70}^{-1}$ Mpc)$^3$ \\	
$d=\frac{N}{V}$        & Mean volumetric number density                  & 0.008 $h_{70}^{3}~$Mpc$^{-3}$ \\
$A$	               & Projected area in the sky                       & (12.65 deg)$^2$ \\	
$d_{surf}=\frac{N}{A}$ & Mean surface number density (sky projection)    & 76.9 deg$^{-2}$ \\
$v_i, a_i$             & Local volume or projected area of \emph{VT} cell & each galaxy \\
$d_{i}$                & Local \emph{VT} (surface or volume) density            & each galaxy \\
$D_C$                  & Comoving distance of the galaxy or system       & each galaxy or system \\
$D_E$                  & Euclidean distance between two galaxies or two nodes (edge size) & each pair \\
$D_{ske}$              & Euclidean distance of galaxy from filament skeleton & each galaxy \\

\hline \hline
\multicolumn{2}{c}{Amount of systems} \\
\hline
N$_{Cl}$               & Richness of the supercluster (nr. of Abell/ACO clusters) & 21 \\ 
$N_{cut}=\frac{N_{gal}}{f}$ & Nr. of extracted \emph{HC} groups          & 1\,140, 768 \\
N$_{HC}$               & Nr. of detected \emph{HC} groups ($N_j \ge 3$) found by GSyF algorithm & 1\,015 \\
N$_{nodes}$            & Nr. of nodes found by GFiF algorithm which remained in the \emph{MST}rees & 359 \\
N$_{FoG}$              & Nr. of systems (that survived the FoG filter)   & 255 \\
N$_{sys}$              & Nr. of systems with $N_{mem} > N_{min}$ in box which matched nodes & 116 \\
N$_{sfil}$             & Nr. of systems embedded in the filaments        & 61 \\
N$_{spair}$            & Nr. of systems forming pairs connected by isolated bridges & 26 \\
N$_{sout}$             & Nr. of systems not forming filaments or pairs   & 29 \\

\hline \hline
\multicolumn{2}{c}{Amount of filaments, isolated bridges and tendrils} \\
\hline
N$_{fil}$              & Nr. of filament candidates of the supercluster  & 4 \\ 
N$_{ske}$              & Nr. of detected filaments in the box (linked bridges) & 9 \\ 
N$_{brid}$            & Nr. of detected isolated bridges between two systems only & 17 \\ 
N$_{tend}$             & Nr. of detected tendrils (not bridges or filaments) & 18 \\

\hline \hline
\multicolumn{2}{c}{Properties of the systems} \\
\hline
$P_j(x)$               & Gaussian model for detected \emph{HC} group     & each \emph{HC} group \\
$N_j$                  & Richness the detected \emph{HC} group           & each \emph{HC} group \\
$C_j$                  & Centroid position of the detected \emph{HC} group & each \emph{HC} group \\
$\sigma_j$             & Compactness (covariance) of the detected \emph{HC} group & each \emph{HC} group \\
C$_{BMG}$              & Position of Brightest \emph{HC} group Member Galaxy & each \emph{HC} group \\
C$_{FoG}$              & Centroid position of the detected FoG system    & each system\\
$N_{mem}$              & Richness (nr. of galaxies) of the detected FoG system & each system \\
$N_{min}$              & Minimum nr. of galaxies for systems in filaments at different $z$ & 10 \\
$R_{h}$                & Harmonic radius of the detected FoG system      & each system \\
$R_{vir}$              & Virial radius of the detected FoG system        & each system \\
$M_{vir}$              & Virial mass of the detected FoG system          & each system \\
$v_{LOS}$              & Robust line-of-sight velocity of the detected FoG system & each system \\
$\sigma_{v}$           & Robust velocity dispersion of the detected FoG system & each system \\

\hline \hline
\multicolumn{2}{c}{Properties of the filaments} \\
\hline
N$_{edges}$            & Nr. of edges that survived filter 1             & 334 \\
N$_{links}$            & Nr. of edges that survived filters 1 and 2      & 316 \\ 
N$_{trees}$            & Nr. of trees after \emph{MST}                   & 17 \\ 
N$_{nod}$              & Nr. of nodes in the filament (or in the skeleton) & each filament (skeleton) \\
$\ell_{fil}$           & Length of filament skeleton                     & each filament \\
$R_{fil}$              & Mean radius of the filament                     & each filament \\
$d_{fil}$              & Mean galaxy number density inside the filament  & each filament \\
N$_{gfil}$             & Nr. of galaxies hosted in the filaments of the box & 2\,568 \\
$V_{fil}$              & Volume occupied by the filaments of the box     & (1.16 $h_{70}^{-1}$ Mpc)$^3$ \\

\hline \hline
\multicolumn{2}{c}{GSyF and GFiF parameters} \\
\hline
$d_{bas}$              & Projected number density baseline               & 15.7 deg$^{-2}$ \\
$\delta_i$             & Local density contrast                          & each galaxy \\
$N_{gal}$              & Nr. of galaxies above the baseline $d_{bas}$    & 6\,842 \\
R$_a$, S$_a$, M$_a$    & Parameters of iterative process for FoG correction & each \emph{HC} group \\
$f$                    & Segmentation parameter (OPTIMIZATION)           & 6, 16 \\
$BC$                   & Bhattacharyya coefficient (edge weight)         & each edge \\
$D_{max}$              & Linking length (first filter) (OPTIMIZATION)    & 8 $h_{70}^{-1}~$Mpc \\
$d_{edge}$             & Edge cylinder density (second filter)           & each edge \\
$R_{cy}$               & Filament concentric cylinder radius             & each filament \\
\hline
         \end{tabular}
         }
      }
\end{table*}

\section{Validation of the methods}
\label{valid}
\subsection{Checking the identified systems of galaxies}
\label{valid_syst}

In order to validate our GSyF algorithm we compared the list of identified 
systems to different cluster and group catalogs in the region of SDSS. 

For MSCC\,310, for instance, GSyF detected 122 systems with ten or more 
galaxies and another 113 systems with $5 \leq N_{mem} < 10$.
A match was considered positive if the projected positions of the system in 
the two compared catalogs were not farther than 1 $h_{70}^{-1}~$Mpc, while 
in redshift space we considered a difference $\Delta z=0.007$ which 
corresponds to $\pm 2\,100$\,km~s$^{-1}$.

For the rich clusters, first we compared our results to the original 
Abell/ACO catalog \citep{ACO1989}, based on the most recent parameter 
measurements for its clusters \citep[e.g.][]{chow2014}. 
Also, we compared the detected systems against the 
central galaxy position provided by the \textit{Brightest 
Cluster Galaxy} catalog \citep[][hereafter L14]{Lauer2014}.
Regarding catalogs based on the SDSS spectroscopic 
sample we compared with the C4 cluster catalog 
\citep{Miller2005}, based on the SDSS-DR2.
For less rich clusters and groups we compared our systems with the 
\textit{Multi-scale Probability Mapping clusters/groups} catalog 
\citep[MSPM,][]{Smith2012} and the \cite{Tempel2011} catalog 
(hereafter T11), carried over the SDSS-DR7 and -DR8 respectively.  
For the comparisons we used all systems detected by GSyF down to a richness 
of 5 galaxies.

By using the tolerance cylinder described above,
19 of the 37 Abell/ACO clusters inside the MSCC\,310 box were detected 
as systems of richness above 5 galaxies with our method (51\%), 
while the equivalent number was 26 (76\%) for the 34 clusters in C4. 
There are 11 clusters {in the L14 catalog} embedded in the 
volume and 8 (73\%) of them have GSyF counterparts.
However, by increasing the aperture to 2 $h_{70}^{-1}$~Mpc, we 
increased the detection of Abell clusters to 29/37 (78\%), C4 
clusters to 33/34 (97\%) and L14 catalog to 100\%, see Table 
\ref{counterparts}.
The increase of $20-30$\% in cluster matches by using a larger aperture 
size can be related to the fact that the mean separation of 
member galaxies 
increases for lower richness systems, and the determination of the cluster  
center then is subject to this separation, see Table \ref{counterparts}.
For example, A1452 and A1507\,B have a GSyF counterpart located at  
$\sim$1.5\Mpc projected distance and $\Delta\sigma_{v}$ 
of $\sim 630$ km~s$^{-1}$ and 120 km~s$^{-1}$ respectively (See Table 
\ref{systemsMSCC310}, systems No. 12 and 14), while their C4 
counterparts are 0.7 and 0.4\Mpc away 
respectively.

\begin{figure*}[!t]
   \centering
	\includegraphics[width=13cm]{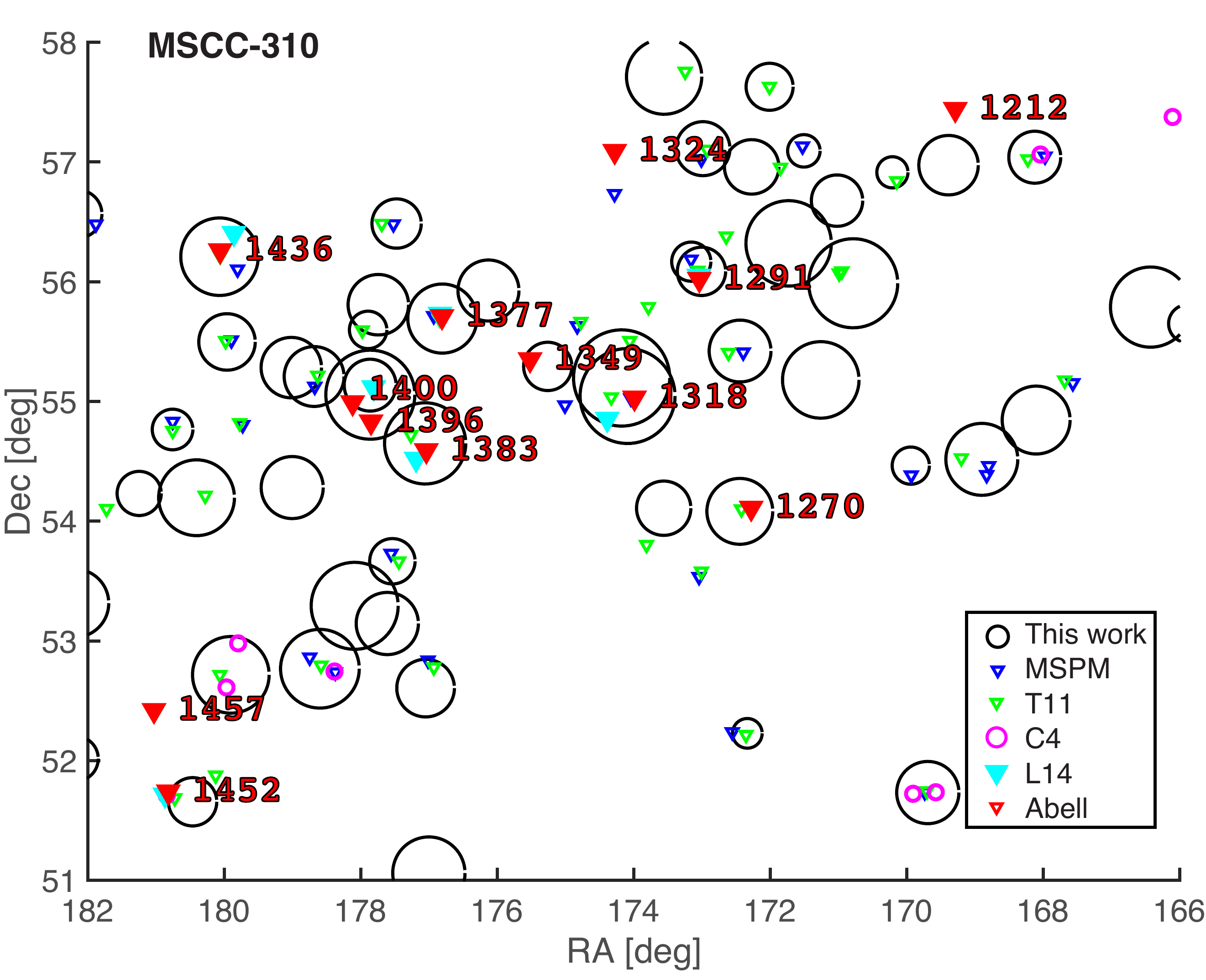}\\
	\includegraphics[width=13cm]{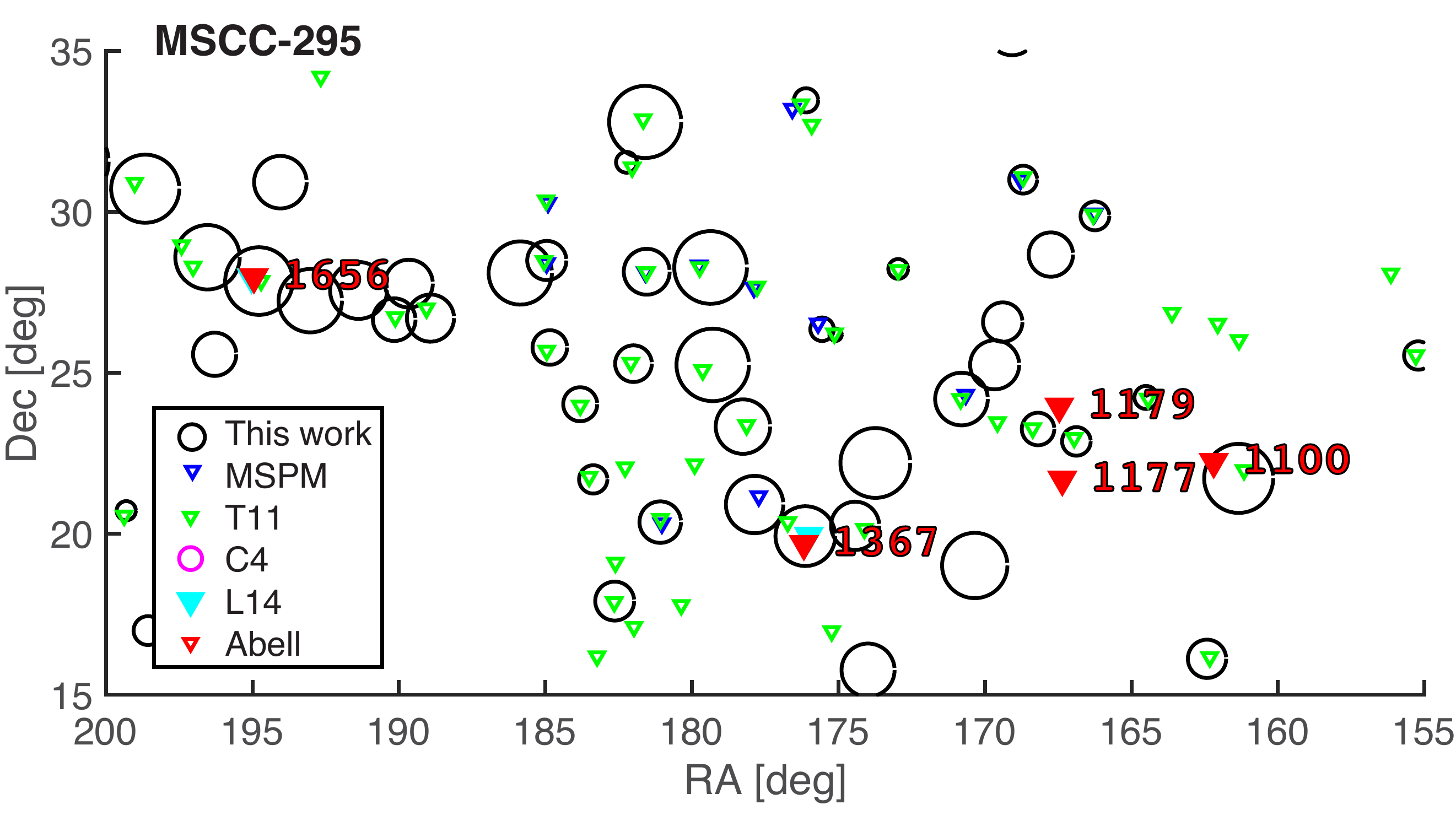}\\
	\caption{Projected distribution in the sky of systems 
	detected by GSyF. Top: Systems detected for the MSCC\,310 (UMa 
        supercluster). Bottom:  Systems detected for the MSCC\,295 (Coma 
        supercluster). The system radii are shown as circles of $r=R_{vir}$.
	For comparisons, the position of systems reported by MSPM, T11, 
        C4, L14 and Abell catalogs are depicted by color points: blue, 
        green, pink, cyan and red, respectively.}
	\label{systemsMSCC310_fig} 
\end{figure*}

\begin{figure}[!t]
	\includegraphics[width=8cm]{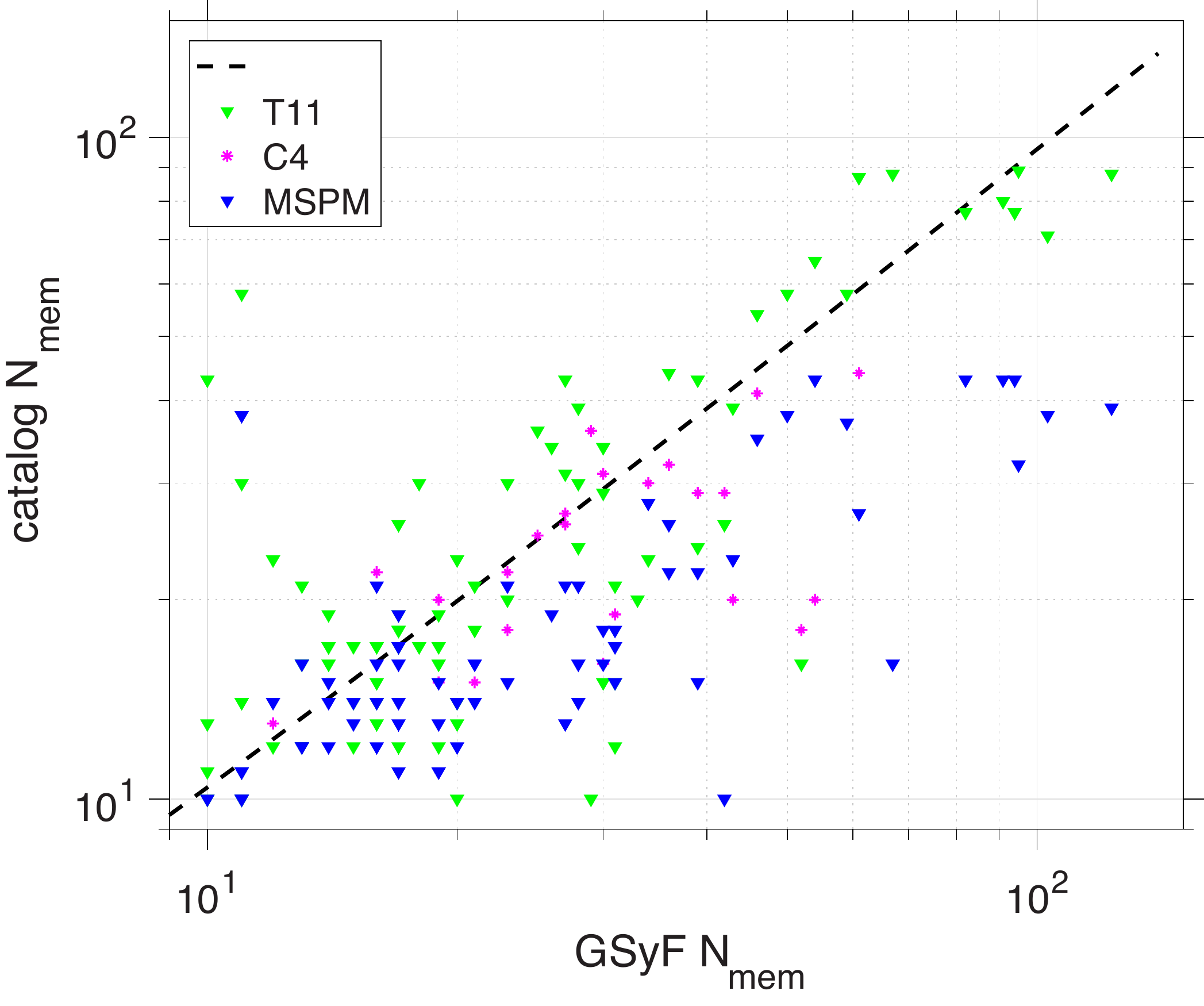}
	\caption{Comparison of MSCC\,310 supercluster GSyF system richness 
against the richness measured by other catalogs for the matching systems. 
Symbol colors are the same as the ones in Figure \ref{systemsMSCC310_fig}.  
The dashed line represents the identity.}
	\label{systemsMSCC310_richness}
\end{figure}

{Concerning the} less massive systems, there are 315 
groups detected by T11 and 213 groups listed in 
the MSPM catalog with richness larger or equal to 5 galaxies for the 
MSCC\,310 volume.
Our algorithm detected systems that correspond to 61\% {(79\%) 
of the T11 groups and 67\% (78\%) of the MSPM groups, within an 
aperture of 1 $h_{70}^{-1}$ Mpc (2 $h_{70}^{-1}$ Mpc)} 
(see Table \ref{counterparts}).
This is acceptable for our purposes since we have 
constructed GSyF to find the clusters that present FoG effect, although 
we can clearly go farther towards poorer systems with it.

The region of the UMa supercluster has been previously studied by 
\cite{Krause2013}. 
These authors identified 31 galaxy systems in the MSCC\,310 area
with a number of galaxies between 15 and 94 galaxies. 
We found that our GSyF systems match with 24 (77\%)
of these clusters within an aperture of 3\Mpcp, of which 10 are Abell 
clusters.

\begin{table*}
  \caption{\label{counterparts} GSyF systems detected by other catalogs 
                                 for the MSCC\,310 supercluster.}
  \begin{center}
     {\small \begin{tabular}[]{c c c c c c c}
     \hline \hline
	 \multicolumn{7}{c}{Aperture = 1\Mpc} \\
     \hline
Other catalog & Number     & Fraction & Number     & Fraction              & \multicolumn{2}{c}{separation} \\
    & \multicolumn{2}{c}{$N_{mem}>5$} & \multicolumn{2}{c}{$N_{mem}>10$ } & \Mpc & $\Delta\sigma_v$ \\
     \hline		
Abell & 19/37   &  51\% & 17/37  &  46\% & 0.45 & 295 \\
C4    & 26/34   &  76\% & 24/34  &  71\% & 0.43 & 100 \\
L14   & 8/11    &  73\% & 8/11   &  73\% & 0.50 & 340 \\
T11   & 192/315 &  61\% & 73/105 &  70\% & 0.33 & 230 \\
MSPM  & 142/213 &  67\% & 63/79  &  80\% & 0.34 & 145 \\
     \hline \hline				
     \multicolumn{7}{c}{Aperture = 2\Mpc} \\
     \hline
Other catalog & Number     & Fraction & Number     & Fraction              & \multicolumn{2}{c}{separation} \\
    & \multicolumn{2}{c}{$N_{mem}>5$} & \multicolumn{2}{c}{$N_{mem}>10$ } & \Mpc & $\Delta\sigma_v$ \\
     \hline
Abell & 29/37   &  78\% & 24/37  &  65\% & 0.82 & 300 \\
C4    & 33/34   &  97\% & 32/34  &  94\% & 0.77 & 166 \\
L14   & 11/11   & 100\% & 11/11  & 100\% & 0.80 & 430 \\
T11   & 249/315 &  79\% & 85/105 &  81\% & 0.64 & 320 \\
MSPM  & 167/213 &  78\% & 68/79  &  86\% & 0.55 & 193 \\
     \hline
     \end{tabular}
     }
  \end{center}
\end{table*}

The systems detected in the main portion of the MSCC\,310 supercluster
are depicted on a sky projected distribution, in Figure 
\ref{systemsMSCC310_fig} (top panel), by black circles with radius equal 
to the measured virial radius. 
The system positions from the Abell, C4, L14, MSPM and T11 catalogs are 
depicted respectively as red, pink, cyan, blue and green points. 
We also observe that the system membership number detected by GSyF is in 
agreement, for most of the cases, with the number of members for the same 
systems detected by T11, C4 and MSPM (see Figure \ref{systemsMSCC310_richness}).
{Qualitatively one can observe in this figure that 
the richness from T11 is in better agreement with our measurements, while 
MSPM estimates a richness slightly lower than both ours and T11.}

A similar analysis can be done for the other superclusters in our sample.
For example, for the \textit{Coma} supercluster (MSCC\,295, 
Figure \ref{systemsMSCC310_fig}, bottom panel),
the GSyF algorithm detected, in total, 115 systems. 
Of these, we found that A1656, the richest one, is 
composed of 579 galaxies. 
The estimated virial radius and mass are, respectively, $1.96~h_{70}^{-1}$~Mpc 
and $7.7 \times 10^{14}$ M$_\odot$.
The second richest cluster, A1367, has 243 galaxies, while its radius 
and mass are, respectively, $1.73~h_{70}^{-1}$~Mpc and  
$5.3\times 10^{14}$ M$_\odot$.
These estimations are in good agreement with those measured by 
\cite{Rines2003}.
The complete catalog of systems for each volume is available 
online\footnote{https:$//$gitlab.com$/$iris.santiagob89$/$LSS\_structures}.

\subsection{Checking the filament skeletons}
\label{validation}

\begin{figure}
	\centering
	\includegraphics[width=1\linewidth]{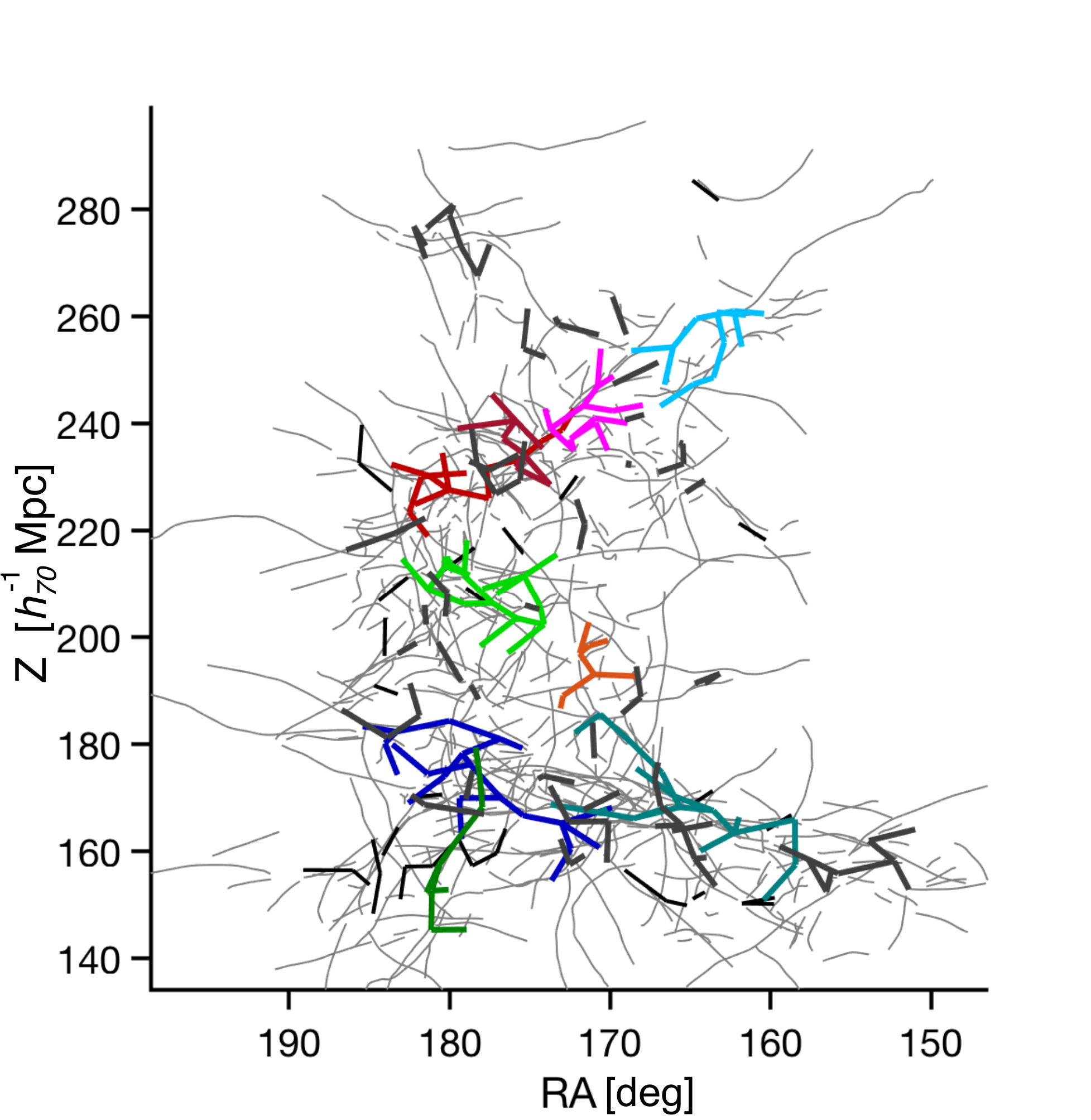}
	\caption{Comparison of GFiF filaments for the MSCC\,310 supercluster 
          to the T14 filaments in the same region for the SDSS-DR8. 
          Gray lines are T14 filaments. Colored lines depict filaments 
          identified in this work.}
	\label{Filament_proj_Tempel}
\end{figure}

We compared the filaments obtained using the GFiF algorithm for the 
MSCC\,310 volume with those presented by \cite{Tempel2014} 
(hereafter, T14) as extracted from their Table 2. 
We transformed the T14 \textit{survey coordinates} filament positions 
(see T14, equation 1) to our rectangular space and cosmology.
There are about 630 T14 filaments that lie in the sampled volume of
MSCC\,310 supercluster. 
These filaments have a mean length of 9\Mpc while the 
largest one has a length of 48\Mpcp.
As a comparison, the filament skeletons detected by GFiF
have a mean length of 42\Mpc and the largest one has a length of 
62\Mpcp. 
We found 40\% match between our detected filaments and 
T14 and 80\% match with our isolated bridges and tendrils.
The mean difference between the medial axis of T14 filaments matching the nearest filament/tendril detected by us is $\sim$ 1.5\Mpcp.
T14 filaments are represented by a sequence of  
points forming a line. 
Then, the calculated separation was taken to be the distance from the T14 filament points to the edges of our filaments.
Our filaments are depicted over T14 filaments in Figure 
\ref{Filament_proj_Tempel}.
As can be seen, GFiF detects the most prominent (dense)
filaments among the ones in T14.

\subsection{Comparison with KDE density maps}
 
\begin{figure}
  \centering
    \includegraphics[width=1\linewidth]{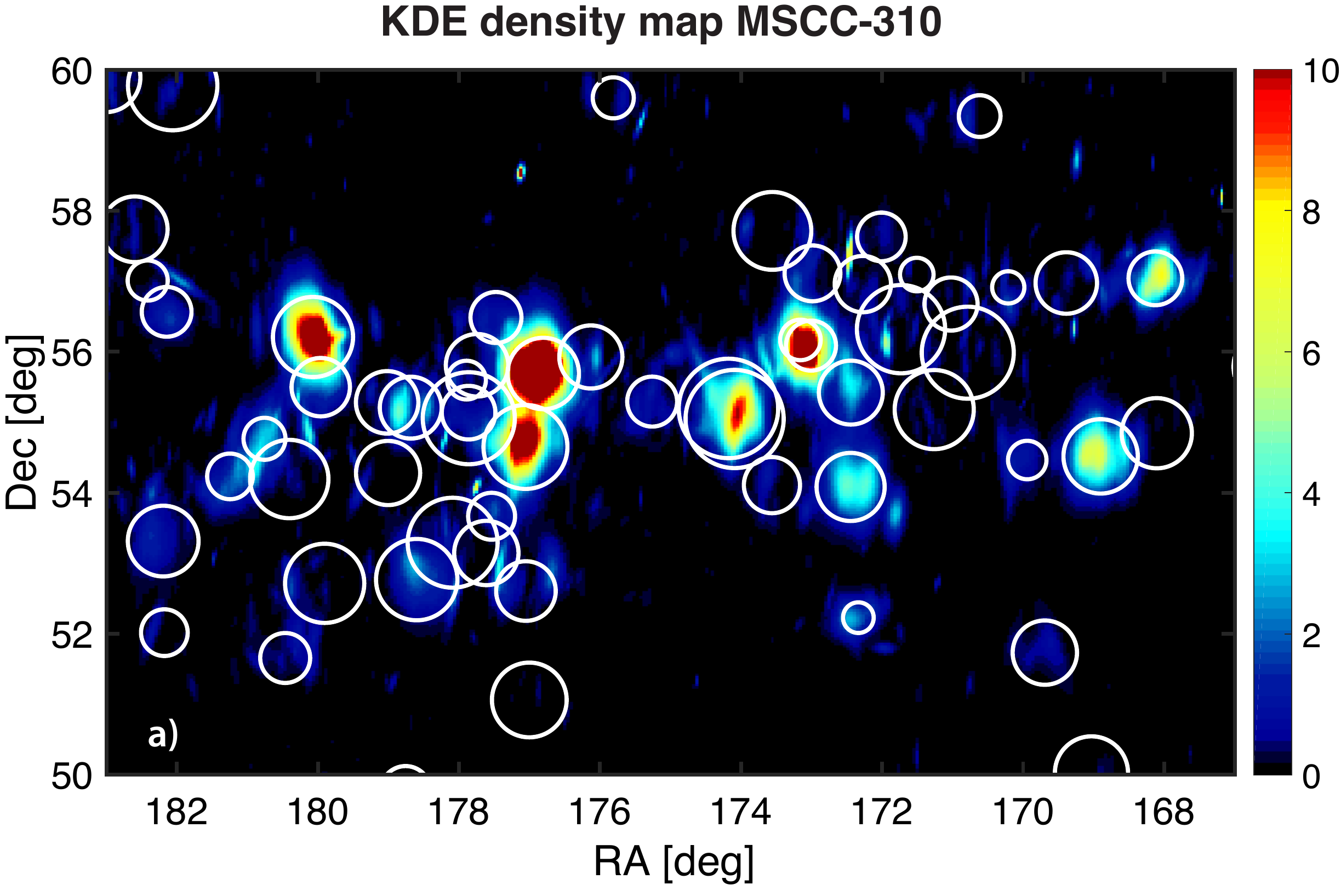} \\
    \includegraphics[width=1\linewidth]{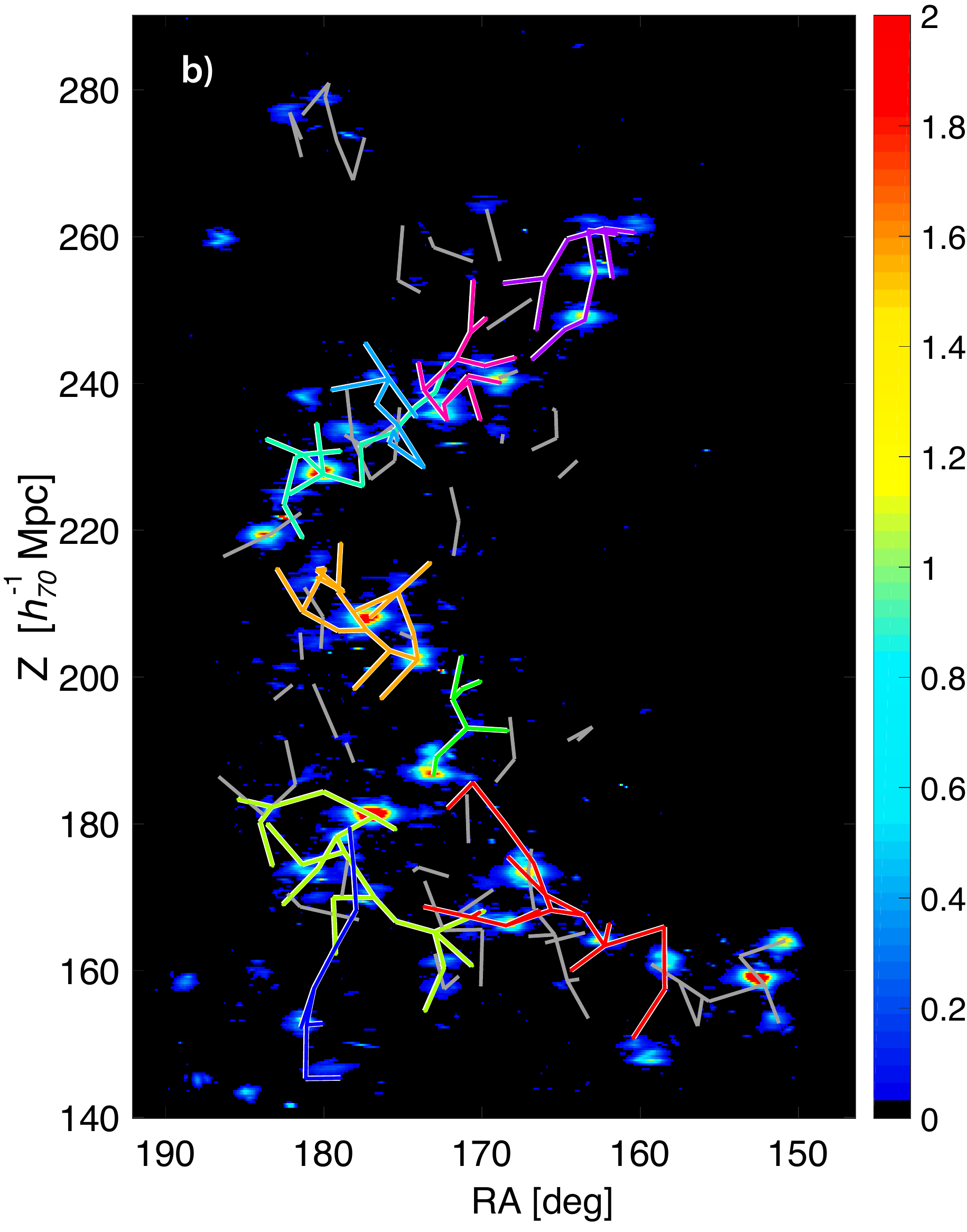}
  \caption{(Top): RA\,$\times$\,Dec. projected density map as measured from 
    3D-\emph{KDE} with 1$\Sigma$ in terms of the density contrast.  
    The GSyF systems are represented by white circles with radius scaled 
    to the estimated $R_{vir}$.
    (Bottom): RA\,$\times$\,Z projection.
    The filaments detected in this work are overlaid in color.	
    Density is represented following the color scale displayed on 
    the right, where denser regions are redder and less dense 
    zones are bluer.}
	\label{clust_kde}
\end{figure}

Another test we did in order to validate the results from GSyF and GFiF 
algorithms was to apply an independent analysis to the galaxies 
in the MSCC\,310 volume in order to corroborate the 
densities, nominally the \emph{KDE} method, as described in section 
\ref{KDE}.
A quantitative comparison in the space between the 3D \emph{KDE} and the 
skeleton structures is left for upcoming works. 
Therefore, we restricted our analysis to 2D projections (density maps) of 
the 3D \emph{KDE}, (XY, XZ, YZ). 
For this analysis we used kernels of size 1 $\Sigma$ (see section \ref{KDE}). 
Since each kernel is created based on the \emph{VT} cell, we used $d_{bas}$ 
as baseline density.
We selected those regions for which  $d_{kde} > d_{bas}$ in the 
RA$\times$Dec projected density map. 
Afterwards, we compared the position of the density peak of each region 
against the centroids of the 122 GSyF systems.
We found that 93 GSyF systems with $N_j \ge 10$ (76\%)
match density peaks above $3~d_{bas}$. 
The remaining 29 GSyF systems (24\%) are identified with density 
peaks {in the range $(1-3)~d_{bas}$}.	
Moreover, we observe that the filament edges connect these density peaks 
forming chains of overdensity regions.
In Figure \ref{clust_kde} (top panel) we show the systems detected by GSyF 
represented by circles of $r=R_{vir}$ over the galaxy density distribution 
as measured using \emph{KDE} in a RA$\times$Dec projection. 
In the bottom panel of Figure \ref{clust_kde} we show the filaments overlaid 
on the KDE density map for the MSCC\,310 volume. 
The density maps are set in terms of the mean number density.

\section{Filament Properties}
\label{disc}

\subsection{Main properties of the filaments}

In a similar way as described for MSCC\,310, we applied the GFiF algorithm 
to the 46 superclusters of our sample, detecting a total of 144 filaments
in 40 superclusters which are listed in Table \ref{MSCC_filaments}. 
This table also lists the parameters used or measured by GFiF: 
column 2 notes the segmentation parameter $f$, while
column 3 presents the number of detected \emph{HC} groups in the supercluster box. 
Column 4 shows the linking length ($D_{max}$) used to connect \emph{HC} groups.
The process of filtering the connections can be followed through columns
5 to 10, which show, respectively, the number of detected edges, the 
number of filtered links, the number of trees
detected after applying \emph{MST} and the final number of filaments, 
N$_{ske}$, number of isolated bridges, N$_{brid}$, and tendrils, N$_{tend}$.
Column 11 lists the minimum richness we considered for GSyF systems 
to be taken as ends of the bridges.
Column 12 to 14 present, respectively, the fraction of these systems included 
in the GFiF filaments, in isolated bridges and the ones not connected by 
bridges.
Finally, columns 15 to 17 show the number of galaxies hosted in the 
GFiF filaments, {N$_{gfil}$}, and the filling factors calculated as 
 $V_{fil}/V$ and {N$_{gfil}$}$/N$.
The list of detected filaments for all studied supercluster volumes can be 
consulted in Table \ref{table8}, in the same format as the one 
presented in Table \ref{filaments_table}. 
The MSCC\,75, MSCC\,76, MSCC\,264, MSCC\,441, MSCC\,579 and MSCC\,586 
superclusters have not been evaluated with GFiF due to the sparseness 
of the SDSS coverage in these sky areas.

\begin{table*}
	\caption{\label{MSCC_filaments} Summary of the properties of the filaments detected by 
          GFiF for the superclusters in Table \ref{table1}. 
	  }
	\small\addtolength{\tabcolsep}{-1pt}
	\begin{center}
		{\small \begin{tabular}[]{c r r r r r r r r r r r r r r  r  r}
				\hline\hline                 

MSCC & $f$ & $N_{cut}$ & $D_{max}$ & N$_{edges}$ & N$_{links}$ & N$_{trees}$ & N$_{ske}$ & N$_{brid}$ & N$_{tend}$ & $N_{min}$ & \multicolumn{3}{c}{\tiny{Fraction (\%) of systems in}}              & N$_{gfil}$ & \multicolumn{2}{c}{\tiny{Filling factor (\%)}} \\ 
ID   &     &           &           &             &             &             &           &            &            &           & \tiny{filaments}     & \tiny{bridges}        & \tiny{isolated}      &            & ${V_{fil}}/{V}$ & ${N_{gfil}}/{N}$ \\
     &     &           &           &             &             &             &           &            &            &           & N$_{sfil}$/N$_{sys}$ & N$_{spair}$/N$_{sys}$ & N$_{sout}$/N$_{sys}$ &            &                      &  \\
(1)  & (2) & (3)       & (4)       & (5)         & (6)         & (7)         & (8)       & (9)        & (10)       & (11)      & (12)                 & (13)                  & (14)                 & (15)       & (16)                 & (17) \\ 
\hline		
55 & 10 & 81 & 6 & 7 & 6 & 2 & 2 & 0 & 0 & 10 &63.6 & 0.0 & 36.4 & 121 & 0.2 & 14.9 \\ 
72 & 16 & 121 & 8 & 29 & 28 & 6 & 4 & 1 & 1 & 8 &56.7 & 6.7 & 36.7 & 829 & 1.0 & 42.7 \\ 
175 & 29 & 86 & 14 & 36 & 29 & 4 & 4 & 0 & 0 & 6 &57.7 & 0.0 & 42.3 & 507 & 0.9 & 20.2 \\ 
184 & 27 & 78 & 15 & 29 & 13 & 2 & 2 & 0 & 0 & 6 &57.1 & 0.0 & 42.9 & 218 & 0.6 & 10.4 \\ 
211 & 16 & 93 & 12 & 29 & 22 & 2 & 1 & 0 & 1 & 5 &100.0 & 0.0 & 0.0 & 233 & 1.0 & 15.7 \\ 
219 & 27 & 71 & 17 & 65 & 36 & 3 & 2 & 2 & -1 & 5 &72.2 & 22.2 & 5.6 & 279 & 1.1 & 14.6 \\ 
222 & 15 & 124 & 16 & 84 & 43 & 5 & 2 & 0 & 3 & 4 &44.4 & 0.0 & 55.6 & 97 & 0.2 & 5.2 \\ 
223 & 15 & 52 & 18 & 31 & 19 & 2 & 1 & 0 & 1 & 4 &75.0 & 0.0 & 25.0 & 11 & 0.1 & 1.4 \\ 
229 & 16 & 116 & 19 & 93 & 44 & 8 & 1 & 1 & 6 & 4 &37.5 & 25.0 & 37.5 & 32 & 0.1 & 1.7 \\ 
236 & 32 & 270 & 9 & 124 & 94 & 11 & 7 & 2 & 2 & 18 &55.3 & 8.5 & 36.2 & 1600 & 0.8 & 18.5 \\ 
238 & 26 & 320 & 18 & 215 & 115 & 20 & 6 & 3 & 11 & 6 &39.0 & 10.2 & 50.8 & 544 & 0.4 & 6.5 \\ 
248 & 21 & 60 & 20 & 47 & 20 & 3 & 1 & 0 & 2 & 5 &50.0 & 0.0 & 50.0 & 159 & 0.9 & 12.6 \\ 
266 & 22 & 44 & 16 & 20 & 12 & 1 & 1 & 0 & 0 & 5 &71.4 & 0.0 & 28.6 & 132 & 0.7 & 13.8 \\ 
272 & 16 & 86 & 6 & 14 & 14 & 3 & 2 & 0 & 1 & 8 &90.0 & 0.0 & 10.0 & 453 & 0.9 & 32.8 \\ 
277 & 15 & 183 & 11 & 77 & 62 & 6 & 2 & 0 & 4 & 5 &62.5 & 0.0 & 37.5 & 611 & 1.0 & 22.2 \\ 
278 & 16 & 495 & 7 & 378 & 256 & 20 & 5 & 1 & 14 & 19 &58.1 & 4.7 & 37.2 & 2144 & 1.6 & 27.1 \\ 
283 & 23 & 101 & 16 & 32 & 21 & 5 & 3 & 0 & 2 & 4 &44.4 & 0.0 & 55.6 & 263 & 0.5 & 11.3 \\ 
295 & 14 & 1022 & 5 & 478 & 398 & 38 & 4 & 5 & 29 & 26 &43.1 & 19.6 & 37.3 & 2992 & 1.0 & 20.9 \\ 
310 & 16 & 768 & 8 & 334 & 273 & 34 & 9 & 7 & 18 & 10 &52.7 & 12.7 & 34.5 & 2817 & 1.0 & 22.9 \\ 
311 & 25 & 211 & 10 & 49 & 37 & 5 & 4 & 0 & 1 & 7 &43.6 & 0.0 & 56.4 & 1118 & 0.8 & 21.2 \\ 
314 & 14 & 40 & 8 & 13 & 7 & 2 & 2 & 0 & 0 & 7 &50.0 & 0.0 & 50.0 & 112 & 0.7 & 20.1 \\ 
317 & 15 & 56 & 17 & 56 & 31 & 3 & 2 & 1 & 0 & 5 &61.5 & 15.4 & 23.1 & 63 & 0.5 & 7.5 \\ 
323 & 22 & 151 & 16 & 46 & 29 & 5 & 2 & 2 & 1 & 4 &26.3 & 10.5 & 63.2 & 239 & 0.5 & 7.2 \\ 
333 & 19 & 104 & 11 & 39 & 28 & 6 & 3 & 1 & 2 & 7 &45.8 & 8.3 & 45.8 & 282 & 0.6 & 14.3 \\ 
335 & 23 & 135 & 13 & 108 & 65 & 8 & 3 & 0 & 5 & 8 &52.9 & 0.0 & 47.1 & 478 & 1.1 & 15.4 \\ 
343 & 11 & 244 & 7 & 51 & 33 & 5 & 3 & 1 & 1 & 7 &34.5 & 6.9 & 58.6 & 335 & 0.4 & 12.5 \\ 
360 & 38 & 58 & 20 & 54 & 23 & 4 & 3 & 0 & 1 & 6 &47.6 & 0.0 & 52.4 & 218 & 0.8 & 9.9 \\ 
386 & 9 & 362 & 7 & 165 & 120 & 20 & 4 & 2 & 14 & 9 &60.6 & 12.1 & 27.3 & 636 & 0.7 & 19.5 \\ 
407 & 16 & 70 & 18 & 49 & 20 & 3 & 1 & 0 & 2 & 4 &50.0 & 0.0 & 50.0 & 101 & 0.7 & 9.0 \\ 
414 & 9 & 1211 & 6 & 462 & 386 & 47 & 15 & 12 & 20 & 9 &42.4 & 16.7 & 41.0 & 2232 & 0.8 & 20.5 \\ 
419 & 15 & 115 & 11 & 39 & 22 & 5 & 3 & 3 & -1 & 5 &28.6 & 17.1 & 54.3 & 254 & 0.3 & 14.7 \\ 
422 & 18 & 59 & 19 & 35 & 10 & 1 & 1 & 0 & 0 & 4 &37.5 & 0.0 & 62.5 & 11 & 0.0 & 1.0 \\ 
430 & 20 & 80 & 12 & 35 & 22 & 5 & 4 & 1 & 0 & 6 &58.3 & 8.3 & 33.3 & 186 & 0.5 & 11.6 \\ 
440 & 15 & 229 & 10 & 56 & 36 & 8 & 1 & 2 & 5 & 5 &20.7 & 13.8 & 65.5 & 184 & 0.2 & 5.3 \\ 
454 & 15 & 380 & 6 & 164 & 117 & 15 & 5 & 4 & 6 & 13 &45.9 & 13.1 & 41.0 & 1516 & 1.2 & 26.6 \\ 
457 & 21 & 194 & 9 & 80 & 68 & 7 & 6 & 0 & 1 & 8 &66.7 & 0.0 & 33.3 & 1525 & 1.9 & 37.5 \\ 
460 & 22 & 159 & 14 & 102 & 65 & 6 & 4 & 1 & 1 & 5 &65.9 & 4.5 & 29.5 & 895 & 1.4 & 25.6 \\ 
463 & 16 & 529 & 8 & 230 & 183 & 27 & 11 & 8 & 8 & 8 &47.2 & 12.8 & 40.0 & 2228 & 1.1 & 26.3 \\ 
474 & 15 & 495 & 5 & 245 & 209 & 20 & 7 & 3 & 10 & 16 &48.0 & 12.0 & 40.0 & 1918 & 0.9 & 25.8 \\ 
484 & 22 & 60 & 16 & 13 & 9 & 1 & 1 & 0 & 0 & 4 &44.4 & 0.0 & 55.6 & 109 & 0.6 & 8.3 \\ 
\hline
\end{tabular}
}
\end{center}
\end{table*}

\begin{figure}[h!]
	\includegraphics[width=1\linewidth]{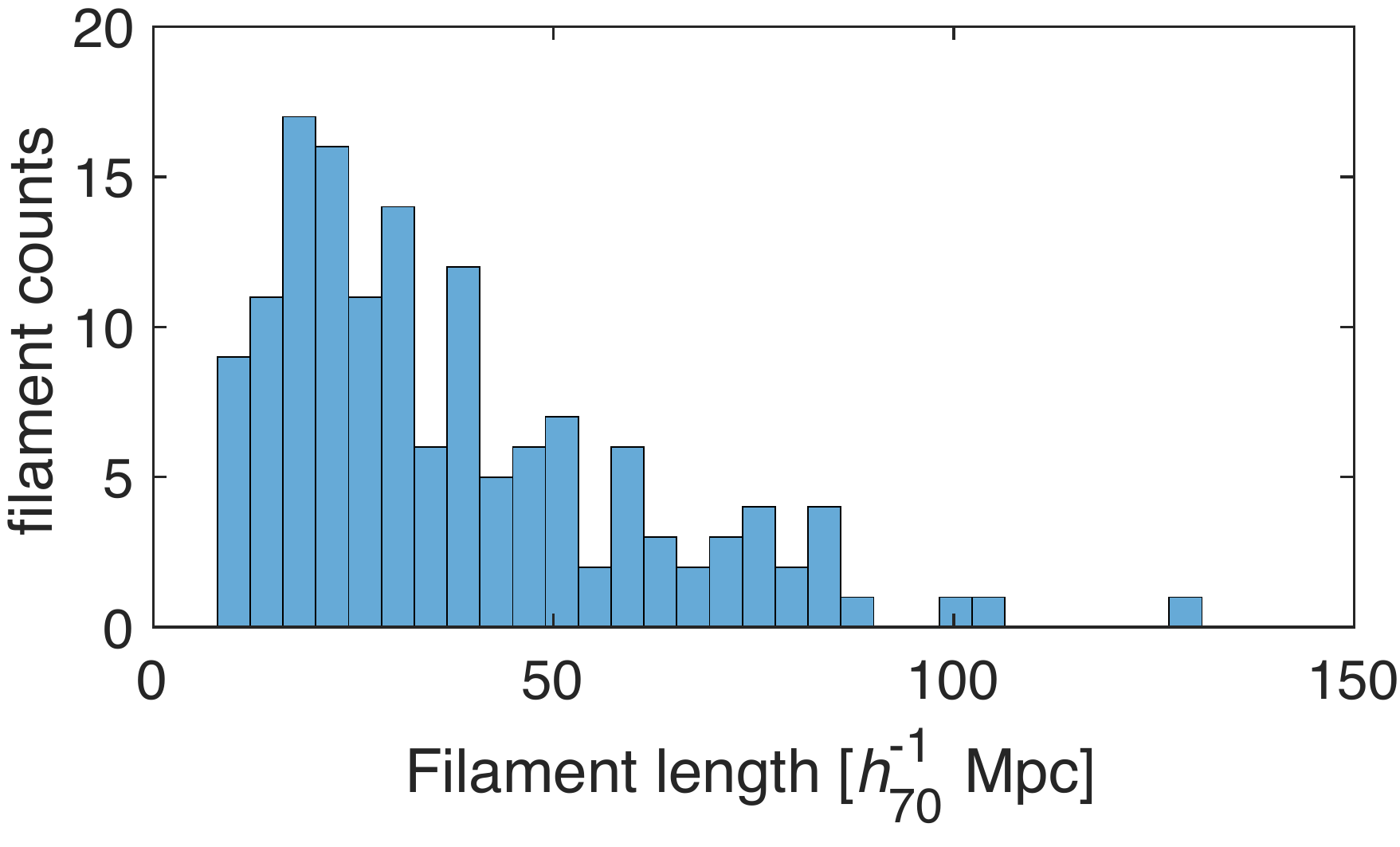}\\
	\caption{Distribution of filament skeleton length for the 144 
           filaments detected by GFiF. 
           The length used corresponds to the longest path between the 
           systems at the extremity of the filament. See Table \ref{table8}. 
           }
\label{longitude}
\end{figure}

The filament skeletons detected by GFiF have lengths between 9 and 130\Mpcp.
Figure \ref{longitude} depicts the length distribution for all the detected 
filaments. 
The distribution shows that the majority of the structure lengths range 
from 10 up to 40\Mpcp. 
There are two structures longer than 100\Mpcp. 
A 130\Mpc long filament is located in MSCC\,323, containing the Abell 
clusters A1449\,B and A1532\,A, the second in MSCC\,335, of 105 \Mpcp, 
containing A1478\,A, A1480\,B, and A1486\,A.
Excluding these two particular cases, we observe that the mean length of the 
filaments is about 37\Mpc while the median corresponds to 29\Mpcp.

\subsection{Distribution of galaxies along the filaments}
\label{density_profile}

\begin{figure}
	\centering
	\includegraphics[width=1\linewidth]{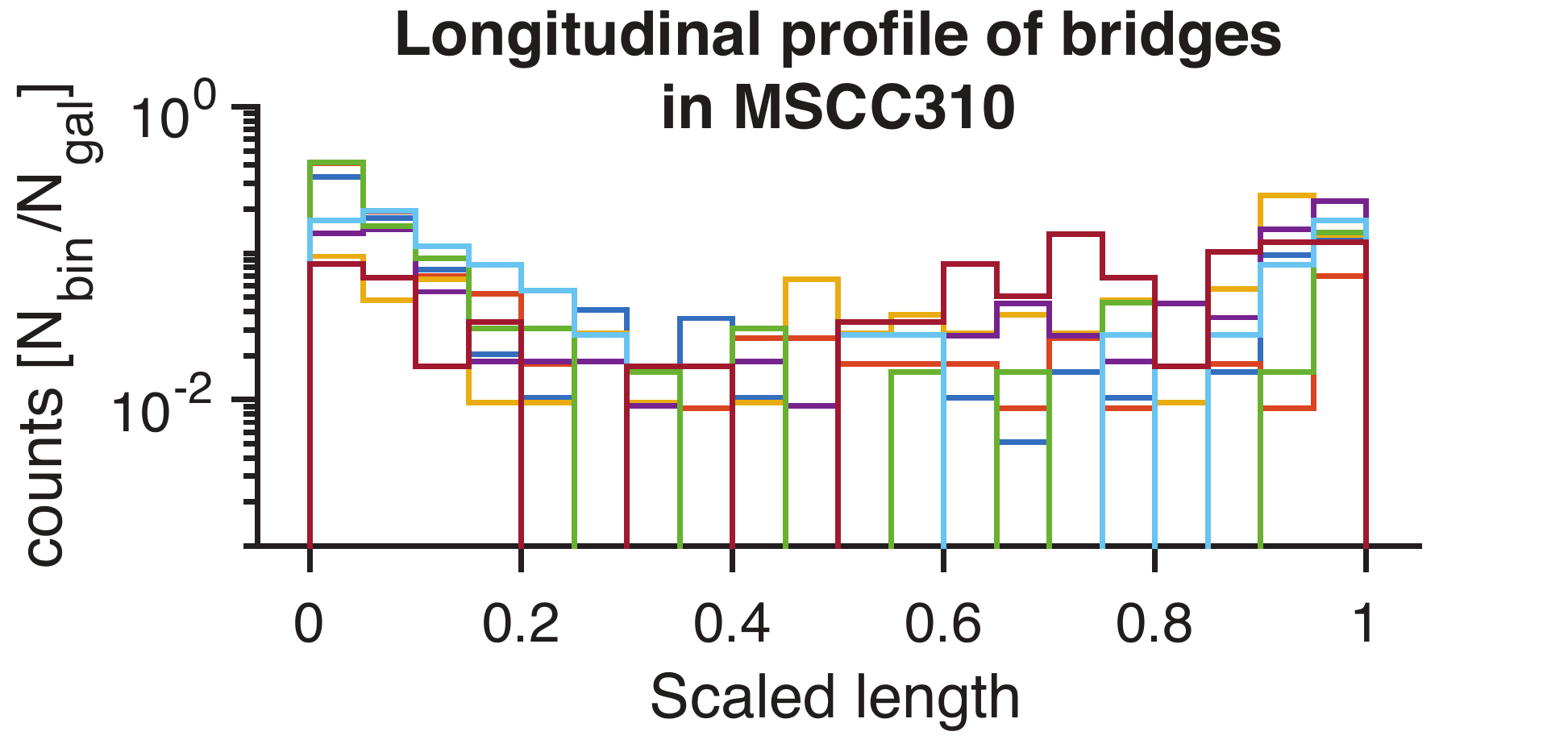}\\
	\caption{Distribution of galaxies along bridges connecting pairs 
          of systems for the nine MSCC\,310 filaments. All bridges are 
          scaled to length 1.0}
	\label{scaled_longitudinal_distro}
\end{figure}

\begin{figure}
	\centering
	\includegraphics[width=1\linewidth]{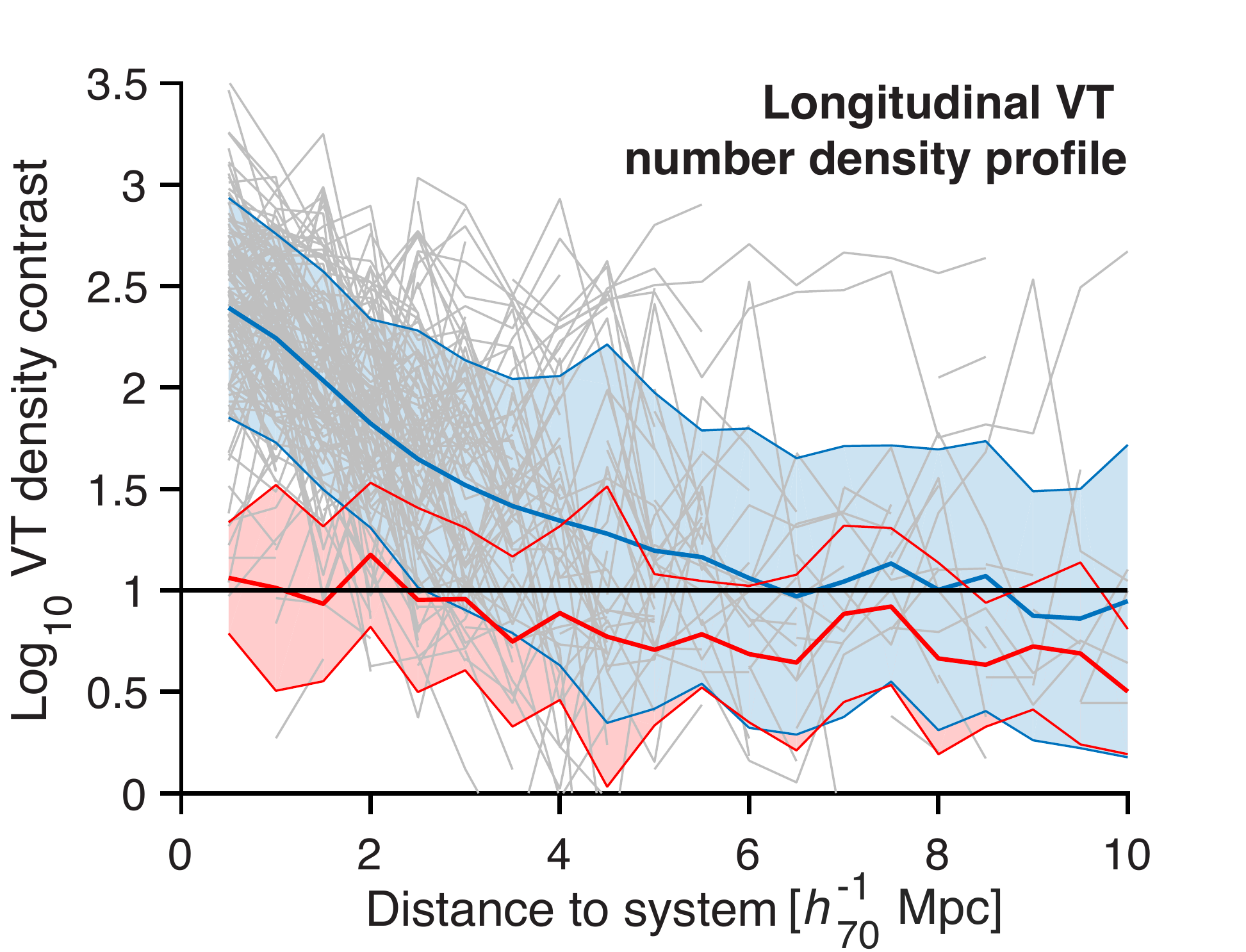}\\
	\caption{Longitudinal \emph{VT} density distribution for galaxies in all 
           bridges of filaments detected by GFiF. Profiles are considered 
           from the system center to the middle of the bridge. 
           The thick blue line depicts the mean longitudinal profile for 
           bridges including galaxies in systems. The thick red line 
           corresponds to the mean longitudinal profile for all filaments 
           excluding galaxies belonging to systems within $1.5\,R_{vir}$. 
           Blue and red shaded areas are the dispersion about the stacked 
           profile.}
	\label{longitudinal_prof}
\end{figure}

In order to evaluate the environment within the filaments, we extracted  
longitudinal profiles of number density.	
In Figure \ref{scaled_longitudinal_distro} we show the longitudinal 
distribution of galaxies for all {bridges, from one 
extreme to the other (ending systems)}, in the supercluster MSCC\,310. 
We observe that the density of galaxies is higher {near 
the ends} of the bridges, as expected, and decreases through 
the midpoint between systems.
Then we proceed to extract density profiles for bridges, from the systems
to the midpoint, by counting the galaxies that lie within a 
cylinder of radius 1\Mpc with medial {skeleton} set by the 
bridge skeleton.
The galaxies are counted in {slices} of size $\Delta d = 0.5$\Mpc 
along the skeleton axes.
We also calculated the longitudinal profiles after {excluding the 
galaxies belonging to} systems (considered at 1.5 $R_{vir}$) from their 
bridges, in order to determine pure filament profiles.
These profiles allow to evaluate the mean density contrast of the filaments 
as compared with the background density.
In Figure \ref{longitudinal_prof} we show the longitudinal number density 
profile for all filaments detected in our sample. 
The stacked longitudinal profile including galaxies in systems is depicted 
by a blue line. The dispersion about the stacked profile is represented by 
a blue shaded area.
The pure profiles (excluding the systems' galaxies)
is represented by the red line, and its corresponding dispersion by 
a red shaded area. 
As can be seen, the mean density contrast along the 
filament is $\sim$ 10, that is, the filament is about 10 times denser than background.

\subsection{Transversal density profiles}

For the calculation of transversal density profiles, we excluded the galaxies 
located in systems, within a radius of 1.5 $R_{vir}$.
The density profile is calculated as described in section 
\ref{profile_extraction}.
The cylinder radius $R_{cy}$ was set from 0 up to 10\Mpc 
in steps of $\Delta R_{cy} = 0.5$\Mpcp.

\begin{figure}
  \centering
  \includegraphics[width=1\linewidth]{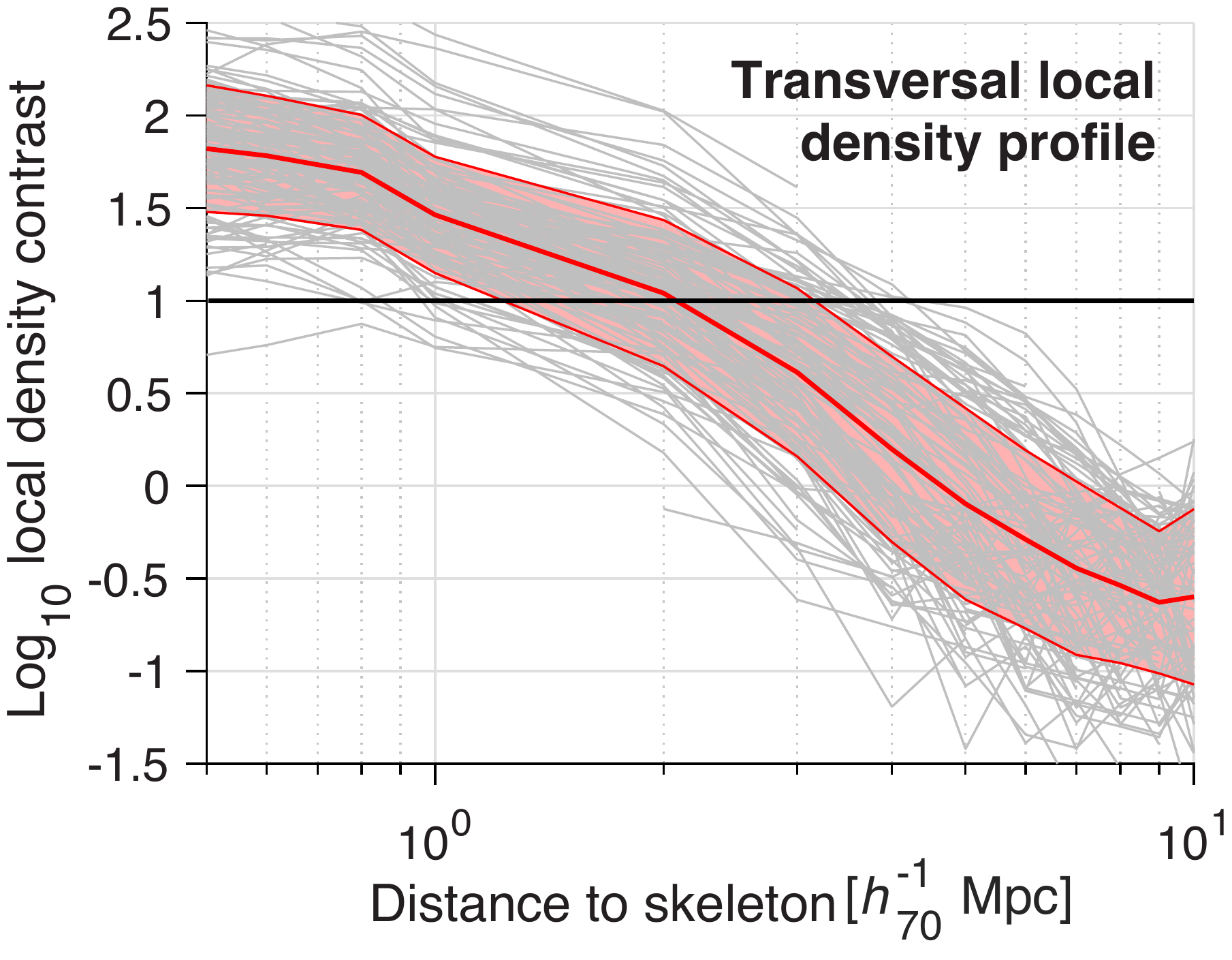}\\
  \includegraphics[width=1\linewidth]{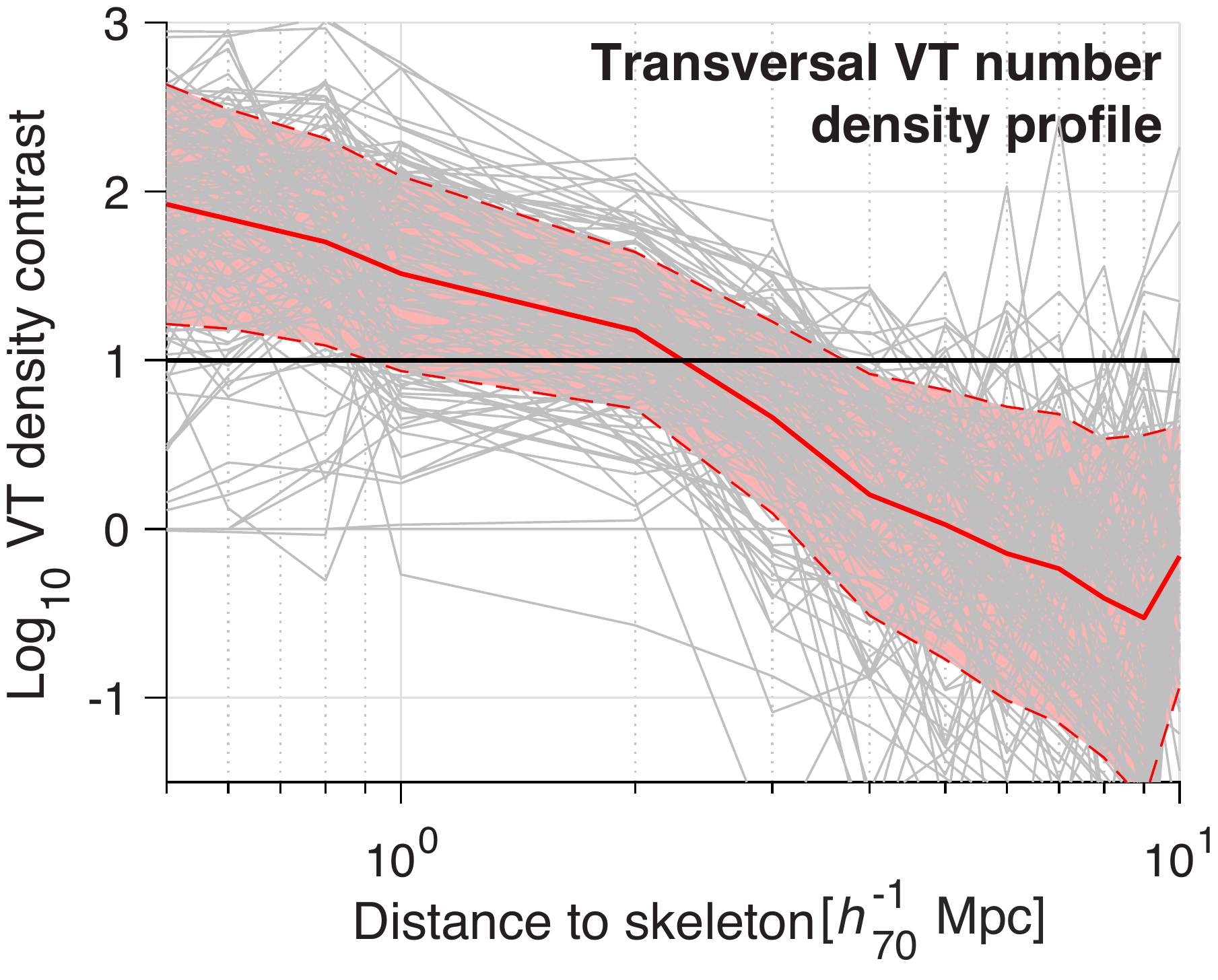}
    \caption{Stacked number density profiles for the 144 filaments 
       identified by GFiF.  
       {Individual profiles are represented by thin gray lines. 
       Top: The red lines corresponds to the mean local density (stacked) 
       profile. 
       Bottom: Mean \emph{VT} density stacked profile. 
       The solid line indicates the mean profile while the shaded area 
       represents the dispersion of the profile.}
       Solid black line depicts the density contrast of $10 \times d$.}
	\label{density_prof}
\end{figure}

We computed the galaxy number density profile for filaments in two ways. 
First we counted the number of galaxies within concentric cylinders and 
divided them by the volume within the cylinders. 
We call this the local number density profile.
For the second, we employed the number densities calculated using the \emph{VT}  
$d_i$, as described in Sec. \ref{VT}. 
Then we measure the mean \emph{VT} number density within concentric cylinders.  
The local number density and \emph{VT} number density profiles are scaled in 
density contrast and stacked together.
Figure \ref{density_prof} shows the stacked profile for all filaments 
detected by GFiF, respectively for local densities (top panel) and \emph{VT} 
densities (bottom panel). 
The first aspect to note is that local number density profiles are smoother 
although, in general, both mean profiles are similar. 
We can observe, in both local and \emph{VT} density profiles, that overdensity 
extends up to 5\Mpcp. 
At about 3\Mpcp, the overdensity reaches a value around 3,
while the typical characteristic density contrast of 10 is reached closer 
to 2\Mpcp.

Finally, we use the density profiles to estimate the mean radius of the 
filaments $R_{fil}$. 
This was achieved by considering the intersection point at which the local
density profile crosses the $10 \times d$ line, as 
indicated in Figure \ref{density_prof} by the black solid line. 
The mean radius as well as the mean density of each filament is noted in 
Table \ref{table8}.

\begin{figure}[h!]	
	\includegraphics[width=1\linewidth]{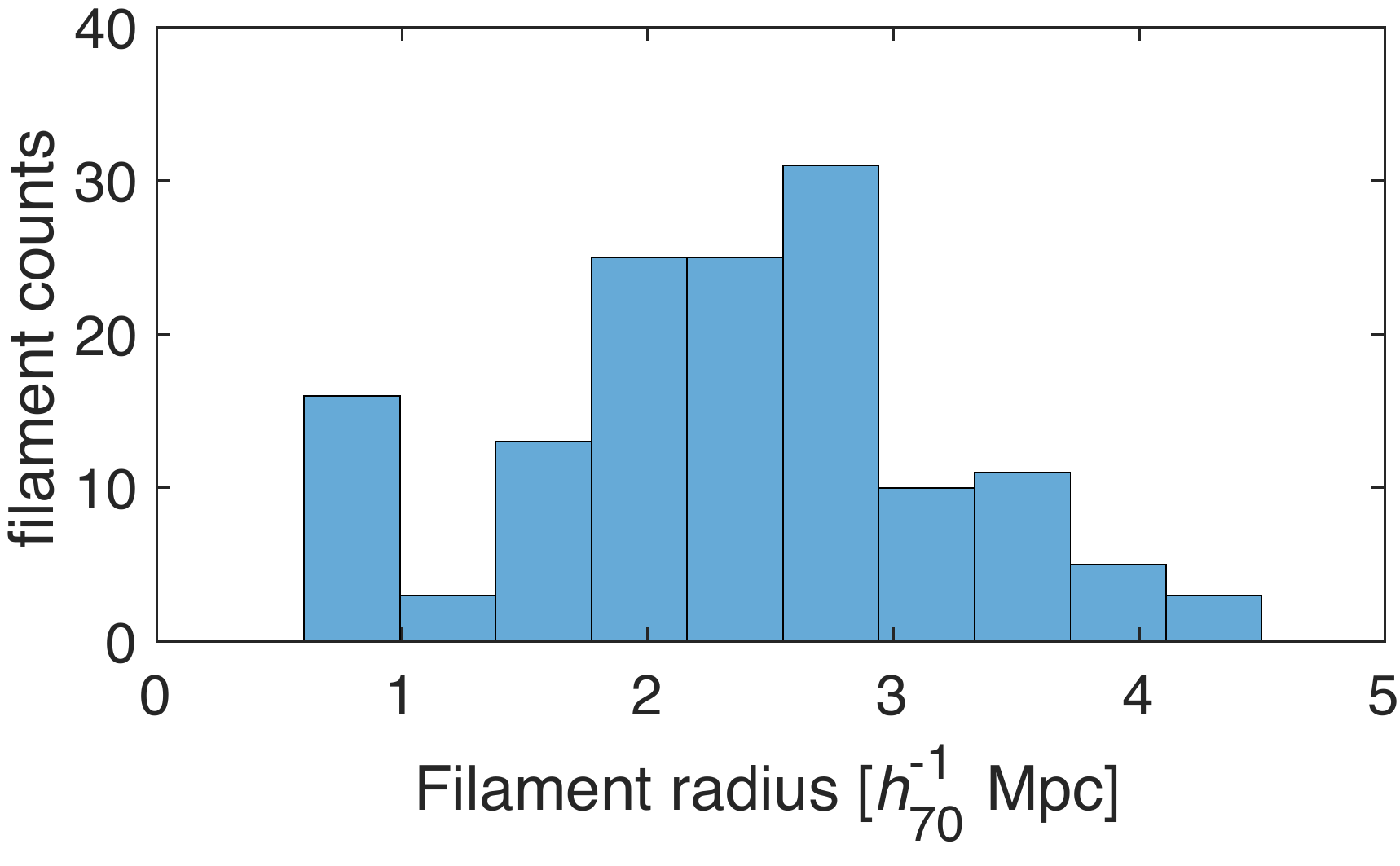}\\
	\includegraphics[width=1\linewidth]{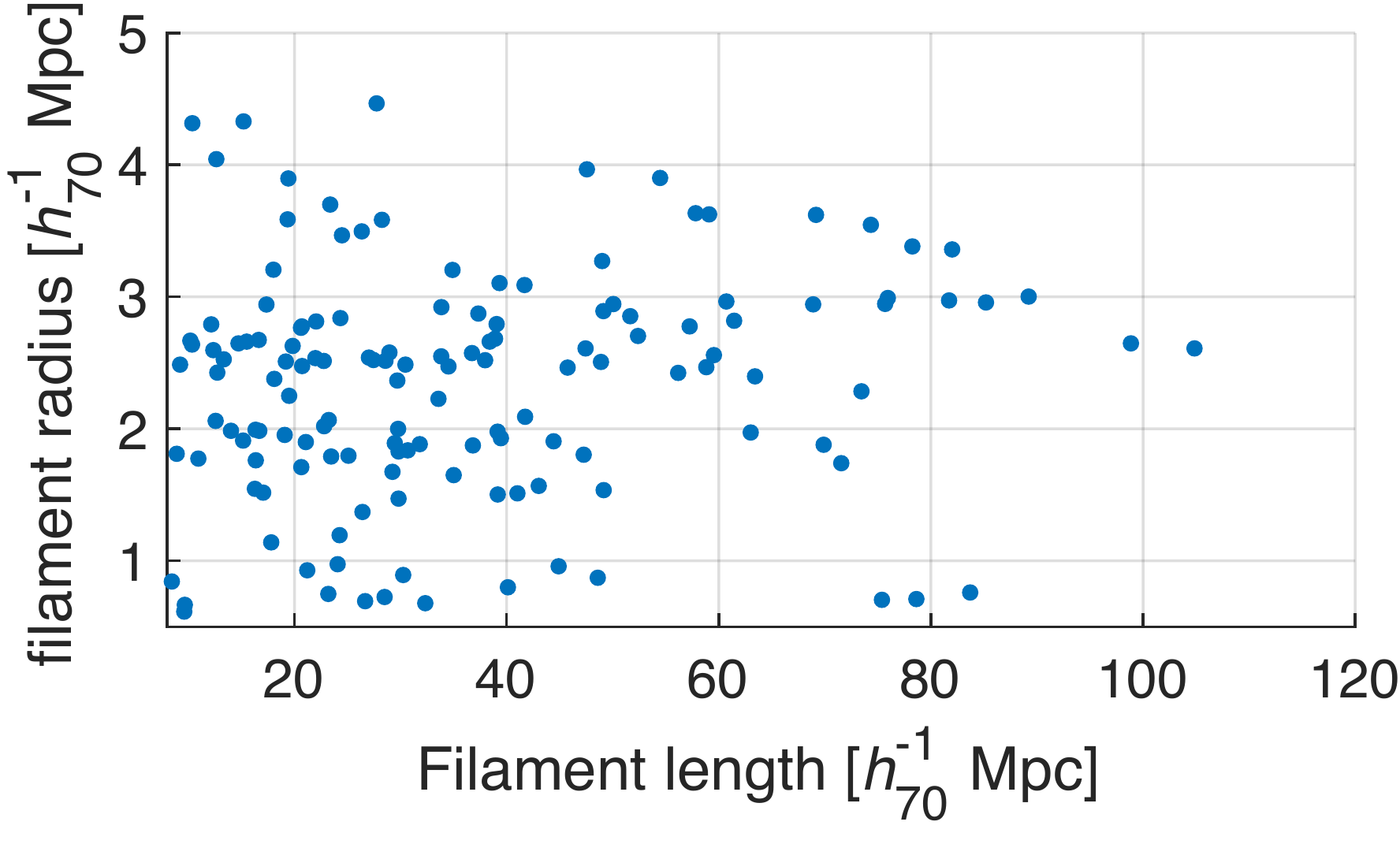}
	\caption{
		Top: distribution of radius of filaments in our sample. 
		Bottom, Comparison of filament length and 
                radius for the 144 
		filaments detected by GFiF. The length used corresponds to 
                the longest path between a pair of systems, 
                {that is, the skeleton length}. 
                See Table \ref{table8}.
	        }
	\label{radius}
\end{figure}

Figure \ref{radius} (top panel) presents the radii  distribution for all 
the filaments. 
The filament radii range from 0.6 to 4.5\Mpc with a mean value 
of 2.4\Mpcp. 
The bottom panel of the figure depicts the filament radius as a function 
of the filament length. 
We observe that the filament length does not correlate with the filament 
radius.
However, it is important to note that the {radius varies
slightly around the mean value} along the filament path.

\section{Properties of galaxies in filaments}
\label{gal_prop}
\subsection{Stellar mass profile}

We constructed a galaxy stellar mass profile for all 
filaments by using the masses from MPA-JHU group 
\citep{Brichmann,Kauffmann03,Tremonti} described in Section \ref{galaxies}. 
First, we weighted the mass by the average mass of the 
volume under analysis to remove the redshift dependence of the stellar 
mass \citep{Chen2017}.
This weighting is equivalent to a normalization of the stellar mass and 
allows to carry out a stacking procedure in order to increase the signal of 
the profiles.
This mass profile is extracted as described in 
Section \ref{profile_extraction}.

\begin{figure}
	\centering
	\includegraphics[width=1\linewidth]{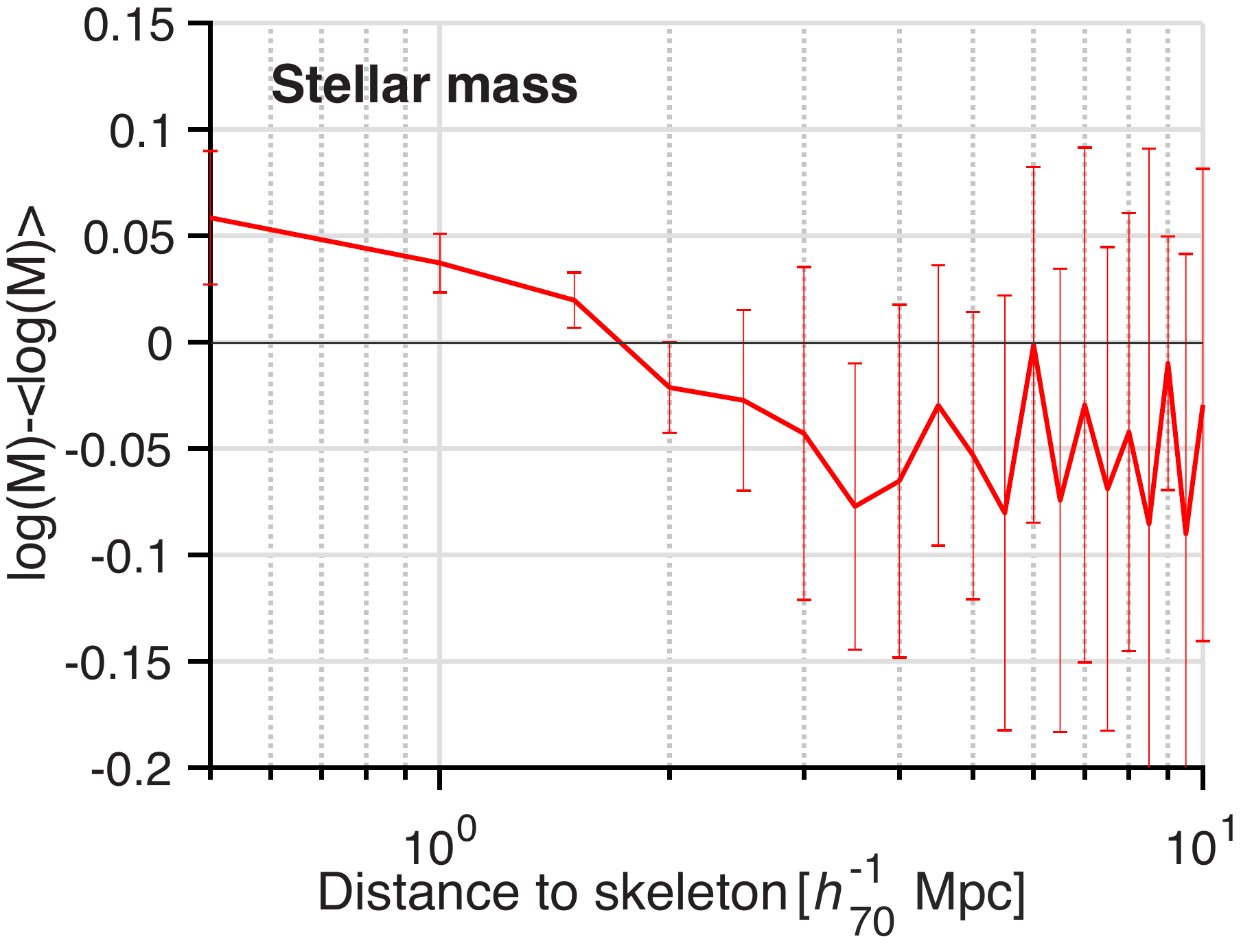}
	\caption{Stacked {transversal stellar} mass profile 
for the 144 filaments detected by GFiF. Errors correspond to the variance 
of the stacked profiles.}
	\label{mass_prof}
\end{figure}

Figure \ref{mass_prof} shows the stacked stellar mass profile for all filaments. 
The variance of the stacked profile is depicted by the error bars.
We observe that, statistically, the stellar masses of the filament galaxies are 
larger than the average mass up to about 2\Mpcp, while beyond 3\Mpc they 
tend to be 10\% smaller. {This region farther than 3\Mpc
probably represents the dispersed population of the supercluster, 
associated to the more extended sheet component. Thus, our results 
indicate} that the stellar mass correlates with the distance to the filament 
{skeleton}, {being larger (up to 25\%) near the
skeleton than far from it}.
These results are in good agreement with the results presented by 
\citet{Chen2017} for MGS sample from DR7 \citep{DR7}.
Our results are also compatible with those presented by 
{\citet{Alpaslan2016} and} \cite{Kraljic2018},
for the GAMA spectroscopic survey, {who find similar 
trends} for the filaments found {at redshifts 
$z < 0.09$ and $0.03 \leq z \leq 0.25$, respectively}.

\subsection{Morphological type }

\begin{figure}
	\centering
	\includegraphics[width=1\linewidth]{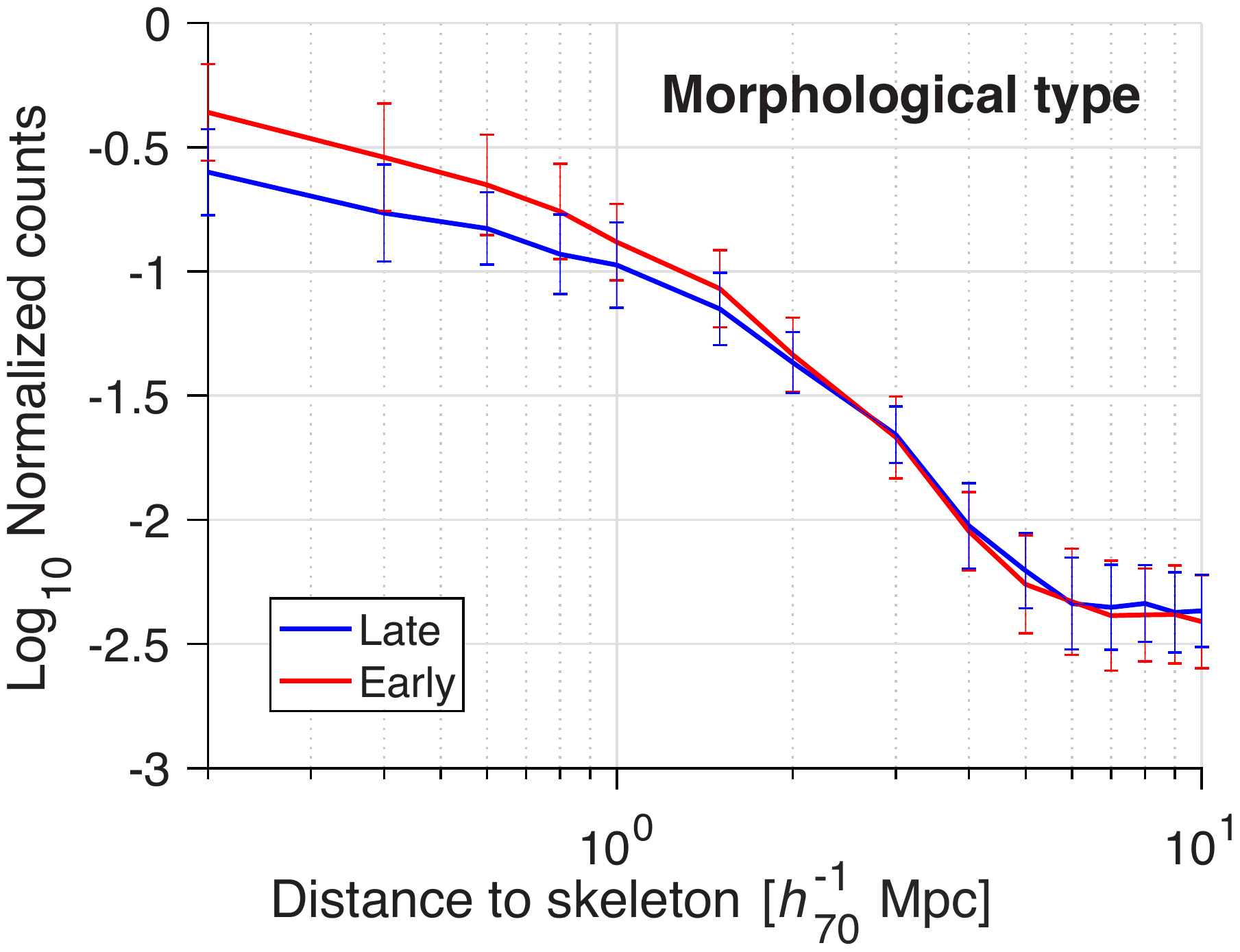}\\
		\includegraphics[width=1\linewidth]{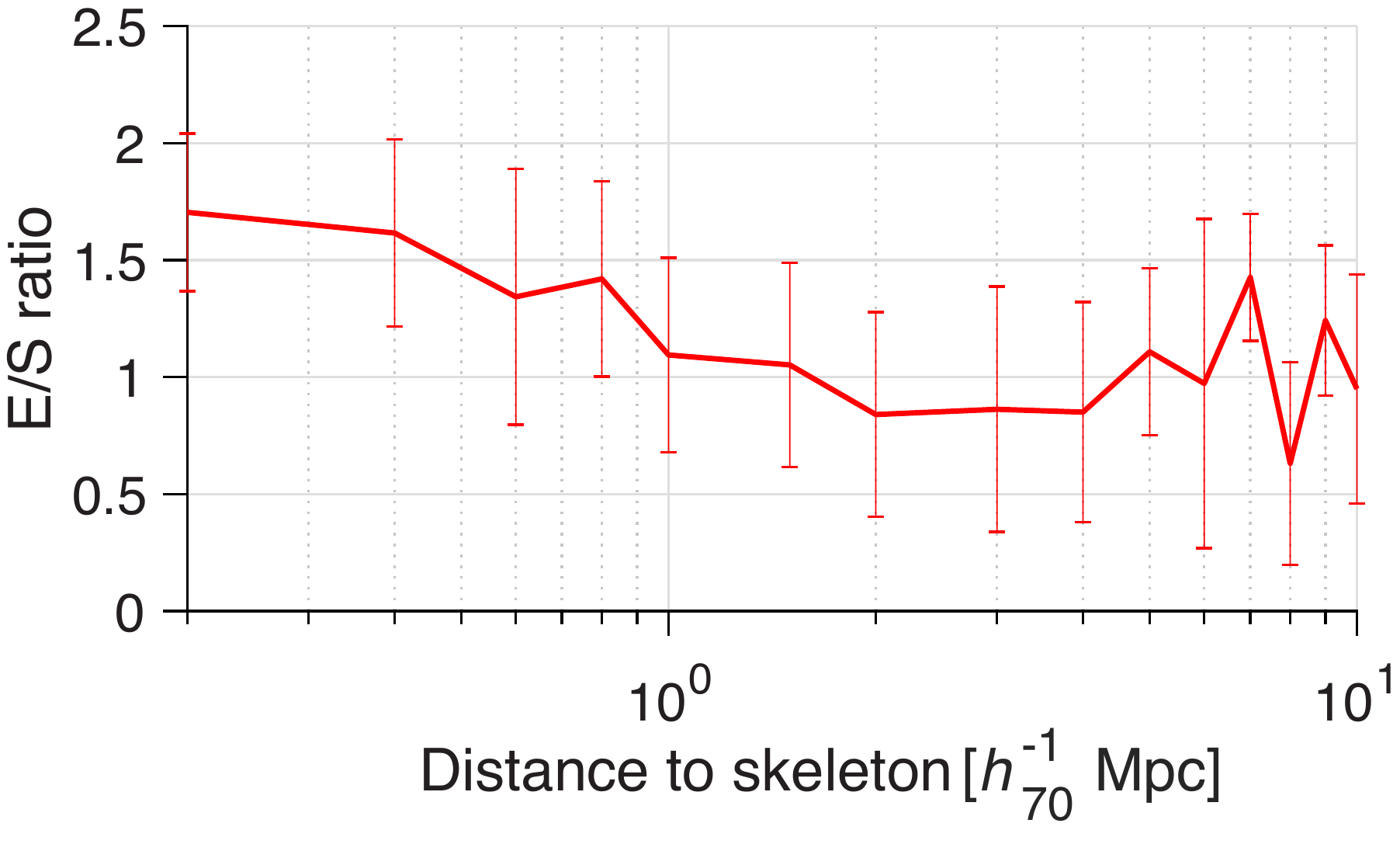}
	\caption{Top: stacked transversal morphological type profiles for the 
           144 filaments detected by GFiF.
	   The error bars correspond to the variance of the stacked profiles.
           Bottom: early-late type ratio as function of the distance to the 
           filament {skeleton}.}
	\label{mtype_prof}
\end{figure}

In order to analyze if there is some morphological 
trend in the population of filament galaxies (as may be expected from 
the morphology-density relation), we also constructed 
morphology profiles based on morphological classifications
by \cite{Huertas-Company}.
They classify the galaxies in four morphological types. 
For our analysis we used the probability 
\,$p$(Early) = $p$(E) $+\,p$(S0)\,
that classifies galaxies in early type as $p$(Early)$\;>\,0.5$ 
and late type as $p$(Early)$\;<\,0.5$.
Then we computed the distribution of both galaxy types as a 
function of the distance to the filament skeleton. 
The distributions were normalized so they can be compared and stacked 
for all filaments in our sample in a similar way as a profile extraction, 
{again excluding galaxies in systems. The result is} shown in 
Figure \ref{mtype_prof}. 

Our results show that the fraction of early type galaxies is higher 
than the one for late types near the filament {skeleton} up to 
$\sim2$\Mpcp. 
This effect is more notorious when computed as an early to late type 
ratio (Figure \ref{mtype_prof}, bottom panel).
We observe that at distances smaller than 2\Mpcp, the fraction 
of early types reaches almost twice the fraction of late types. 
At larger distances (that is, towards the dispersed supercluster population) 
the fractions tend to be similar (E/S ratio $\sim1$).
A two sample Kolmogorov-Smirnov test applied to the 
distributions of early and late types in Fig. \ref{mtype_prof} reveals 
that, for the first bins, they are significantly different
(p-value lower than 0.1).
Our results are consistent with those presented by \cite{Kuutma2017} for 
the \cite{Huertas-Company} sample --   they also observe that early type 
galaxies are more abundant near the filament {skeleton}.

\subsection{Activity type}

For the analysis with respect to activity type, we used the activity 
classification from the MPA-JHU group, \citep{Brichmann,Kauffmann03,Tremonti} 
described in Section \ref{galaxies}.
We computed the distribution of the different galaxy activity populations 
as a function of the filament {skeleton} distance. 
All distributions are normalized for all filaments and stacked together.

\begin{figure}
	\centering
	\includegraphics[width=1\linewidth]{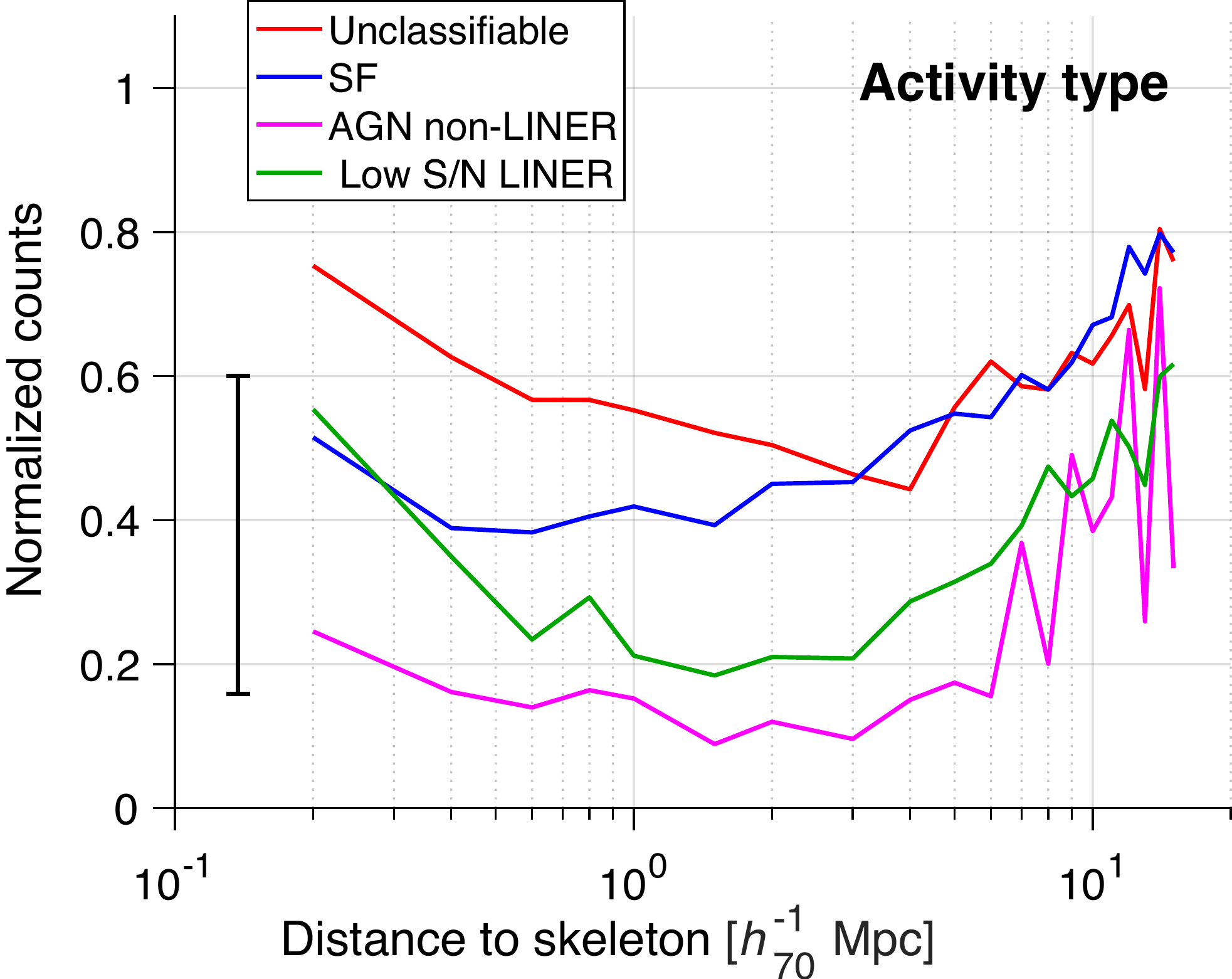}\\
	\caption{Stacked transversal activity type profiles for the 144 
          filaments detected by GFiF. 
          The black bar on the left represents the typical errors on the 
          stacked profiles, not overlaid for clarity.}
	\label{activity_prof}
\end{figure}

Figure \ref{activity_prof} divides galaxies into four activity 
groups: AGNs, SF galaxies, LINERs and non-active galaxies (unclassified). 
The error bars were not displayed over the lines for clarity -- note that 
they are very large, implying that we have to interpret 
this figure with caution. Another effect to take into account is 
that the fractions are averaged over all the filaments (at different redshifts).
The most evident tendency we can see in these distributions is a decrease 
in the activity as long as the galaxies ``approach'' the filament, although 
the fractions for the dispersed component are very noisy.
Inside the filaments, the tendency is to have more {passive} 
galaxies, implying again smaller fractions of AGNs and SF galaxies. 
However, the fraction of LINERs also increases towards the filament 
skeletons, possibly indicating a post-activity phase for the galaxies. 
Deeper analyses are necessary to give a clear picture of the effect of the 
filament environment on the activity of galaxies.

\section{Conclusions}
\label{concl}

{In this paper we studied the bridges and filaments 
of galaxies in the environment of superclusters of galaxies.
We developed two algorithms, the \emph{Galaxy Systems-Finding} algorithm, 
GSyF, and the \emph{Galaxy Filaments-Finding} algorithm, GFiF, respectively 
to detect systems of galaxies (clusters and groups), 
aiming especially to correct for the finger-of-God effect, 
and to identify the elongated structures just mentioned. 
These algorithms were applied to a sample of SDSS galaxies with 
spectroscopic redshifts in rectangular boxes enclosing 
46 superclusters of galaxies selected from MSCC catalog in a redshift range  
from $0.02$ to $0.15$.}

GSyF and GFiF employ a set of different classic pattern recognition methods.
Both of them are probabilistic in the sense that they define systems and 
filaments as a function of the relative position and orientation of the 
Gaussian groups, which are detected with a \emph{Hierarchical Clusterization} 
method. 
For GSyF, the membership of the Gaussian groups is refined by using a 
virial approximation, allowing {to discern gravitationally bound 
systems of galaxies from} misdetections. 
For GFiF, these measurements are used to define a general tree from 
which we extract independent structures based on density criteria.
Although the \emph{HC} algorithm needs to be optimized for the number of 
clustering groups,
this can be automatized based on the density function characterizing the 
survey. 
The structures are represented by a filament skeleton that allows to 
measure and trace quantitatively the filament path.

We show (section \ref{valid_syst}) that the systems detected by our 
methodology are in good agreement with those reported in the 
literature.
Specifically, our comparisons of the systems sample against 
{other cluster and group catalogs} 
(Abell, C4, L14, T11 and MSPM) 
showed a match rate above {78\% for groups with 
richness above 5} galaxies at redshifts $z<0.11$.  
{For systems with richness above 10 galaxies the 
coincidences were slightly higher for the group catalogues (T11 and
MSPM) and slightly lower for the cluster catalogues (Abell and C4)}.
Moreover, the {richness}, velocity dispersion and virial 
radius of systems measured by the GSyF algorithm are in good 
agreement with those 
reported in other system catalogs.
Our GSyF algorithm detected a total of 2\,705 systems in 
{the rectangular boxes enclosing the volumes of 45 of the} 
superclusters in our sample.
Of these, 159 systems with richness above 10 galaxies were not previously 
reported in the literature\footnote{Data on these systems are 
available in the electronic version of the paper.}.

We also compared, in section \ref{validation}, the results of our 
filaments-finding algorithm with those of \cite{Tempel2014} for the 
same regions.
{We observe that T14 filaments are shorter, more numerous 
and describe sparser and finer structures while GFiF} 
{detects 
larger and denser elongated structures that bridge galaxy systems}.
Our filaments{, in some sense, link} 
several T14 threads {in one larger structure, 
providing a broader} picture of the filament. 
The comparison with isolated bridges and tendrils (a sub-product 
of our algorithm) shows a match of 80\% with T14 filaments and comparable 
filament lengths.

The GFiF algorithm detected a total of 144 filaments 
and 63 isolated bridges {in the rectangular boxes enclosing
the volumes of 40 of the} superclusters in our sample.
The {supercluster filaments we detected have lengths} 
from 9 up to 130\Mpc (mean 37\Mpcp, median 29\Mpcp) while the isolated 
bridges have lengths between 5 and 15\Mpcp. {These values
are consistent with the median bridge length value from \citet{Kraljic2019},
7.9\Mpcp, for the \textsc{horizon-AGN} simulation}.

For most of the cases, the numerical density inside the filaments was
found to be between 5-15 times the mean density.
The radii of the filament skeletons range from 0.6 up to 
4.5\Mpcp, with most between 2 and 3\Mpcp.
These values are consistent with the ones found by \citet{Cautun2013}
by applying the \emph{NEXUS} algorithm to an N-body simulation.

We also compared the properties of the galaxies that inhabit the 
filament as a function of the distance from its skeleton. 
We conclude the following: 
i) The transversal local and \emph{VT} number density profiles for pure filaments 
show that, at distances up to 5\Mpcp, the filaments have a positive 
overdensity respect to the background density inside de boxes. 
ii) At distances of about 3\Mpcp, the density contrast reaches a value of 
3, a limit that matches the range where typically the environmental effects 
studied in section 9 seem to apply. 
iii) The mean density contrast of the filaments, 10, is reached closer
to 2\Mpcp, a limit that we used as reference for estimating the radius
of the filaments.

Our analysis regarding the stellar masses, morphological type 
and activity type show correlations between these galaxy properties and 
the distance from the filament {skeleton}.
We arrive to the following conclusions:
i) Inside 3\Mpc from the filament skeleton the galaxy stellar masses  
increase up to about 25\%. 
This result leads to two hypotheses: 
(a) the mass growth of the galaxies is sensitive to the environment, or 
(b) the dynamical evolution brings massive galaxies into the potential 
well of the filaments.
This result confirms several analyses which suggest that stellar masses 
are sensitive to the environment   
\citep{Alpaslan2015,Alpaslan2016,Poudel2016,Chen2017,Malavasi2017,Kraljic2018, Musso2018}.
ii) The early to late type galaxy ratio has its maximum at the 
center of the filament and remains above ratio 1:1 up to a distance of
1.5 \Mpcp.
This result is in good agreement with a similar work by 
\cite{Kuutma2017} for the SDSS.
iii) Concerning the activity type, we observed that the fraction of AGNs 
and SF galaxies seem to be higher outside the filaments (in the 
supercluster dispersed component), showing a decrease as the galaxies 
approach these structures. 
Inside the filaments, the fractions of non-active galaxies and LINERs 
increase, indicating a possible post-activity phase. 
A similar result for the star-forming galaxies has been observed by 
\cite{Kraljic2018}, for the GAMA spectroscopic survey.

The GSyF and GFiF algorithms can be used to search for this 
kind of structures in different surveys, using spectroscopic or 
photometric redshifts. 
We plan to apply them to other galaxy databases, like the 
ones that are becoming available for the southern celestial hemisphere, 
and also, to galaxy surveys that reach deeper redshifts.
{Both algorithms and catalogs can be obtained electronically 
upon request}.

\begin{acknowledgements}
I.S.-B. thanks CONACyT and DAIP for funding this research. 
This project was partially financed by DAIP funding CIIC 205/2019.
{The authors are grateful to the anonymous referee 
for the important comments that helped to improve the paper}.
All the pattern recognition computing, statistics and graphics have 
been made using MATLAB$\textcopyright$.
Part of this work was carried out with the computational facility 
TITAN at the Institut de Recherche en Astrophysique et Plan\'etologie, 
Toulouse, France.
This work has made use of NASA's Astrophysics Data 
System Bibliographic Services.
Funding for SDSS-III has been provided by the Alfred P. Sloan Foundation, 
the Participating Institutions, the National Science Foundation, and 
the U.S. Department of Energy Office of Science. The SDSS-III web site 
is http://www.sdss3.org/.
SDSS-III is managed by the Astrophysical Research Consortium for the 
Participating Institutions of the SDSS-III Collaboration including the 
University of Arizona, the Brazilian Participation Group, Brookhaven 
National Laboratory, Carnegie Mellon University, University of Florida, 
the French Participation Group, the German Participation Group, Harvard 
University, the Instituto de Astrofisica de Canarias, the Michigan 
State/Notre Dame/JINA Participation Group, Johns Hopkins University, 
Lawrence Berkeley National Laboratory, Max Planck Institute for 
Astrophysics, Max Planck Institute for Extraterrestrial Physics, 
New Mexico State University, New York University, Ohio State University, 
Pennsylvania State University, University of Portsmouth, Princeton 
University, the Spanish Participation Group, University of Tokyo, 
University of Utah, Vanderbilt University, University of Virginia, 
University of Washington, and Yale University.
\end{acknowledgements}

\bibliographystyle{aa.bst}
\bibliography{biblio.bib}

\onecolumn
	\small \addtolength{\tabcolsep}{-1pt}
\begin{longtable}{l r r c c c c c r r r}
	\caption{ Main properties of the filaments extracted through GFiF.} 
        \label{table8} \\
\hline\hline                 
Fil. & N$_{sfil}$ & N$_{gfil}$ & \multicolumn{3}{c}{redshift} & $d_{fil}$                 & $R_{fil}$           & \multicolumn{2}{c}{N$_{nod}$} & $\ell_{fil}$ \\
ID   & systems    & gals.      &     [mean, &min, &max]       & [$h_{70}^{3}~$Mpc$^{-3}$] & [$h^{-1}_{70}~$Mpc] & filament  & skeleton          & [$h^{-1}_{70}~$Mpc] \\
(1)  & (2)        & (3)        & (4) & (5) & (6)              & (7)                       & (8)                 & (9)       & (10)              & (11)\\ 
\hline		
	\endfirsthead
	
	\multicolumn{11}{c}%
	{{\bfseries \tablename\ \thetable{} -- continued from previous page}} \\
\hline\hline                 
Fil. & N$_{sfil}$ & N$_{gfil}$ & \multicolumn{3}{c}{redshift} & $d_{fil}$                 & $R_{fil}$           & \multicolumn{2}{c}{N$_{nod}$} & $\ell_{fil}$ \\
ID   & systems    & gals.      &     [mean, &min, &max]       & [$h_{70}^{3}~$Mpc$^{-3}$] & [$h^{-1}_{70}~$Mpc] & filament  & skeleton          & [$h^{-1}_{70}~$Mpc] \\
(1)  & (2)        & (3)        & (4) & (5) & (6)              & (7)                       & (8)                 & (9)       & (10)              & (11)\\ 
\hline		
	\endhead
	
	\hline \multicolumn{11}{r}{{Continued on next page}} \\ \hline
	\endfoot
	
	\hline \hline
	\endlastfoot
MSCC\,55-F1 & 4 & 119 & 0.0619 & 0.0529 & 0.0706 & 0.1570 & 4.31 & 3 & 2 & 10.5\\ 
MSCC\,55-F2 & 3 & 2 & 0.0585 & 0.0527 & 0.0680 & 0.1911 & 0.60 & 2 & 2 & 9.7\\ 
MSCC\,72-F1 & 5 & 257 & 0.0790 & 0.0722 & 0.0868 & 0.5636 & 3.57 & 6 & 6 & 28.4\\ 
MSCC\,72-F2 & 4 & 217 & 0.0801 & 0.0718 & 0.0866 & 0.3688 & 3.46 & 5 & 5 & 24.6\\ 
MSCC\,72-F3 & 5 & 215 & 0.0777 & 0.0717 & 0.0839 & 0.3043 & 3.89 & 6 & 5 & 19.5\\ 
MSCC\,72-F4 & 3 & 140 & 0.0850 & 0.0787 & 0.0917 & 0.2012 & 3.49 & 4 & 4 & 26.5\\ 
MSCC\,175-F1 & 6 & 156 & 0.0912 & 0.0841 & 0.0996 & 0.1541 & 3.08 & 8 & 6 & 41.8\\ 
MSCC\,175-F2 & 3 & 159 & 0.0978 & 0.0885 & 0.1032 & 0.2260 & 3.19 & 3 & 3 & 35.0\\ 
MSCC\,175-F3 & 3 & 87 & 0.0937 & 0.0883 & 0.1001 & 0.1714 & 1.79 & 5 & 4 & 47.4\\ 
MSCC\,175-F4 & 3 & 105 & 0.0923 & 0.0867 & 0.0995 & 0.2184 & 2.86 & 3 & 3 & 37.5\\ 
MSCC\,184-F1 & 5 & 157 & 0.1071 & 0.0994 & 0.1171 & 0.0488 & 2.88 & 8 & 5 & 49.3\\ 
MSCC\,184-F2 & 3 & 61 & 0.0980 & 0.0911 & 0.1029 & 0.0838 & 3.09 & 3 & 3 & 39.5\\ 
MSCC\,211-F1 & 3 & 233 & 0.1205 & 0.1100 & 0.1297 & 0.0465 & 4.46 & 13 & 8 & 27.9\\ 
MSCC\,219-F1 & 10 & 275 & 0.1125 & 0.1063 & 0.1207 & 0.1169 & 2.96 & 17 & 6 & 81.8\\ 
MSCC\,219-F2 & 3 & 4 & 0.1235 & 0.1183 & 0.1287 & 0.0347 & 0.79 & 3 & 3 & 40.2\\ 
MSCC\,222-F1 & 3 & 89 & 0.1422 & 0.1311 & 0.1522 & 0.0418 & 1.87 & 14 & 8 & 70.0\\ 
MSCC\,222-F2 & 5 & 8 & 0.1349 & 0.1238 & 0.1460 & 0.0461 & 0.70 & 11 & 8 & 78.8\\ 
MSCC\,223-F1 & 3 & 11 & 0.1367 & 0.1286 & 0.1472 & 0.0245 & 0.86 & 12 & 8 & 48.7\\ 
MSCC\,229-F1 & 3 & 32 & 0.1445 & 0.1361 & 0.1514 & 0.0389 & 1.96 & 9 & 6 & 63.1\\ 
MSCC\,236-F1 & 5 & 231 & 0.0324 & 0.0228 & 0.0415 & 0.3382 & 1.49 & 16 & 9 & 39.3\\ 
MSCC\,236-F2 & 4 & 324 & 0.0411 & 0.0361 & 0.0461 & 0.2409 & 2.65 & 7 & 6 & 38.5\\ 
MSCC\,236-F3 & 5 & 286 & 0.0331 & 0.0274 & 0.0399 & 0.4889 & 2.62 & 8 & 5 & 20.0\\ 
MSCC\,236-F4 & 3 & 150 & 0.0296 & 0.0255 & 0.0364 & 0.3971 & 2.24 & 5 & 4 & 19.6\\ 
MSCC\,236-F5 & 3 & 188 & 0.0354 & 0.0286 & 0.0422 & 0.4448 & 2.47 & 5 & 5 & 20.8\\ 
MSCC\,236-F6 & 3 & 318 & 0.0330 & 0.0276 & 0.0373 & 2.3032 & 2.66 & 2 & 2 & 16.8\\ 
MSCC\,236-F7 & 3 & 103 & 0.0333 & 0.0297 & 0.0391 & 0.4359 & 1.53 & 4 & 3 & 16.4\\ 
MSCC\,238-F1 & 5 & 174 & 0.1190 & 0.1075 & 0.1309 & 0.0558 & 2.95 & 11 & 7 & 85.3\\ 
MSCC\,238-F2 & 6 & 201 & 0.0915 & 0.0802 & 0.1010 & 0.0928 & 2.98 & 14 & 10 & 76.1\\ 
MSCC\,238-F3 & 3 & 155 & 0.1054 & 0.0979 & 0.1115 & 0.0782 & 2.64 & 11 & 8 & 99.0\\ 
MSCC\,238-F4 & 3 & 3 & 0.0957 & 0.0896 & 0.1024 & 0.1467 & 0.71 & 3 & 3 & 28.6\\ 
MSCC\,238-F5 & 3 & 4 & 0.1016 & 0.0934 & 0.1099 & 0.0285 & 0.67 & 3 & 2 & 32.5\\ 
MSCC\,238-F6 & 3 & 7 & 0.1096 & 0.1040 & 0.1156 & 0.0398 & 0.88 & 4 & 3 & 30.4\\ 
MSCC\,248-F1 & 3 & 159 & 0.1256 & 0.1161 & 0.1355 & 0.0895 & 3.35 & 10 & 7 & 82.1\\ 
MSCC\,266-F1 & 5 & 132 & 0.1282 & 0.1188 & 0.1344 & 0.0366 & 2.94 & 9 & 8 & 75.8\\ 
MSCC\,272-F1 & 6 & 374 & 0.0752 & 0.0694 & 0.0808 & 0.8405 & 2.93 & 6 & 4 & 17.5\\ 
MSCC\,272-F2 & 3 & 79 & 0.0757 & 0.0714 & 0.0814 & 0.1245 & 2.66 & 4 & 3 & 10.3\\ 
MSCC\,277-F1 & 10 & 384 & 0.1124 & 0.1031 & 0.1208 & 0.1268 & 2.99 & 21 & 11 & 89.3\\ 
MSCC\,277-F2 & 5 & 227 & 0.1053 & 0.0956 & 0.1135 & 0.0805 & 2.93 & 16 & 10 & 69.0\\ 
MSCC\,278-F1 & 7 & 956 & 0.0328 & 0.0251 & 0.0398 & 0.9871 & 2.41 & 36 & 14 & 56.3\\ 
MSCC\,278-F2 & 3 & 292 & 0.0322 & 0.0273 & 0.0361 & 0.4963 & 1.70 & 24 & 13 & 20.8\\ 
MSCC\,278-F3 & 4 & 145 & 0.0319 & 0.0266 & 0.0386 & 0.3810 & 1.18 & 14 & 7 & 24.4\\ 
MSCC\,278-F4 & 8 & 629 & 0.0254 & 0.0221 & 0.0298 & 1.0335 & 1.86 & 27 & 10 & 36.9\\ 
MSCC\,278-F5 & 3 & 122 & 0.0348 & 0.0297 & 0.0387 & 0.4401 & 1.76 & 7 & 6 & 11.1\\ 
MSCC\,283-F1 & 4 & 165 & 0.1339 & 0.1245 & 0.1470 & 0.0701 & 3.96 & 7 & 4 & 47.7\\ 
MSCC\,283-F2 & 5 & 94 & 0.1364 & 0.1284 & 0.1465 & 0.0655 & 3.26 & 6 & 5 & 49.1\\ 
MSCC\,283-F3 & 3 & 4 & 0.1357 & 0.1296 & 0.1496 & 0.0397 & 0.96 & 2 & 2 & 24.2\\ 
MSCC\,295-F1 & 7 & 1020 & 0.0230 & 0.0160 & 0.0284 & 2.1243 & 1.89 & 37 & 28 & 44.6\\ 
MSCC\,295-F2 & 7 & 1289 & 0.0228 & 0.0158 & 0.0299 & 2.2694 & 2.37 & 34 & 12 & 18.2\\ 
MSCC\,295-F3 & 4 & 497 & 0.0232 & 0.0189 & 0.0283 & 0.5824 & 1.36 & 44 & 17 & 26.5\\ 
MSCC\,295-F4 & 4 & 186 & 0.0218 & 0.0188 & 0.0255 & 0.6254 & 1.83 & 10 & 9 & 30.8\\ 
MSCC\,310-F1 & 10 & 499 & 0.0609 & 0.0518 & 0.0689 & 0.4180 & 2.81 & 18 & 11 & 61.6\\ 
MSCC\,310-F2 & 8 & 523 & 0.0502 & 0.0443 & 0.0588 & 0.5210 & 2.84 & 22 & 11 & 51.8\\ 
MSCC\,310-F3 & 7 & 407 & 0.0481 & 0.0427 & 0.0528 & 0.3486 & 2.50 & 19 & 10 & 49.0\\ 
MSCC\,310-F4 & 8 & 313 & 0.0656 & 0.0585 & 0.0710 & 0.4263 & 2.46 & 14 & 10 & 59.0\\ 
MSCC\,310-F5 & 7 & 325 & 0.0700 & 0.0642 & 0.0774 & 0.5303 & 2.60 & 13 & 7 & 47.6\\ 
MSCC\,310-F6 & 6 & 243 & 0.0725 & 0.0651 & 0.0791 & 0.2986 & 1.97 & 12 & 9 & 39.3\\ 
MSCC\,310-F7 & 4 & 219 & 0.0551 & 0.0479 & 0.0619 & 0.4062 & 2.76 & 7 & 5 & 20.7\\ 
MSCC\,310-F8 & 4 & 164 & 0.0546 & 0.0485 & 0.0617 & 0.2118 & 2.22 & 9 & 7 & 33.7\\ 
MSCC\,310-F9 & 4 & 124 & 0.0464 & 0.0437 & 0.0528 & 0.1464 & 1.13 & 9 & 6 & 17.9\\ 
MSCC\,311-F1 & 13 & 607 & 0.0813 & 0.0753 & 0.0892 & 0.2812 & 3.54 & 14 & 9 & 74.5\\ 
MSCC\,311-F2 & 5 & 254 & 0.0847 & 0.0769 & 0.0937 & 0.3235 & 3.69 & 4 & 3 & 23.5\\ 
MSCC\,311-F3 & 3 & 138 & 0.0888 & 0.0821 & 0.0937 & 0.1900 & 4.32 & 2 & 2 & 15.3\\ 
MSCC\,311-F4 & 3 & 119 & 0.0826 & 0.0746 & 0.0901 & 0.2447 & 3.58 & 2 & 2 & 19.5\\ 
MSCC\,314-F1 & 4 & 46 & 0.0819 & 0.0769 & 0.0893 & 0.1201 & 1.88 & 4 & 4 & 29.6\\ 
MSCC\,314-F2 & 3 & 66 & 0.0777 & 0.0715 & 0.0844 & 0.1610 & 2.78 & 2 & 2 & 12.3\\ 
MSCC\,317-F1 & 4 & 55 & 0.1331 & 0.1224 & 0.1417 & 0.0282 & 1.73 & 15 & 10 & 71.7\\ 
MSCC\,317-F2 & 4 & 8 & 0.1187 & 0.1102 & 0.1264 & 0.0528 & 0.69 & 10 & 7 & 75.5\\ 
MSCC\,323-F1 & 7 & 223 & 0.1380 & 0.1262 & 0.1540 & 0.0436 & 3.73 & 14 & 11 & 129.3\\ 
MSCC\,323-F2 & 3 & 16 & 0.1383 & 0.1309 & 0.1535 & 0.0261 & 1.66 & 5 & 4 & 29.4\\ 
MSCC\,333-F1 & 3 & 82 & 0.0760 & 0.0723 & 0.0833 & 0.1205 & 2.36 & 5 & 4 & 29.8\\ 
MSCC\,333-F2 & 4 & 117 & 0.0803 & 0.0752 & 0.0880 & 0.1519 & 2.83 & 5 & 4 & 24.5\\ 
MSCC\,333-F3 & 4 & 83 & 0.0794 & 0.0729 & 0.0835 & 0.2052 & 2.46 & 5 & 5 & 34.6\\ 
MSCC\,335-F1 & 10 & 345 & 0.0779 & 0.0705 & 0.0843 & 0.1260 & 2.60 & 21 & 11 & 105.0\\ 
MSCC\,335-F2 & 4 & 22 & 0.0665 & 0.0613 & 0.0731 & 0.0650 & 0.75 & 9 & 8 & 83.8\\ 
MSCC\,335-F3 & 4 & 111 & 0.0762 & 0.0712 & 0.0814 & 0.1113 & 2.45 & 6 & 4 & 45.9\\ 
MSCC\,343-F1 & 4 & 115 & 0.0822 & 0.0786 & 0.0874 & 0.1394 & 2.51 & 7 & 6 & 28.7\\ 
MSCC\,343-F2 & 3 & 95 & 0.0798 & 0.0733 & 0.0876 & 0.2638 & 2.50 & 3 & 3 & 19.3\\ 
MSCC\,343-F3 & 3 & 125 & 0.0824 & 0.0738 & 0.0876 & 0.1991 & 2.65 & 6 & 5 & 15.6\\ 
MSCC\,360-F1 & 4 & 182 & 0.1050 & 0.0967 & 0.1159 & 0.1367 & 2.95 & 9 & 6 & 60.8\\ 
MSCC\,360-F2 & 3 & 31 & 0.1074 & 0.1004 & 0.1152 & 0.0870 & 1.52 & 6 & 4 & 49.3\\ 
MSCC\,360-F3 & 3 & 5 & 0.1033 & 0.0973 & 0.1093 & 0.0416 & 1.46 & 3 & 2 & 29.9\\ 
MSCC\,386-F1 & 7 & 346 & 0.0734 & 0.0657 & 0.0805 & 0.4163 & 2.91 & 18 & 10 & 34.0\\ 
MSCC\,386-F2 & 4 & 115 & 0.0708 & 0.0636 & 0.0765 & 0.1969 & 1.92 & 10 & 8 & 39.6\\ 
MSCC\,386-F3 & 5 & 129 & 0.0627 & 0.0591 & 0.0705 & 0.1760 & 1.56 & 13 & 10 & 43.2\\ 
MSCC\,386-F4 & 4 & 46 & 0.0617 & 0.0594 & 0.0664 & 0.1703 & 0.92 & 7 & 6 & 21.3\\ 
MSCC\,407-F1 & 5 & 101 & 0.1388 & 0.1254 & 0.1468 & 0.0177 & 3.61 & 10 & 6 & 59.2\\ 
MSCC\,414-F1 & 4 & 420 & 0.0626 & 0.0537 & 0.0691 & 0.6269 & 2.51 & 19 & 9 & 38.1\\ 
MSCC\,414-F2 & 6 & 243 & 0.0628 & 0.0548 & 0.0667 & 0.3793 & 2.54 & 12 & 7 & 34.0\\ 
MSCC\,414-F3 & 6 & 240 & 0.0750 & 0.0688 & 0.0809 & 0.5198 & 2.53 & 12 & 7 & 27.1\\ 
MSCC\,414-F4 & 5 & 196 & 0.0641 & 0.0585 & 0.0675 & 0.3705 & 2.52 & 10 & 5 & 13.5\\ 
MSCC\,414-F5 & 4 & 243 & 0.0748 & 0.0684 & 0.0810 & 0.4251 & 2.53 & 10 & 8 & 22.1\\ 
MSCC\,414-F6 & 3 & 126 & 0.0657 & 0.0613 & 0.0701 & 0.3368 & 2.01 & 7 & 6 & 22.9\\ 
MSCC\,414-F7 & 3 & 118 & 0.0616 & 0.0576 & 0.0647 & 0.2329 & 1.97 & 10 & 5 & 14.1\\ 
MSCC\,414-F8 & 3 & 72 & 0.0657 & 0.0603 & 0.0722 & 0.1197 & 1.82 & 7 & 6 & 29.9\\ 
MSCC\,414-F9 & 4 & 116 & 0.0663 & 0.0594 & 0.0726 & 0.3001 & 2.50 & 5 & 5 & 22.9\\ 
MSCC\,414-F10 & 4 & 107 & 0.0546 & 0.0518 & 0.0590 & 0.4163 & 1.99 & 6 & 6 & 29.9\\ 
MSCC\,414-F11 & 5 & 147 & 0.0616 & 0.0589 & 0.0657 & 0.1262 & 1.64 & 14 & 8 & 35.1\\ 
MSCC\,414-F12 & 3 & 14 & 0.0762 & 0.0696 & 0.0801 & 0.1126 & 0.66 & 4 & 4 & 9.8\\ 
MSCC\,414-F13 & 4 & 66 & 0.0640 & 0.0614 & 0.0674 & 0.1692 & 1.94 & 4 & 4 & 19.2\\ 
MSCC\,414-F14 & 4 & 93 & 0.0727 & 0.0686 & 0.0758 & 0.4275 & 1.78 & 7 & 6 & 23.6\\ 
MSCC\,414-F15 & 3 & 31 & 0.0607 & 0.0563 & 0.0640 & 0.1070 & 1.80 & 3 & 3 & 9.0\\ 
MSCC\,419-F1 & 3 & 130 & 0.1139 & 0.1084 & 0.1210 & 0.1849 & 4.03 & 2 & 2 & 12.8\\ 
MSCC\,419-F2 & 3 & 88 & 0.1099 & 0.1044 & 0.1182 & 0.1816 & 3.20 & 2 & 2 & 18.1\\ 
MSCC\,419-F3 & 4 & 36 & 0.1127 & 0.1080 & 0.1191 & 0.0693 & 1.79 & 5 & 4 & 25.2\\ 
MSCC\,422-F1 & 3 & 11 & 0.1424 & 0.1321 & 0.1528 & 0.0291 & 0.95 & 8 & 6 & 45.0\\ 
MSCC\,430-F1 & 4 & 151 & 0.0975 & 0.0880 & 0.1066 & 0.1391 & 2.78 & 7 & 4 & 39.2\\ 
MSCC\,430-F2 & 4 & 23 & 0.0937 & 0.0867 & 0.0999 & 0.1684 & 1.50 & 5 & 4 & 41.1\\ 
MSCC\,430-F3 & 3 & 12 & 0.0942 & 0.0891 & 0.1015 & 0.0691 & 1.89 & 2 & 2 & 21.2\\ 
MSCC\,430-F4 & 3 & 0 & 0.1024 & 0.0973 & 0.1041 & 0.0225 & 0.00 & 2 & 2 & 18.3\\ 
MSCC\,440-F1 & 6 & 184 & 0.1173 & 0.1074 & 0.1284 & 0.1550 & 2.93 & 8 & 6 & 50.2\\ 
MSCC\,454-F1 & 13 & 687 & 0.0389 & 0.0334 & 0.0466 & 2.7338 & 2.39 & 23 & 13 & 63.5\\ 
MSCC\,454-F2 & 6 & 498 & 0.0446 & 0.0377 & 0.0500 & 2.3032 & 2.57 & 20 & 9 & 29.1\\ 
MSCC\,454-F3 & 3 & 146 & 0.0520 & 0.0464 & 0.0557 & 4.0214 & 2.48 & 3 & 3 & 9.3\\ 
MSCC\,454-F4 & 3 & 146 & 0.0490 & 0.0425 & 0.0551 & 1.5093 & 2.42 & 3 & 3 & 12.8\\ 
MSCC\,454-F5 & 3 & 39 & 0.0438 & 0.0424 & 0.0460 & 1.1194 & 0.83 & 3 & 3 & 8.6\\ 
MSCC\,457-F1 & 19 & 908 & 0.0786 & 0.0709 & 0.0877 & 0.4021 & 3.37 & 23 & 11 & 78.4\\ 
MSCC\,457-F2 & 8 & 256 & 0.0758 & 0.0673 & 0.0830 & 0.1594 & 2.69 & 12 & 8 & 52.5\\ 
MSCC\,457-F3 & 6 & 179 & 0.0846 & 0.0768 & 0.0896 & 0.2454 & 2.48 & 7 & 5 & 30.6\\ 
MSCC\,457-F4 & 3 & 92 & 0.0829 & 0.0757 & 0.0881 & 0.1691 & 2.80 & 3 & 3 & 22.2\\ 
MSCC\,457-F5 & 3 & 77 & 0.0756 & 0.0676 & 0.0785 & 0.0982 & 1.97 & 3 & 3 & 16.8\\ 
MSCC\,457-F6 & 3 & 13 & 0.0816 & 0.0779 & 0.0872 & 0.0649 & 0.74 & 3 & 3 & 23.3\\ 
MSCC\,460-F1 & 12 & 536 & 0.1142 & 0.1067 & 0.1242 & 0.1862 & 3.61 & 17 & 9 & 69.3\\ 
MSCC\,460-F2 & 6 & 214 & 0.1139 & 0.1043 & 0.1237 & 0.1195 & 3.62 & 9 & 6 & 57.9\\ 
MSCC\,460-F3 & 6 & 82 & 0.1080 & 0.1003 & 0.1159 & 0.0695 & 2.55 & 6 & 5 & 59.7\\ 
MSCC\,460-F4 & 5 & 63 & 0.1218 & 0.1155 & 0.1304 & 0.0522 & 2.51 & 7 & 5 & 27.6\\ 
MSCC\,463-F1 & 9 & 489 & 0.0722 & 0.0652 & 0.0788 & 0.3732 & 2.77 & 16 & 10 & 57.4\\ 
MSCC\,463-F2 & 6 & 352 & 0.0765 & 0.0699 & 0.0839 & 0.3129 & 2.67 & 15 & 10 & 39.0\\ 
MSCC\,463-F3 & 11 & 313 & 0.0768 & 0.0696 & 0.0842 & 0.2787 & 2.27 & 13 & 11 & 73.6\\ 
MSCC\,463-F4 & 3 & 234 & 0.0657 & 0.0585 & 0.0743 & 0.4589 & 2.77 & 5 & 4 & 20.8\\ 
MSCC\,463-F5 & 5 & 192 & 0.0655 & 0.0600 & 0.0730 & 0.4039 & 2.08 & 10 & 8 & 41.9\\ 
MSCC\,463-F6 & 7 & 233 & 0.0837 & 0.0775 & 0.0894 & 0.2868 & 2.56 & 7 & 6 & 36.9\\ 
MSCC\,463-F7 & 4 & 8 & 0.0688 & 0.0654 & 0.0739 & 0.0878 & 0.68 & 8 & 7 & 26.8\\ 
MSCC\,463-F8 & 4 & 119 & 0.0663 & 0.0628 & 0.0698 & 0.4765 & 2.59 & 6 & 4 & 12.5\\ 
MSCC\,463-F9 & 4 & 99 & 0.0706 & 0.0660 & 0.0764 & 0.2416 & 1.87 & 6 & 5 & 31.9\\ 
MSCC\,463-F10 & 3 & 83 & 0.0701 & 0.0651 & 0.0754 & 0.3595 & 2.05 & 4 & 3 & 12.7\\ 
MSCC\,463-F11 & 3 & 106 & 0.0802 & 0.0748 & 0.0861 & 0.1864 & 2.64 & 4 & 4 & 14.8\\ 
MSCC\,474-F1 & 3 & 881 & 0.0369 & 0.0300 & 0.0444 & 2.7548 & 2.63 & 13 & 5 & 10.5\\ 
MSCC\,474-F2 & 5 & 407 & 0.0340 & 0.0296 & 0.0387 & 1.3043 & 2.06 & 13 & 7 & 23.4\\ 
MSCC\,474-F3 & 3 & 147 & 0.0368 & 0.0318 & 0.0412 & 0.2423 & 1.98 & 9 & 8 & 16.5\\ 
MSCC\,474-F4 & 3 & 176 & 0.0321 & 0.0284 & 0.0388 & 0.7372 & 1.75 & 7 & 5 & 16.5\\ 
MSCC\,474-F5 & 3 & 114 & 0.0337 & 0.0292 & 0.0372 & 0.3332 & 1.51 & 10 & 6 & 17.2\\ 
MSCC\,474-F6 & 4 & 193 & 0.0365 & 0.0324 & 0.0423 & 0.5697 & 1.90 & 6 & 6 & 15.3\\ 
MSCC\,474-F7 & 3 & 0 & 0.0379 & 0.0328 & 0.0406 & 0.1777 & 0.00 & 10 & 7 & 28.3\\ 
MSCC\,484-F1 & 4 & 109 & 0.1361 & 0.1243 & 0.1501 & 0.0130 & 3.89 & 7 & 6 & 54.6\\  
	\hline
	 
\end{longtable}
\tablefoot{\\
(1) ID of the filament; \\
(2) number of systems detected by GSyF linked by the filament; \\
(3) number of galaxies attributed to the filament; \\
(4) to (6) mean, min. and max. redshift of the filament; \\
(7) mean galaxy number density inside the filament; \\
(8) mean transversal radius of the filament measured at $10 \times d$; \\
(9) and (10) number of nodes that constitute the filament and its 
central skeleton; \\
(11) length of the filament skeleton.
}
	

\end{document}